\documentclass[a4paper]{IEEEtran}

\usepackage{amsmath,amssymb,cite,dsfont}
\usepackage{mathtools}	
\usepackage{pict2e}	
\usepackage{curve2e}	
\usepackage{graphicx}
\usepackage{stmaryrd}	
\usepackage{multirow}	
\usepackage{booktabs}	
\usepackage{stfloats}	
\usepackage{pgfplots}
\usepackage[caption=false,font=footnotesize]{subfig}

\usepackage{tikz}
\usetikzlibrary{positioning,calc,fit,shapes.geometric}

\definecolor{light-gray}{gray}{0.75}
\definecolor{dark-green}{RGB}{0,99,0}

\usepackage{hyperref}
\hypersetup{
    colorlinks,%
    citecolor=black,%
    filecolor=black,%
    linkcolor=black,%
    urlcolor=black,%
    pdfauthor={German Bassi},%
    pdftitle={Capacity Bounds for a Class of Interference Relay Channels}%
}

\newcommand{\imgdir}{img}
\graphicspath{{\imgdir/}}

\newtheorem{theorem}{Theorem}
\newtheorem{lemma}{Lemma}
\newtheorem{proposition}{Proposition}
\newtheorem{corollary}{Corollary}
\newtheorem{remark}{Remark}
\newtheorem{definition}{Definition}

\newcommand{\ve}[1]{\ensuremath{\underline{#1}}}

\newcommand{\I}[2]{\ensuremath{I(#1;#2)}}
\newcommand{\IC}[3]{\ensuremath{I(#1;#2\vert #3)}}
\renewcommand{\H}[1]{\ensuremath{h(#1)}}
\newcommand{\HC}[2]{\ensuremath{h(#1\vert #2)}}
\newcommand{\C}[1]{\ensuremath{\mathsf{C}\!\left[ #1 \right]}}
\newcommand{\SNR}[1]{\ensuremath{S_{#1}}}

\newcommand{\mkv}{-\!\!\!\!\minuso\!\!\!\!-}
\newcommand{\cond}{\,\vert\,}

\newcommand{\cX}{{\mathcal X}}
\newcommand{\cY}{{\mathcal Y}}

\newcommand{\bE}{{\mathbb E}}
\newcommand{\bN}{{\mathbb N}}

\newcommand{\norm}[1]{\left\lVert#1\right\rVert}
\newcommand{\abs}[1]{\left\lvert#1\right\rvert}

\newcommand{\typ}[2]{T_\delta^{#1}(#2)}
\newcommand{\typc}[2]{T_{\delta'}^{#1}(#2)}

\newcommand{\toas}[1]{\xrightarrow[#1]{}}

\begin{document}

\sloppy

\title{Capacity Bounds for a Class of Interference\\ Relay Channels}

\author{
  \IEEEauthorblockN{Germ{\'a}n~Bassi, Pablo~Piantanida and Sheng~Yang}\\
\thanks{This work was partially supported by the ANR grant (FIREFLIES) INTB 0302
01, and the Celtic European project SHARING.
The material in this paper was presented in part at the 51st Annual Allerton
Conference on Communication, Control, and Computing, Oct. 2013, and at the 2014
IEEE International Symposium on Information Theory, Jun. 2014.}
\thanks{The authors are with the Laboratoire des Signaux et Syst{\`e}mes (L2S,
UMR8506) CNRS-CentraleSup{\'e}lec-Universit{\'e} Paris Sud, 91192
Gif-sur-Yvette, France
(e-mail: german.bassi@centralesupelec.fr, {pablo.piantanida}
@centralesupelec.fr, sheng.yang@centralesupelec.fr).}
}

\maketitle

\begin{abstract}
The capacity of a class of Interference Relay Channels~(IRC) --the Injective
Semideterministic IRC where the relay can only observe one of the sources-- is
investigated.
We first derive a novel outer bound and two inner bounds which are based on a
careful use of each of the available cooperative strategies together with the
adequate interference decoding technique.
The outer bound extends Telatar and Tse's work while the inner bounds contain
several known results in the literature as special cases.
Our main result is the characterization of the capacity region of the Gaussian
class of IRCs studied within a fixed number of bits per dimension --constant
gap.
The proof relies on the use of the different cooperative strategies in specific
SNR regimes due to the complexity of the schemes.
As a matter of fact, this issue reveals the complex nature of the Gaussian IRC
where the combination of a single coding scheme for the Gaussian relay and
interference channel may not lead to a good coding scheme for this problem,
even when the focus is only on capacity to within a constant gap over all
possible fading statistics.
\end{abstract}


\section{Introduction}

\PARstart{C}{ellular} networks have reached practical limits in many dense urban
areas while data traffic and the number of users seem to be continuously
increasing. Interference has become one of the most crucial problems in cellular
networks where users must compete for the available resources, e.g., an
improvement in terms of data rate for one of them may  be detrimental to the
performance of another user. Although the existence of a large amount of users
in cellular networks has driven communication channels from being noise-limited
to interference-limited, it can also be exploited to boost the overall network
throughput by means of user cooperation. 

In order to provision a new communication infrastructure, network operators are
rethinking conventional cellular system topologies to consider a new paradigm
called heterogeneous networks.  This consists of planned macro base station (BS)
deployments that typically transmit at high power overlaid with several low
power nodes such as: relay and pico BSs, distributed antennas, and femto BSs.
These lower power nodes are deployed to further increase the coverage of the
network, especially when terminals are far away from the macro BS. Fixed relays
are infrastructure equipment that connect wirelessly to the BS and these relays
aid in the signal transmission between the macro BS and the mobile users by
receiving and retransmitting messages. Indeed, these relays may offer a flexible
option where backhauls are not available. In order to assess the benefits of
this strategy, an information-theoretic analysis of cooperation through relaying
in interference-limited environments should be carried out. Nonetheless, each
one of these two fundamental problems --relaying and interference-- appears to
be rather involved and unfortunately only partial results are available in the
literature.

\subsection{Related Work}

Perhaps the simplest model of a communication network with interference is the
Interference Channel~(IC), whose capacity region --even without a relay-- is
still an open problem. The largest known  achievable rate region is due to Han
and Kobayashi~\cite{han_new_1981} and it is based on the idea of interference
decoding via ``rate-splitting'' at the sources, also referred to as
``Han-Kobayashi scheme''. This scheme has been shown by
Etkin-Tse-Wang~\cite{etkin_gaussian_2008} to achieve within $1$ bit per complex
dimension to the capacity region of the Gaussian IC. The important feature
behind the notion of ``constant gap'' is that it guarantees an uniform gap
between the inner and the outer bound over all channel coefficients and hence
all possible fading statistics. This result hinges on a new upper-bounding
technique that has been later on extended to a more general class of
ICs~\cite{telatar_bounds_2007}, also referred to as ``Injective
Semideterministic IC''~\cite{gamal_network_2011}. 

Another challenging problem is the Relay Channel~(RC), where a relay node helps
the communication between a source-destination pair. Since the seminal work of
Cover and El Gamal~\cite{cover_capacity_1979}, which has introduced the main
cooperative strategies of ``decode-and-forward''~(DF) and
``compress-and-forward''~(CF), there has been a great deal of research on this
topic. Although the capacity of the RC is still unknown in general, the benefits
of cooperation by relaying are rather clear by now, at least in the context of
single source and/or single destination relay
networks~\cite{kramer_cooperative_2005}. An approximation approach to general
networks via deterministic channels was introduced by
Avestimehr-Diggavi-Tse~\cite{avestimehr_wireless_2011}. This approach yields a
novel improvement over CF scheme --referred to as ``quantize-map-and-forward"
(QMF)-- that achieves capacity to within a constant gap for unicast additive
white Gaussian noise (AWGN) networks with an arbitrary number of relays. As a
matter of fact, both DF and CF schemes can perform within the same constant gap
to the capacity of the Gaussian RC, regardless of the channel
parameters~\cite{avestimehr_wireless_2011,chang_gaussian_2010} and thus of the
fading statistics. More recently, Lim~\emph{et al.}~\cite{lim_noisy_2011}
generalized the QMF approach to arbitrary memoryless multicast networks via the
``noisy network coding" (NNC) scheme. Relay nodes based on NNC scheme send the
same --long-- message over many blocks of equal length and the descriptions at
the relays do not require binning while their indices are non-uniquely decoded
at the destination. 

In wireless networks with multiple source nodes that communicate simultaneously
to several destinations, ``interference'' becomes the central issue, and the
different roles that relays can play to enhance the reliability in such
scenarios are not well understood yet. In this paper, we consider the simplest
scenario where interference and relaying appear together, that is the
Interference Relay Channel~(IRC). The problem itself is not
new~\cite{sahin_achievable_2007} and the research on this topic has been growing
during the past years. In~\cite{tian_gaussian_2011}, among other works, the
authors proposed inner bounds on the capacity region of the IRC based on the
standard CF scheme while DF-based schemes are also studied
in~\cite{chaaban_generalized_2012}. It is worth mentioning here that these
coding schemes do not use ``joint decoding'' at the destination to recover all
transmit messages and the compression indices. The idea of NNC was later on
extended to the IRC in~\cite{kang_new_2013} by adding rate-splitting. Besides
these works, capacity of the \emph{physically degraded} IRC in the \emph{strong
interference} regime was determined in~\cite{maric_relaying_2012} by assuming
that the relay node can only observe one of the two source encoders. Several
variations of this problem have also been investigated, e.g., the cognitive IRC
where the relay has non-casual knowledge of the sources' messages was treated
in~\cite{sahin_cancelation_2007,rini_outer_2010}. Additionally, the IRC with an
``out-of-band relay'', i.e., the relay operates over an orthogonal band with
respect to the underlying IC, was also studied in~\cite{sahin_interference_2011,
tian_symmetric_2012, razaghi_birds_2013, zhou_incremental_2012,
simeone_codebook_2011}. Capacity results were
obtained in~\cite{simeone_codebook_2011} for an IRC with oblivious relaying  in
which the relay is unaware of the codebook used by the source encoders.

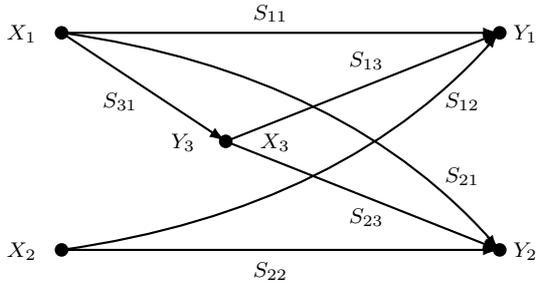
\begin{figure}[t]
\centering
\footnotesize
\setlength{\unitlength}{1.8mm}
\begin{picture}(40,20)
\thicklines
\put(0,17.5){$X_1$}
\put(4,18){\circle*{1}\vector(1,0){32}\vector(3,-2){12}}
\qbezier(4,18)(25,15)(36,2)
\put(35.1,3){\vector(.8,-1){.85}} 
\put(0,1.5){$X_2$}
\put(4,2){\circle*{1}\vector(1,0){32}}
\qbezier(4,2)(25,5)(36,18)
\put(35.1,17){\vector(.8,1){.85}} 
\put(12,9.5){$Y_3$}
\put(18.5,9.5){$X_3$}
\put(16,10){\circle*{1}\vector(5,2){20}\vector(5,-2){20}}
\put(37,17.5){$Y_1$}
\put(36,18){\circle*{1}}
\put(37,1.5){$Y_2$}
\put(36,2){\circle*{1}}
\put(7,12.5){\SNR{31}}
%
\put(18,19){\SNR{11}}
\put(25,15.5){\SNR{13}}
\put(32,12.5){\SNR{12}}
\put(18,0){\SNR{22}}
\put(25,4){\SNR{23}}
\put(32,7){\SNR{21}}
\end{picture}
\caption{The Gaussian IRC where the values \SNR{ij} represent the SNR
between nodes $j$ and $i$. \label{fig:IN-Gaussian_model}}
\end{figure}

The interference channel with cooperation at either the transmitter or receiver
end, or both has also been investigated. In the extreme regimes where the relay
can be thought of being collocated with the transmitters or the receivers, the
IRC becomes a virtual multi-antenna  IC with transmitter or receiver
cooperation. The benefits of such a system have been studied
in~\cite{host-madsen_capacity_2006}. Additionally, constant-gap results
regardless of channel conditions were provided
in~\cite{prabhakaran_interf_source_2011, prabhakaran_interf_dest_2011,
wang_interf_trans_2011, wang_interf_receiver_2011}, while capacity results in
strong interference regime were determined in~\cite{maric_capacity_2007} for the
case of transmitter cooperation. Recently, in the case of unilateral source
cooperation, improved outer bounds were reported in~\cite{cardone_new_2014}.

\subsection{Contribution and Outline}

In this paper we focus on a simplified version of the two-user
IRC~\cite{bassi_allerton_2013} which still captures the rather complex interplay
between interference and relaying. This is the two-user IC with a relay node
which can only observe one of the source encoders. Although this is not the most
general two-user IRC, we shall see that it still captures the central issue of
interference and relaying and hence, we seek to provide some useful insights
into the understanding  of this complex problem. In particular, for the class of
Gaussian IRCs shown in Fig.~\ref{fig:IN-Gaussian_model}, we aim at determining
the underlying SNR regimes together with the adequate coding schemes and
decoding technique that are needed to achieve capacity to within a
\emph{constant gap}.

Our results involve a novel outer bound for the considered class of IRCs --the
Injective Semideterministic IRC-- and two inner bounds based on rate-splitting
and different relaying strategies (building on DF and CF schemes) with the
adequate interference decoding technique. Although the use of DF and CF schemes
in the context of the IRC is not new~\cite{sahin_achievable_2007,
tian_gaussian_2011, chaaban_generalized_2012, kang_new_2013,
maric_relaying_2012}, our aim is to provide a set of simple but powerful enough
strategies in order to characterize the capacity region of Gaussian IRCs to
within a constant gap, as previously stated.
In this regard, our main contributions with respect to the literature are the
introduction of \emph{partial} DF, where the relay forwards only part of the
source's message, and the use of different decoding strategies in the CF scheme
which helps us obtain a compact expression of the inner bound.

The main outcome of this work is the characterization within a constant gap of
the capacity of the aforementioned Gaussian IRC.
We show that, for any channel realization, at least one of the proposed schemes 
achieves the capacity region to within a constant gap.
More precisely, it is shown that when the source-to-relay channel is stronger 
than the source-to-destination channel full DF scheme is recommended (this 
regime includes the capacity result in~\cite[Thm. 3]{maric_relaying_2012}). As 
the strength of the source-to-relay channel reduces, it is preferable to 
partially decode the message and thus partial DF scheme is required. Finally, 
when the source-to-relay channel is weaker than the interfering channel from the 
source to the other destination, CF scheme together with different ways of 
decoding is needed instead.

This paper is organized as follows. Section~\ref{sec:Problem} presents the
problem definition while the outer bound and the two inner bounds are deferred
to Sections~\ref{sec:Outer} and~\ref{sec:Inner}, respectively. The constant
gap results are shown in Section~\ref{sec:Gap}. Finally, all proofs are
relegated to the appendices.

\subsection*{Notation and Conventions}

Given two integers $i$ and $j$, the expression $[i: j]$ denotes the set $\{i,
i+1, \ldots, j\}$, whereas for real values $a$ and $b$, $[a, b]$ denotes the
closed interval between $a$ and $b$.
Lowercase letters such as $x$ and $y$ are mainly used to represent realizations
of random variables, whereas capital letters such as $X$ and $Y$ stand for the
random variables in itself. Bold capital letters such as $\pmb{H}$ and
$\pmb{Q}$ represent matrices, while calligraphic letters such as $\mathcal{X}$
and $\mathcal{Y}$ are reserved for sets.
The probability distribution~(PD) of the random vector $X^n$, $p_{X^n}(x^n)$, is
succinctly written as $p(x^n)$ without subscript when it can be understood from
the argument $x^n$.
Given three random variables $X$, $Y$, and $Z$, if its joint PD can be
decomposed as $p(xyz) = p(y) p(x\vert y) p(z\vert y)$, then they form a Markov
chain, denoted by $X \mkv Y \mkv Z$.
Differential entropy is denoted by $\H{\cdot}$ and the mutual information,
$\I{\cdot}{\cdot}$. The expression $\C{x}= \frac{1}{2}\log_2(1+x)$ stands for
the capacity of a Gaussian channel with SNR of value $x$.
Definitions and properties of strongly typical sequences and delta-convention
are provided in Appendix~\ref{app:typical}.

\section{Problem Definition}
\label{sec:Problem}

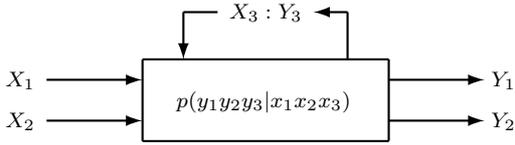
\begin{figure}[t!]
\centering
\footnotesize
\setlength{\unitlength}{1.8mm}
\begin{picture}(38,12)
\thicklines
\put(0,5){$X_1$}
\put(3,5.5){\vector(1,0){7}}
\put(0,2){$X_2$}
\put(3,2.5){\vector(1,0){7}}
\put(12.5,3.5){$p(y_1 y_2 y_3\vert x_1 x_2 x_3)$}
\put(10,1){\line(0,1){6}\line(1,0){18}}
\put(28,7){\line(-1,0){18}\line(0,-1){6}}
\put(28,5.5){\vector(1,0){7}}
\put(35.5,5){$Y_1$}
\put(28,2.5){\vector(1,0){7}}
\put(35.5,2){$Y_2$}
\put(16.35,10){$X_3 : Y_3$}
\put(25,10.5){\line(0,-1){3.5}\vector(-1,0){2.5}}
\put(13,10.5){\line(1,0){2.5}\vector(0,-1){3.5}}
\end{picture}
\caption{Interference Relay Channel (IRC) model. \label{fig:PD-IRC_model}}
\end{figure}

The IRC consists of two source encoders, two destinations and one relay node.
Encoder~$k$ wishes to send a message~$\tilde{m}_k\in\mathcal{\tilde{M}}_{n,k}
\triangleq \left\{ 1,\ldots,M_{n,k} \right\}$ to destination~$k$, $k\in\{1,2\}$,
with the help of the relay. The IRC, depicted in Fig.~\ref{fig:PD-IRC_model}, is
modeled as a memoryless channel without feedback defined by a conditional
probability distribution (PD):
\begin{equation*}
p(y_1,y_2 ,y_3\vert x_1, x_2 ,x_3):\, \mathcal{X}_1\times \mathcal{X}_2\times
\mathcal{X}_3 \longmapsto \mathcal{Y}_1\times \mathcal{Y}_2\times\mathcal{Y}_3
\end{equation*}
where $x_k\in\mathcal{X}_k$ and $y_k \in\mathcal{Y}_k$, $k\in\{1,2\}$, are the
input at source~$k$ and output at destination~$k$, respectively, whereas $x_3\in
\mathcal{X}_3$ and $y_3\in\mathcal{Y}_3$ are the input and output at the relay,
respectively. The relaying functions are defined as a sequence of mappings
$\left\{ \phi_{i} : \mathcal{Y}_3^{i-1} \mapsto \mathcal{X}_3 \right\}_{i=1}^n$.

As it was previously stated, throughout the paper we deal with a specific type
of IRC in which only one of the sources is connected to the relay, i.e.,
\begin{equation} 
p(y_1, y_2, y_3\vert x_1, x_2, x_3) = p(y_3\vert x_1, x_3) p(y_1, y_2\vert
x_1, x_2, x_3, y_3). \label{eq:PD-general_pdf}
\end{equation}
Unless it is noted otherwise, this is a basic assumption of our model.

We also recall that a pair of rates $(R_1,R_2)$ is said to be achievable for an
IRC if for every $\epsilon>0$ there exists a block length $n$ and encoders
$\textrm{enc}_k : \mathcal{\tilde{M}}_{n,k} \mapsto \mathcal{X}_k^n$, $M_{n,k} \ge
2^{n(R_k-\epsilon)}$, $k\in\{1,2\}$, and decoder $\textrm{dec}_k :
\mathcal{Y}_k^n \mapsto \mathcal{\tilde{M}}_{n,k}$, $k\in\{1,2\}$, such that 
\begin{align*}
\frac{1}{M_{n,1} M_{n,2}} \sum_{\tilde{m}_1, \tilde{m}_2} \mathbb{P}\big\{ 
  &\big( \textrm{dec}_1(Y_1^n), \textrm{dec}_2(Y_2^n)\big) \neq (\tilde{m}_1,
\tilde{m}_2) \cond \\
& X_1^n = \textrm{enc}_1(\tilde{m}_1), X_2^n = \textrm{enc}_2(\tilde{m}_2)
\big\} \le \epsilon.
\end{align*}

\begin{figure}[t!]
\centering
\footnotesize
\setlength{\unitlength}{1.8mm}
\begin{picture}(42,20)
\thicklines
\put(0,18){$(X_1 X_3)$}
\put(6,18.5){\vector(1,0){23}}
\put(10,18.5){\circle*{.5}\line(0,-1){12}}
\put(10,6.5){\vector(1,0){3}}
\put(2,1){$X_2$}
\put(6,1.5){\vector(1,0){23}}
\put(9,1.5){\circle*{.5}\line(0,1){12}}
\put(9,13.5){\vector(1,0){4}}
\put(38.5,17){$Y_1$}
\put(37.5,2){$(Y_2 Y_3)$}
\put(13,12){\line(1,0){9}\line(0,1){3}}
\put(22,15){\line(-1,0){9}\line(0,-1){3}}
\put(14.6,13){$p(s_2\vert x_2)$}
\put(26.5,13.5){\line(-1,0){4.5}\line(0,1){3}}
\put(26.5,16.5){\vector(1,0){2.5}}
\put(23.5,14.5){$S_2$}
\put(29,19){\line(1,0){4}\line(0,-1){3}}
\put(33,16){\line(-1,0){4}\line(0,1){3}}
\put(30,17){$f_1$}
\put(33,17.5){\vector(1,0){4}}
\put(13,8){\line(1,0){9}\line(0,-1){3}}
\put(22,5){\line(-1,0){9}\line(0,1){3}}
\put(13.6,6){$p(\ve{s_1}\vert x_1x_3)$}
\put(26.5,6.5){\line(-1,0){4.5}\line(0,-1){3}}
\put(26.5,3.5){\vector(1,0){2.5}}
\put(23.5,4.6){$\ve{S_1}$}
\put(29,1){\line(1,0){4}\line(0,1){3}}
\put(33,4){\line(-1,0){4}\line(0,-1){3}}
\put(30,2){$f_2$}
\put(33,2.5){\vector(1,0){4}}
\end{picture}
\caption{Injective Semideterministic IRC (IS-IRC) model.
\label{fig:PD-IS_IRC_model}}
\end{figure}
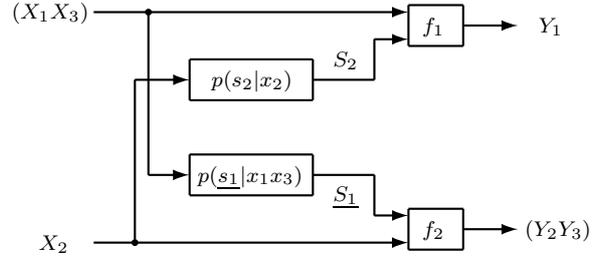

\begin{definition}[Injective Semideterministic IRC]
In this paper, we shall focus on the class of IRCs referred to as the Injective
Semideterministic IRC (IS-IRC), as shown in Fig.~\ref{fig:PD-IS_IRC_model},
which is an extension of that introduced in~\cite{telatar_bounds_2007} for the
IC. In this model, the randomness of the channel is captured by the interference
signals $S_1$, $S_2$ and $S_3$. For sake of clarity, we will denote the pair
$(S_1 S_3)$ as the vector $\ve{S_1}$.

The conditional PD of the interference signals may be decomposed as follows,
$p(\ve{s_1} s_2 \vert x_1 x_2 x_3) = p(\ve{s_1}\vert x_1 x_3) p(s_2\vert x_2)$,
and the outputs of the channel are deterministic functions of $(X_1, X_2, X_3,
\ve{S_1}, S_2)$. Specifically, we have that $Y_1 = f_1(X_1, X_3, S_2)$, $Y_2 =
f_2'(X_2, S_1)$, and $(Y_2 Y_3) = f_2(X_2, \ve{S_1})$, where $f_1$, $f_2'$, and
$f_2$ are functions that, for every $(x_1, x_2, x_3)$,
\begin{align*}
f_1(x_1, x_3, \cdot\,) &:\ \mathcal{S}_2 \to \mathcal{Y}_1,  \quad s_2 \mapsto
  f_1(x_1, x_3, s_2), \\
f_2'(x_2, \cdot\,) &:\ \mathcal{S}_1 \to \mathcal{Y}_2, \quad s_1 \mapsto
f_2'(x_2, s_1), \\
f_2(x_2, \cdot\,) &:\ \ve{\mathcal{S}_1} \to \mathcal{Y}_2\times\mathcal{Y}_3,
 \quad \ve{s_1} \mapsto f_2(x_2, \ve{s_1})
\end{align*}%
are invertible.
\end{definition}

\vspace*{1mm}
\begin{remark}
Since the relay only observes the first source, its input $X_3$ cannot depend on
$X_2$. Therefore, $X_3$ is regarded as desired signal at $Y_1$ and as
interference at $Y_2$, which motivates us to model this class of IRCs as
depicted in Fig.~\ref{fig:PD-IS_IRC_model}. It comes as no surprise that the
pair $(X_1 X_3)$ should be taken as a whole. However, as it is shown later in
the derivation of the outer bound, it is also convenient to put the pair $(Y_2
Y_3)$ together.
\end{remark}
\vspace*{1mm}


A special case of the IS-IRC is the real Gaussian model, as it is shown in
Fig.~\ref{fig:IN-Gaussian_model}, and defined by
\begin{subequations}\label{eq:PD-Gaussian_case}
\begin{align}
Y_1 &= h_{11} X_1 + h_{12} X_2 + h_{13} X_3 + Z_1, \\
Y_2 &= h_{21} X_1 + h_{22} X_2 + h_{23} X_3 + Z_2, \\
Y_3 &= h_{31} X_1 + Z_3,
\end{align}
\end{subequations}%
where each noise process $Z_k \sim \mathcal{N}(0,N_k)$, $k \in \{1,2,3\}$, is
independent of each other, and each input has an average power constraint
$\bE[|X_k|^2]\leq P_k$, $k \in \{1,2,3\}$.
The link between node $j$ and $i$ has a fixed channel coefficient $h_{ij}$, and
the SNR associated to it is denoted $\SNR{ij} \triangleq |h_{ij}|^2 P_j /N_i$.
In this model, the interference signals are
\begin{equation}\label{eq:PD-Gaussian_aux}
\ve{S_1}\!=\!
 \begin{bmatrix}
  S_1 \\
  S_3
 \end{bmatrix}\!=\!%
 \begin{bmatrix}
  h_{21} X_1 + h_{23} X_3 + Z_2 \\
  h_{31} X_1 + Z_3
 \end{bmatrix} \textnormal{and}\
S_2 = h_{12} X_2 + Z_1.
\end{equation}%
Therefore, results for the IS-IRC can be applied straightforwardly to the
Gaussian case.

\section{Outer Bound}
\label{sec:Outer}

In this section, we develop an outer bound for the IS-IRC model described in
Section~\ref{sec:Problem}. The model in Fig.~\ref{fig:PD-IS_IRC_model} is
provided to help the reader understand the genie-aided technique used in the
derivation of the bounds. It would be worth to emphasize that this model by no
means assumes that the relay has previous knowledge of any message nor that
$X_3$ or $Y_3$ are collocated with $X_1$ or $Y_2$ as it could be wrongly
interpreted based on the aforementioned figure.

Let $\mathcal{P}_1$ be the set of all joint PDs that can be factored as:
\begin{equation}
p(q)p(x_1 x_3\vert q) p(x_2\vert q) p(\ve{v_1} v_2\vert x_1 x_2 x_3 q),
\label{eq:OB-IS_IRC-pdf}
\end{equation}
where $p(\ve{v_1} v_2\vert x_1 x_2 x_3 q) = p_{\ve{S_1}\vert X_1
X_3}(\ve{v_1}\vert x_1 x_3) p_{S_2\vert X_2}(v_2 \vert x_2)$, i.e., $(\ve{V_1}
V_2)$ is a conditionally independent copy of $(\ve{S_1} S_2)$ given $(X_1 X_2
X_3)$. Let us recall that $V_1$ represents the first component of $\ve{V_1}$.

\vspace*{1mm}
\begin{theorem}[outer bound]
\label{th:OB-IS_IRC}
Given a specific $P_1\in \mathcal{P}_1$, let $\mathcal{R}_{o}(P_1)$ be the
region of nonnegative rate pairs $(R_1,R_2)$ satisfying
\begin{subequations}\label{eq:OB-IS_IRC}
\begin{flalign}
R_1 &\leq \IC{X_1}{Y_1 Y_3}{X_2 X_3 Q}, \label{eq:OB-IS_IRC1} \\
R_1 &\leq \IC{X_1 X_3}{Y_1}{X_2 Q}, \label{eq:OB-IS_IRC2} \\
R_2 &\leq \IC{X_2}{Y_2}{X_1 X_3 Q}, \label{eq:OB-IS_IRC3} \\
%
R_1\!+\!R_2 &\leq \IC{X_1 X_3}{Y_1}{V_1 X_2 Q} +\!\IC{X_1 X_2 X_3}{Y_2}{Q},
\label{eq:OB-IS_IRC4}\\
R_1\!+\!R_2 &\leq \IC{X_1 X_2 X_3}{Y_1}{V_1 Q} +\!\IC{X_1 X_2 X_3}{Y_2}{V_2 Q},
\!\!\!\!\!\!\label{eq:OB-IS_IRC5}\\
R_1\!+\!R_2 &\leq \IC{X_1 X_2 X_3}{Y_1}{Q} +\!\IC{X_2}{Y_2}{X_1 V_2 X_3 Q},
\displaybreak[2]\label{eq:OB-IS_IRC6}\\
%
R_1\!+\!R_2 &\leq \IC{X_1}{Y_1 Y_3}{V_1 X_2 X_3 Q} +\!\IC{X_1 X_2 X_3}{Y_2}{Q},
\!\!\!\!\!\!\!\!\!\label{eq:OB-IS_IRC7}\\
R_1\!+\!R_2 &\leq \IC{X_1 X_2}{Y_1 Y_3}{V_1 X_3 Q} +\!\IC{X_1 X_2 X_3}{Y_2}{V_2
Q}, \label{eq:OB-IS_IRC8}\\
R_1\!+\!R_2 &\leq \IC{X_1 X_2}{Y_1 Y_3}{X_3 Q} +\!\IC{X_2}{Y_2}{X_1 V_2 X_3 Q},
\!\!\!\!\!\!\displaybreak[2]\label{eq:OB-IS_IRC9}\\
%
R_1\!+\!R_2 &\leq \IC{X_1}{Y_1 Y_3}{\ve{V_1} X_2 X_3 Q} +\!\IC{X_1 X_2}{Y_2
Y_3}{X_3 Q}, \label{eq:OB-IS_IRC10}\\
R_1\!+\!R_2 &\leq \IC{X_1 X_2}{Y_1 Y_3}{\ve{V_1} X_3 Q}\!+\!\IC{X_1 X_2}{Y_2
Y_3}{V_2 X_3 Q}\!, \displaybreak[2]\label{eq:OB-IS_IRC11}
\end{flalign}\vspace{-.75cm} 
\begin{flalign}
\!\!
2R_1\!+\!R_2 &\leq \IC{X_1 X_3}{Y_1}{V_1 X_2 Q} + \IC{X_1 X_2 X_3}{Y_1}{Q} 
\nonumber\\
&\quad + \IC{X_1 X_2 X_3}{Y_2}{V_2 Q}, \label{eq:OB-IS_IRC12}\\
\!\!
2R_1\!+\!R_2 &\leq \IC{X_1 X_3}{Y_1}{V_1 X_2 Q} + \IC{X_1 X_2}{Y_1 Y_3}{X_3 Q}
\nonumber\\
&\quad + \IC{X_1 X_2 X_3}{Y_2}{V_2 Q}, \label{eq:OB-IS_IRC13}\\
\!\!
2R_1\!+\!R_2 &\leq \IC{X_1}{Y_1 Y_3}{V_1 X_2 X_3 Q} + \IC{X_1 X_2 X_3}{Y_1}{Q}
\nonumber\\
&\quad + \IC{X_1 X_2 X_3}{Y_2}{V_2 Q}, \label{eq:OB-IS_IRC14}\\
%
\!\!
2R_1\!+\!R_2 &\leq \IC{X_1}{Y_1 Y_3}{V_1 X_2 X_3 Q} + \IC{X_1 X_2}{Y_1 Y_3}{X_3
Q} \nonumber\\
&\quad + \IC{X_1 X_2 X_3}{Y_2}{V_2 Q}, \label{eq:OB-IS_IRC15}\\
\!\!
2R_1\!+\!R_2 &\leq \IC{X_1}{Y_1 Y_3}{\ve{V_1} X_2 X_3 Q} +\IC{X_1 X_2
X_3}{Y_1}{Q} \nonumber\\
&\quad + \IC{X_1 X_2}{Y_2 Y_3}{V_2 X_3 Q}, \label{eq:OB-IS_IRC16}\\
\!\!
2R_1\!+\!R_2 &\leq \IC{X_1}{Y_1 Y_3}{\ve{V_1} X_2 X_3 Q} + \IC{X_1 X_2}{Y_1 
Y_3}{X_3 Q}\nonumber\\
&\quad + \IC{X_1 X_2}{Y_2 Y_3}{V_2 X_3 Q}, \label{eq:OB-IS_IRC17}
\displaybreak[2]\\
%
\!\!
R_1\!+\!2R_2 &\leq \IC{X_1 X_2 X_3}{Y_1}{V_1 Q} + \IC{X_2}{Y_2}{X_1 V_2 X_3 Q}
\nonumber\\
&\quad + \IC{X_1 X_2 X_3}{Y_2}{Q}, \label{eq:OB-IS_IRC18}\\
\!\!
R_1\!+\!2R_2 &\leq \IC{X_1 X_2}{Y_1 Y_3}{V_1 X_3 Q} + \IC{X_2}{Y_2}{X_1 V_2
X_3 Q} \nonumber\\
&\quad + \IC{X_1 X_2 X_3}{Y_2}{Q}, \label{eq:OB-IS_IRC19}\\
\!\!
R_1\!+\!2R_2 &\leq \IC{X_1 X_2}{Y_1 Y_3}{\ve{V_1} X_3 Q} + \IC{X_2}{Y_2}{X_1 V_2
X_3 Q} \nonumber\\
&\quad + \IC{X_1 X_2}{Y_2 Y_3}{X_3 Q}. \label{eq:OB-IS_IRC20}
\end{flalign}
\end{subequations}
Then, an outer bound for the IS-IRC is defined by the union of
$\mathcal{R}_{o}(P_1)$ over all PDs $P_1 \in \mathcal{P}_1$, as decomposed
in~\eqref{eq:OB-IS_IRC-pdf}.
\end{theorem}
\begin{IEEEproof}
See Appendix~\ref{sec:AP-Proof_OB}.
\end{IEEEproof}


The real Gaussian model, presented in Section~\ref{sec:Problem}, is a special
case of the IS-IRC. Therefore, according to~\eqref{eq:OB-IS_IRC-pdf}, the
sources' inputs $X_1$ and $X_2$ are
independent, and $X_1$ is arbitrarily correlated to the relay's input $X_3$,
i.e., $E[X_1 X_2] = 0$, $E[X_1 X_3] = \rho\sqrt{P_1 P_3}$ and $E[X_2 X_3] = 0$.
The Gaussian expression of the outer bound is readily found using the
model~\eqref{eq:PD-Gaussian_case} and generating the auxiliaries $\ve{V_1}$ and
$V_2$ according to~\eqref{eq:PD-Gaussian_aux}, but with independent noises. 

The foregoing Gaussian outer bound $\mathcal{R}_{o} =\bigcup_{\rho\in[-1,1]}
\mathcal{R}_{o}(\rho)$ depends on the correlation coefficient $\rho$ between
$X_1$ and $X_3$ and, due to the large number of bounds, only a numerical
maximization results viable. In order to obtain analytical expressions which can
be used later to characterize the gap between inner and outer bounds, we
establish an outer bound on $\mathcal{R}_{o}$. This outer bound is obtained by
maximizing each individual rate constrain in $\mathcal{R}_{o}(\rho)$
independently.

Let us define any of the bounds in $\mathcal{R}_{o}(\rho)$ as $b(\rho)$ and
$\rho_{\max}$ as the value that maximizes that particular bound. Then, it can be
shown that $b(\rho_{\max})= b(0)$ or $b(\rho_{\max}) \leq b(0) +\Delta$, where
$\Delta$ is either $0.5$ or $1$ bit. Therefore, we can simplify the expressions
in the outer bound and avoid the maximization procedure if we use uncorrelated
inputs and enlarge certain bounds, as we see in the following corollary.
A similar observation has also been made in~\cite[Appx.
A]{avestimehr_wireless_2011} and~\cite[(19)]{lim_noisy_2011}.

\vspace*{1mm}
\begin{corollary}[outer bound for the Gaussian case]\label{cl:OB-AG}
$\!\!$An outer bound for the Gaussian IRC
is given by the set of nonnegative rate pairs $(R_1,R_2)$ satisfying
\begin{subequations}\label{eq:OB-Gaussian}
\begin{flalign}
R_1 &\leq \C{ \SNR{11} + \SNR{31} }, \label{eq:OB-Gaussian1}\\
R_1 &\leq \C{ \SNR{11} +\SNR{13} } +\frac{1}{2}, \label{eq:OB-Gaussian2}\\
R_2 &\leq \C{ \SNR{22} }, \label{eq:OB-Gaussian3}\\
R_1\!+\!R_2 &\leq \C{\frac{\SNR{11}\!+\!\SNR{13}\!+\!\delta}{ 1\!+\!\SNR{21} 
\!+\!\SNR{23} }}\!+\!\C{ \SNR{21}\!+\!\SNR{22}\!+\!\SNR{23} }\!+\!\frac{1}{2},\\
R_1\!+\!R_2 &\leq \C{ \SNR{12} +\frac{\SNR{11} +\SNR{13} +\delta}{ 1 +\SNR{21}
+\SNR{23} } } \nonumber\\
&\quad +\C{ \SNR{21} +\SNR{23} +\frac{\SNR{22}}{1 +\SNR{12}}} +\frac{1}{2},\\
R_1\!+\!R_2 &\leq \C{ \SNR{11} +\SNR{12} +\SNR{13} } +\C{ \frac{\SNR{22}}{1 
+\SNR{12}} } +\frac{1}{2}, \displaybreak[2]\\
R_1\!+\!R_2 &\leq \C{ \frac{ \SNR{11} + \SNR{31} }{ 1 +\SNR{21} } } +\C{
\SNR{21} +\SNR{22} +\SNR{23} } +\frac{1}{2}, \\
R_1\!+\!R_2 &\leq \C{ \SNR{12} +\frac{ \SNR{11} +\SNR{31}(1 +\SNR{12}) }{ 1 
+\SNR{21}} } \nonumber\\
&\quad +\C{ \SNR{21} +\SNR{23} +\frac{\SNR{22}}{1 +\SNR{12}}} 
+\frac{1}{2}, \\
R_1\!+\!R_2 &\leq \C{ \SNR{11}\!+\!\SNR{12}\!+\!\SNR{31}( 1\!+\!\SNR{12}) } 
\!+\!\C{ \frac{\SNR{22}}{1\!+\!\SNR{12}} }\!, \label{eq:OB-Gaussian9}\\
R_1\!+\!R_2 &\leq \C{ \frac{ \SNR{11}\!+\!\SNR{31} }{ 1\!+\!\SNR{21}\!+\! 
\SNR{31} } } \!\!+\!\C{ \SNR{21}\!+\!\SNR{22}\!+\!\SNR{31}(1\!+\!\SNR{22}) }\!,
\label{eq:OB-Gaussian10}\displaybreak[2]\\
R_1\!+\!R_2 &\leq \C{ \SNR{12} +\frac{ \SNR{11} +\SNR{31}(1+\SNR{12}) }{ 1 
+\SNR{21} +\SNR{31} } } \nonumber\\
&\quad + \C{ \SNR{21} +\SNR{31} +\frac{ \SNR{22}(1 +\SNR{31}) 
}{ 1 +\SNR{12} } }, \label{eq:OB-Gaussian11}
\end{flalign}\vspace{-.5cm} 
\begin{flalign}
2R_1 +R_2 &\leq \C{ \frac{\SNR{11} +\SNR{13} +\delta}{ 1 +\SNR{21} +\SNR{23}} }
+\C{ \SNR{21} +\SNR{23} +\frac{\SNR{22}}{1 +\SNR{12}}} \nonumber\\
&\quad +\C{ \SNR{11} +\SNR{12} +\SNR{13} } +1,\\
2R_1 +R_2 &\leq \C{ \frac{\SNR{11} +\SNR{13} +\delta}{ 1 +\SNR{21} +\SNR{23}} }
+\C{ \SNR{21} +\SNR{23} +\frac{\SNR{22}}{1 +\SNR{12}}} \nonumber\\
&\quad +\C{ \SNR{11} +\SNR{12} +\SNR{31}( 1 +\SNR{12}) } +\frac{1}{2},\\
2R_1 +R_2 &\leq \C{ \frac{ \SNR{11} + \SNR{31} }{ 1 +\SNR{21} } } +\C{ \SNR{11} 
+\SNR{12} +\SNR{13} } \nonumber\\
&\quad +\C{ \SNR{21} +\SNR{23} +\frac{\SNR{22}}{1 +\SNR{12}}} +1,\\
2R_1 +R_2 &\leq \C{ \frac{ \SNR{11} + \SNR{31} }{ 1 +\SNR{21} } } +\C{ \SNR{11} 
+\SNR{12} +\SNR{31}( 1 +\SNR{12}) } \nonumber\\
&\quad +\C{ \SNR{21} +\SNR{23} +\frac{\SNR{22}}{1 +\SNR{12}}} +\frac{1}{2},\\
2R_1 +R_2 &\leq \C{ \frac{ \SNR{11} + \SNR{31} }{ 1 +\SNR{21} +\SNR{31} } }  + 
\C{ \SNR{11} +\SNR{12} +\SNR{13} } \nonumber\\
&\quad + \C{\SNR{21} +\SNR{31} +\frac{ \SNR{22}(1 +\SNR{31}) }{ 1 +\SNR{12} } }
+\frac{1}{2},\\
2R_1 +R_2 &\leq \C{ \frac{ \SNR{11} + \SNR{31} }{ 1\!+\!\SNR{21}\!+\!\SNR{31} 
}} \!+\! 
\C{ \SNR{11}\!+\!\SNR{12}\!+\!\SNR{31}(1\!+\!\SNR{12}) } \nonumber\\
&\quad + \C{\SNR{21} +\SNR{31} +\frac{ \SNR{22}(1 +\SNR{31}) }{ 1 +\SNR{12} }
}, \displaybreak[2]\\
R_1 +2R_2 &\leq \C{ \SNR{12} +\frac{\SNR{11} +\SNR{13} +\delta}{ 1 +\SNR{21} 
+\SNR{23} }}  +\C{ \frac{\SNR{22}}{1 +\SNR{12}}  } \nonumber\\
&\quad + \C{ \SNR{21} +\SNR{22} +\SNR{23} } +\frac{1}{2}, \displaybreak[2]\\
R_1 +2R_2 &\leq \C{ \SNR{12} +\frac{ \SNR{11} +\SNR{31}(1 +\SNR{12}) }{ 1 
+\SNR{21}} }  +\C{ \frac{\SNR{22}}{1 +\SNR{12}}  } \nonumber\\
&\quad + \C{ \SNR{21} +\SNR{22} +\SNR{23} } +\frac{1}{2}, \\
R_1 +2R_2 &\leq \C{ \SNR{12} +\frac{ \SNR{11} +\SNR{31}(1+\SNR{12}) }{ 1 
+\SNR{21} +\SNR{31} } } +\C{ \frac{\SNR{22}}{1 +\SNR{12}} } \nonumber\\
&\quad + \C{ \SNR{21} +\SNR{22} +\SNR{31}(1 +\SNR{22}) }
\end{flalign}
\end{subequations}
where $\delta \triangleq \left( \sqrt{\SNR{11} \SNR{23}} \pm \sqrt{\SNR{13}
\SNR{21}} \right)^2$.
\end{corollary}
\begin{IEEEproof}
See Appendix~\ref{sec:AP-Proof-OB-decorr}.
\end{IEEEproof}

\vspace*{1mm}
\begin{remark}
If we define the following matrices,
\begin{equation}
 \pmb{H} = \begin{bmatrix}
  h_{11} & h_{13} \\
  h_{21} & h_{23}
 \end{bmatrix}\ \textnormal{and }\
 \pmb{Q} = \frac{1}{\sqrt{N_1 N_2}}\begin{bmatrix}
  P_1 & 0 \\
  0   & P_3
 \end{bmatrix},
\end{equation}
we readily see that $\delta = \det\!\left( \pmb{H} \pmb{Q} \pmb{H}^T \right)$.
Thus, the sign in the expression $\delta$ depends on the sign of the channel
coefficients. If there is an even number of negative coefficients in $\pmb{H}$,
then  $\delta = \left( \sqrt{\SNR{11} \SNR{23}} -\sqrt{\SNR{13} \SNR{21}}
\right)^2$, otherwise $\delta = \left( \sqrt{\SNR{11} \SNR{23}} +\sqrt{\SNR{13}
\SNR{21}} \right)^2$.
\end{remark}

\vspace*{1mm}
\begin{remark}
In the strong interference regime, where each receiver can decode the
interfering message completely without restricting its rate, tighter outer
bounds can be derived, similarly to the IC under strong
interference~\cite[Remark 6.9]{gamal_network_2011}. The sum-rates in the
capacity regions under strong interference~\cite[Thm. 5]{tian_gaussian_2011}
and~\cite[Thm. 2]{maric_relaying_2012}, the former with the assumption of a
potent relay, i.e., $P_3\rightarrow\infty$, are tighter than the ones presented
here, namely~\eqref{eq:OB-Gaussian9}, \eqref{eq:OB-Gaussian10},
\eqref{eq:OB-IS_IRC4}, \eqref{eq:OB-IS_IRC6}, and~\eqref{eq:OB-IS_IRC7}.
\end{remark}

\begin{remark}
Outer bound sum-rates using genie-aided techniques are given in~\cite[Thm.
4]{tian_gaussian_2011} and~\cite[Thm. 4]{maric_relaying_2012}, the former
extending the ``useful'' and ``smart'' genie
from~\cite{annapureddy_gaussian_2009} while the latter using Kramer's
approach~\cite{kramer_outer_2004}.

As it is shown in~\cite{annapureddy_gaussian_2009}, the ``smart'' genie provides
an outer bound that is tighter than Etkin~\emph{et
al.}'s~\cite{etkin_gaussian_2008} under weak interference, thus, the
sum-rate~\cite[Thm. 4]{tian_gaussian_2011} is tighter than the analogous in our
region, namely, \eqref{eq:OB-Gaussian11}.
Additionally, the optimization of parameters in the sum-rate~\cite[Thm.
4]{maric_relaying_2012} can potentially give tight bounds. For example, if
$d_1=h_{21}$, $d_2=d_3=0$, $d_4=\sqrt{N_2}$, and $d_5=h_{23}$ the genie signal
$Y_{1g}$ becomes $V_1=h_{21}X_1+h_{23}X_3+Z_2'$ and it is easy to verify that
the sum-rate~\cite[Thm. 4]{maric_relaying_2012} is tighter
than~\eqref{eq:OB-IS_IRC5}.
\end{remark}

\section{Inner Bounds}
\label{sec:Inner}


In the following, we provide two inner bounds corresponding to two different
relaying strategies, namely, DF and CF. With DF, the relay decodes the message
from the only connected source (partially or completely), re-encodes it, and
transmits it to both destinations. With CF, the relay compresses the received
signal, and sends a compression index associated to it. A previous version of
these schemes was presented in~\cite{bassi_allerton_2013}, but here we show a
more compact expression for the CF scheme and a completely new and improved
version for the DF scheme. Four main ingredients are required: rate-splitting,
binning, and block-Markov coding at the sources, and backward decoding at the
destinations.
In the sequel, we assume the indices $(k,j) \in \left\{ (1,2), (2,1) \right\}$.

In every strategy, to allow cooperation from the relay, the transmission is
split in several blocks. During block $b$, each source~$k$ divides its
message~$\tilde{m}_{kb}$ into two short messages: a common part $m_{kb}$ and a
private part $w_{kb}$. As in the Han-Kobayashi scheme, each receiver decodes the
common part of the interfering message, hence reducing the interference.

The use of DF and CF schemes for IRCs is well-known \cite{sahin_achievable_2007,
tian_gaussian_2011, chaaban_generalized_2012, kang_new_2013,
maric_relaying_2012}, however, our goal is to derive \emph{simple} but
powerful enough strategies in order to characterize the capacity region of the
IRC within a \emph{constant gap}.
The biggest obstacle to obtaining an inner bound with a manageable number of
inequalities is the use of a relaying strategy jointly with rate-splitting to
deal with interference. This issue may be overcome by assuming some special
condition in the model, e.g., symmetric channels~\cite{sahin_achievable_2007,
chaaban_generalized_2012} or strong interference~\cite{maric_relaying_2012}, or
by employing successive decoding of codewords instead of
joint-decoding~\cite{tian_gaussian_2011, chaaban_generalized_2012}. However, we
do not want to rely on these assumptions here.

Additionally, the proposed schemes have some key differences with respect to the
literature. In the DF scheme, the amount of information decoded by the relay is
optimized separately from the rate-splitting used to deal with interference,
which can potentially improve the achievable rates. Moreover, the CF scheme
presented in Section~\ref{ssec:IN-CF} does not force both receivers to decode
the compression index, unlike~\cite{tian_gaussian_2011, kang_new_2013}, which
could reduce the performance of the scheme if there is a large asymmetry among
the channels.

\vspace*{1mm}
\begin{remark}
It is worth noting that the inner bounds stated below apply to general
memoryless IRCs and thus they are not limited to the IS-IRC.
\end{remark}

\subsection{Decode-and-Forward}
\label{ssec:IN-DF}

Each source sends~$B$ messages during~$B+1$ time blocks, and the relay forwards
in block~$b$ what it has decoded from the first source in the previous block. In
this scheme, the \emph{private} message of the first source is split into two
parts and the relay only decodes and retransmits one of them (plus the
\emph{common} message).
At the end of transmission, receiver~$k$ decodes backwardly the private
message~$w_{kb}$ as well as both common messages~$m_{kb}$ and~$m_{jb}$. 

\begin{figure}
 \centering
 \begin{tikzpicture}[->,shorten >=1pt,auto,node distance=2cm,thick]

  \tikzset{green dotted/.style={draw=green!50!black, line width=1pt,
                               dash pattern=on 1pt off 3pt on 5pt off 3pt}};
  \tikzset{blue dotted/.style={draw=blue!75!white, line width=1pt,
                               dash pattern=on 4pt off 4pt}};

  \draw[draw=none, use as bounding box] (-.75,-2) rectangle (6.75,2);

  \node (1) {$V_3^n$};
  \node (2) [right of=1,yshift=1cm] {$X_3^n$};
  \node (3) [below of=2] {$V_1^n$};
  \node (4) [right of=3,yshift=1cm] {$U_1^n$};
  \node (5) [right of=4] {$X_1^n$};

  \path[every node/.style={font=\sffamily\small}]
    (1) edge node {} (2)
    (1) edge node {} (3)
    (2) edge node {} (4)
    (3) edge node {} (4)
    (4) edge node {} (5)
    (2) edge[<->,dash pattern=on 3pt off 3pt] node {} (3);

\draw[green dotted]
  let \p1=(1), \p2=(3),
  \n1={atan2(\y2-\y1,\x2-\x1)}, \n2={veclen(\y2-\y1,\x2-\x1)}
  in ($ (1)!0.5!(3) $) ellipse [%
    x radius=\n2/2+15pt, y radius=0.5cm, rotate=90-\n1 ]
  node [below, inner sep=6mm, rotate=90-\n1]
    {\footnotesize \textcolor{green!50!black}{Common Message}};
\draw[green dotted]
  let \p1=(2), \p2=(5), 
  \n1={atan2(\y2-\y1,\x2-\x1)}, \n2={veclen(\y2-\y1,\x2-\x1)} 
  in ($ (2)!0.5!(5) $) ellipse [%
    x radius=\n2/2+15pt, y radius=0.85cm, rotate=90-\n1 ]
  node [above, inner sep=2mm, rotate=90-\n1]
    {\footnotesize \textcolor{green!50!black}{Private Message}};
\draw[blue dotted]
  let \p1=(1), \p2=(2), 
  \n1={atan2(\y2-\y1,\x2-\x1)}, \n2={veclen(\y2-\y1,\x2-\x1)} 
  in ($ (1)!0.5!(2) $) ellipse [%
    x radius=\n2/2+15pt, y radius=0.5cm, rotate=90-\n1 ]
  node [above, inner sep=6mm, rotate=90-\n1]
    {\footnotesize \textcolor{blue!75!white}{Relay Codebook}};
\draw[blue dotted]
  let \p1=(3), \p2=(4), 
  \n1={atan2(\y2-\y1,\x2-\x1)}, \n2={veclen(\y2-\y1,\x2-\x1)} 
  in ($ (3)!0.5!(4) $) ellipse [%
    x radius=\n2/2+15pt, y radius=0.5cm, rotate=90-\n1 ]
  node [below, inner sep=6mm, rotate=90-\n1]
    {\footnotesize \textcolor{blue!75!white}{Relay Decodes}};
%
\end{tikzpicture}
\caption{Codewords of the relay and the first source. Solid arrows denote
superimposed codewords while dashed arrows denote binning.
\label{fig:IB-DF-codewords}}
\end{figure}
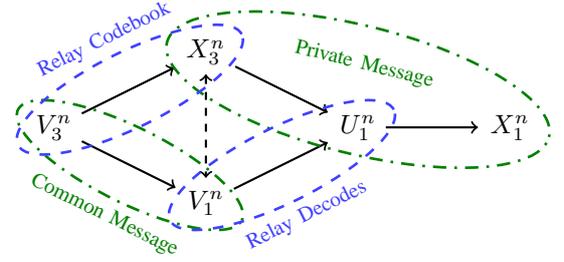

Let $\mathcal{P}_2$ be the set of PDs that factor as
\begin{multline}
p(q) p(x_1 x_3\vert q) p(x_2\vert q) p(v_1\vert x_1 x_3 q) \\ p(u_1\vert x_1 q)
p(v_2\vert x_2 q) p(v_3\vert x_3 q). \label{eq:IB-pDF-pdf}
\end{multline}\vspace{-2mm} 
\begin{theorem}[partial DF scheme] \label{th:IB-pDF}
Given a $P_2 \in \mathcal{P}_2$, let $\mathcal{R}_{\textrm{p-DF}}(P_2)$ be the
region of nonnegative rate pairs $(R_1,R_2)$ satisfying%
\begin{subequations}\label{eq:IB-pDF-A}
\begin{flalign}
R_1 &\leq \IC{U_1}{Y_3}{X_3 Q} + \IC{X_1}{Y_1}{V_1 U_1 V_2 X_3 Q},
\label{eq:IB-pDF-A1}\\
R_1 &\leq \IC{X_1 X_3}{Y_1}{V_2 Q},\\
R_2 &\leq \IC{X_2}{Y_2}{V_1 V_3 Q},\\
R_2 &\leq \IC{V_1 X_2 V_3}{Y_2}{Q} -I_b, \displaybreak[2]\\
R_1\!+\!R_2 &\leq \IC{X_1 X_3}{Y_1}{V_1 V_2 V_3 Q}\!+\!\IC{V_1 X_2 
V_3}{Y_2}{Q},\!\!\!\!\!\\
R_1\!+\!R_2 &\leq \IC{U_1}{Y_3}{V_1 X_3 Q} + \IC{X_1}{Y_1}{V_1 U_1 V_2 X_3 Q} 
\nonumber\\
 &\quad + \IC{V_1 X_2 V_3}{Y_2}{Q} -I_b, \displaybreak[2]\\
R_1\!+\!R_2 &\leq \IC{X_1 V_2 X_3}{Y_1}{V_1 V_3 Q} +\IC{V_1 X_2 V_3}{Y_2}{V_2 
Q},\\
R_1\!+\!R_2 &\leq \IC{U_1}{Y_3}{V_1 X_3 Q} + \IC{X_1 V_2}{Y_1}{V_1 U_1 X_3 Q} 
\nonumber\\
 &\quad + \IC{V_1 X_2 V_3}{Y_2}{V_2 Q} -I_b, \displaybreak[2]\\
R_1\!+\!R_2 &\leq \IC{X_1 V_2 X_3}{Y_1}{Q} + \IC{V_1 X_2 V_3}{Y_2}{V_2 Q} 
-I_b,\\
R_1\!+\!R_2 &\leq \IC{X_1 V_2 X_3}{Y_1}{Q} + \IC{X_2}{Y_2}{V_1 V_2 V_3 Q},\\
R_1\!+\!R_2 &\leq \IC{U_1}{Y_3}{X_3 Q} + \IC{X_1 V_2}{Y_1}{V_1 U_1 X_3 Q} 
\nonumber\\
 &\quad + \IC{X_2}{Y_2}{V_1 V_2 V_3 Q}, \label{eq:IB-pDF-A11}
\end{flalign}\vspace{-.75cm} 
\begin{flalign}
2 R_1\!+\!R_2 &\leq \IC{X_1 X_3}{Y_1}{V_1 V_2 V_3 Q} + \IC{X_1 V_2 X_3}{Y_1}{Q} 
\nonumber\\
 &\quad + \IC{V_1 X_2 V_3}{Y_2}{V_2 Q}, \displaybreak[2]\\
2 R_1\!+\!R_2 &\leq \IC{X_1 X_3}{Y_1}{V_1 V_2 V_3 Q}\!+\!\IC{X_1 V_2}{Y_1}{V_1 
U_1 X_3 Q} \nonumber\\
 &\quad + \IC{U_1}{Y_3}{X_3 Q}\!+\!\IC{V_1 X_2 V_3}{Y_2}{V_2 Q}, 
\label{eq:IB-pDF-A13}\\
2 R_1\!+\!R_2 &\leq \IC{U_1}{Y_3}{V_1 X_3 Q}\!+\!\IC{X_1}{Y_1}{V_1 U_1 V_2 X_3 
Q} -I_b \nonumber\\
 &\quad +\!\IC{X_1 V_2 X_3}{Y_1}{Q}\!+\!\IC{V_1 X_2 V_3}{Y_2}{V_2 Q}, 
\!\!\!\!\\
R_1\!+\!2 R_2 &\leq \IC{X_1 V_2 X_3}{Y_1}{V_1 V_3 Q} + \IC{X_2}{Y_2}{V_1 V_2 
V_3 Q} \nonumber\\
 &\quad + \IC{V_1 X_2 V_3}{Y_2}{Q},\\
R_1\!+\!2 R_2 &\leq \IC{U_1}{Y_3}{V_1 X_3 Q}\!+\!\IC{X_1 V_2}{Y_1}{V_1 U_1 X_3 
Q} -I_b \nonumber\\
 &\quad +\!\IC{X_2}{Y_2}{V_1 V_2 V_3 Q}\!+\!\IC{V_1 X_2 V_3}{Y_2}{Q} \!
\end{flalign}
\end{subequations}
where $I_b \triangleq \IC{X_3}{V_1}{V_3 Q}$.
Then, an achievable region for the IRC is defined by the union of all rate pairs
in $\mathcal{R}_{\textrm{p-DF}}(P_2)$ over all joint PDs $P_2\in\mathcal{P}_2$, 
as defined in~\eqref{eq:IB-pDF-pdf}.
\end{theorem}
\begin{IEEEproof}
The codewords $V_2^n$ and $X_2^n$ convey the common and full messages of the
second source, respectively, with $X_2^n$ superimposed over $V_2^n$. This
representation follows the steps proposed in~\cite{chong_comparison_2006}, due
to its simplicity compared to~\cite{han_new_1981}, though both representations
are equivalent~\cite{chong_han-kobayashi_2008}.

The codebook of the first source, however, is much more involved in order to
allow the relay to cooperate, see Fig.~\ref{fig:IB-DF-codewords}.
The scheme forces the relay to decode the common message of the first source,
i.e., the codeword $V_1^n$, entirely but only a part of the private message.
Thus, unlike the second source, an intermediate layer $U_1^n$ is included
between $V_1^n$ and $X_1^n$.

The indices decoded by the relay are forwarded through superimposed codewords
$V_3^n$ and $X_3^n$, analogous to $V_1^n$ and $U_1^n$. Coherent cooperation is
achieved by superimposing $V_1^n$ and $U_1^n$ over $V_3^n$ and $X_3^n$,
respectively. An additional binning step between the codewords $V_1^n$ and
$X_3^n$ is required to comply with~\eqref{eq:IB-pDF-pdf}, thus the negative term
$I_b$ in~\eqref{eq:IB-pDF-A}.

The region $\mathcal{R}_{\textrm{p-DF}}$~\eqref{eq:IB-pDF-A} is strictly smaller
than the actual partial DF region since we have purposely reduced all the bounds
with \IC{V_1U_1}{Y_3}{X_3} into \IC{U_1}{Y_3}{X_3}, namely,
in~\eqref{eq:IB-pDF-A1}, \eqref{eq:IB-pDF-A11}, and~\eqref{eq:IB-pDF-A13}, in
order to have a more compact expression of the whole region. See
Appendix~\ref{sec:AP-Proof-pDF} for details.
\end{IEEEproof}

If the relay is able to decode the private message of the first source
completely without imposing a restriction on the achievable rate, the
maximization of the previous inner bound would result in $U_1 = X_1$. In this
case, let $\mathcal{P}_3$ be the set of PDs which factor as
\begin{equation}
p(q) p(x_1 x_3\vert q) p(x_2\vert q) p(v_1\vert x_1 x_3 q) p(v_2\vert x_2 q)
p(v_3\vert x_3 q). \label{eq:IB-DF-pdf}
\end{equation}

\begin{corollary}[full DF scheme] \label{th:IB-DF}
$\,$ Given a $P_3 \in \mathcal{P}_3$, let $\mathcal{R}_{\textrm{f-DF}}(P_3)$
be the
region of nonnegative rate pairs $(R_1,R_2)$ satisfying%
\begin{subequations}\label{eq:IB-DF-A}
\begin{align}
R_1 &\leq \IC{X_1}{Y_3}{X_3 Q}, \label{eq:IB-DF-A1}\\
R_1 &\leq \IC{X_1 X_3}{Y_1}{V_2 Q}, \label{eq:IB-DF-A2}\\
R_2 &\leq \IC{X_2}{Y_2}{V_1 V_3 Q}, \label{eq:IB-DF-A3}\\
R_2 &\leq \IC{V_1 X_2 V_3}{Y_2}{Q} -I_b, \displaybreak[2] \label{eq:IB-DF-A4}\\
R_1 +R_2 &\leq \IC{X_1 X_3}{Y_1}{V_1 V_2 V_3 Q} + \IC{V_1 X_2 V_3}{Y_2}{Q},
\label{eq:IB-DF-A5}\\
R_1 +R_2 &\leq \IC{X_1}{Y_3}{V_1 X_3 Q} + \IC{V_1 X_2 V_3}{Y_2}{Q} -I_b,
\label{eq:IB-DF-A6}\\
R_1 +R_2 &\leq \IC{X_1 V_2 X_3}{Y_1}{V_1 V_3 Q} + \IC{V_1 X_2 V_3}{Y_2}{V_2 Q},
\label{eq:IB-DF-A7}\displaybreak[2]\\
R_1 +R_2 &\leq \IC{X_1 V_2 X_3}{Y_1}{Q} + \IC{V_1 X_2 V_3}{Y_2}{V_2 Q} -I_b,
\label{eq:IB-DF-A8}\displaybreak[2]\\
R_1 +R_2 &\leq \IC{X_1 V_2 X_3}{Y_1}{Q} + \IC{X_2}{Y_2}{V_1 V_2 V_3 Q},
\displaybreak[2]\label{eq:IB-DF-A9}\\
2 R_1 +R_2 &\leq \IC{X_1 X_3}{Y_1}{V_1 V_2 V_3 Q} + \IC{X_1 V_2 X_3}{Y_1}{Q} 
\nonumber\\
 &\quad + \IC{V_1 X_2 V_3}{Y_2}{V_2 Q}, \label{eq:IB-DF-A10}\\
2 R_1 +R_2 &\leq \IC{X_1}{Y_3}{V_1 X_3 Q} + \IC{X_1 V_2 X_3}{Y_1}{Q} \nonumber\\
 &\quad + \IC{V_1 X_2 V_3}{Y_2}{V_2 Q} -I_b,\label{eq:IB-DF-A11}\\
R_1 +2 R_2 &\leq \IC{X_1 V_2 X_3}{Y_1}{V_1 V_3 Q} +\IC{X_2}{Y_2}{V_1 V_2 V_3 Q} 
\nonumber\\
 &\quad + \IC{V_1 X_2 V_3}{Y_2}{Q} \label{eq:IB-DF-A12}
\end{align}
\end{subequations}
where $I_b \triangleq \IC{X_3}{V_1}{V_3 Q}$.
Then, an achievable region for the IRC is defined by the union of all rate pairs
in $\mathcal{R}_{\textrm{f-DF}}(P_3)$ over all joint PDs $P_3\in\mathcal{P}_3$, 
as defined in \eqref{eq:IB-DF-pdf}.
\end{corollary}
\begin{IEEEproof}
The region $\mathcal{R}_{\textrm{f-DF}}$~\eqref{eq:IB-DF-A} is not obtained by
setting $U_1 = X_1$ in $\mathcal{R}_{\textrm{p-DF}}$~\eqref{eq:IB-pDF-A}, 
since some additional redundant bounds remain. To easily eliminate these bounds,
one should replace $U_1$ with $X_1$ in the set of partial rates before applying
Fourier-Motzkin elimination in the proof of Theorem~\ref{th:IB-pDF}. See
Appendix~\ref{sec:AP-Proof-3} for details.
\end{IEEEproof}

The keen reader can see the resemblance between the region
$\mathcal{R}_{\textrm{f-DF}}$~\eqref{eq:IB-DF-A} and the Han-Kobayashi
region~\cite{chong_han-kobayashi_2008}, with the addition of bounds regarding
the decoding at the relay or the presence of binning. 

\vspace{1mm}
\begin{remark}\label{rm:IB-DF-cap}
The capacity of the \emph{physically degraded} IRC in the \emph{strong
interference} regime~\cite[Thm. 3]{maric_relaying_2012} is achieved by the full
DF scheme.

The choice of variables $V_k = X_k$ for $k\in [1:3]$ eliminates the private
messages and renders the binning process unnecessary. Then, by using the strong
interference condition $\IC{X_1X_3}{Y_1}{X_2} \leq \IC{X_1X_3}{Y_2}{X_2}$, the
full DF inner bound becomes
\begin{subequations}\label{eq:IB-DF-B}
\begin{align}
R_1 &\leq \IC{X_1}{Y_3}{X_3 Q}, \\
R_1 &\leq \IC{X_1 X_3}{Y_1}{X_2 Q}, \\
R_2 &\leq \IC{X_2}{Y_2}{X_1 X_3 Q}, \\
R_1 +R_2 &\leq \IC{X_1 X_2 X_3}{Y_1}{Q}, \\
R_1 +R_2 &\leq \IC{X_1 X_2 X_3}{Y_2}{Q}.
\end{align}
\end{subequations}
The region~\eqref{eq:IB-DF-B} coincides with the outer bound~\cite[Thm.
2]{maric_relaying_2012} by choosing $U_1=X_3$ and $U_2=X_2$, and considering
that
\begin{enumerate}
 \item the relay is only able to observe the first source, i.e., $p(y_3\vert x_1
x_2 x_3) = p(y_3\vert x_1 x_3)$, and
 \item the IRC is physically degraded, i.e., the Markov chain $(X_1 X_2) \mkv
(X_3 Y_3) \mkv (Y_1 Y_2)$ holds.
\end{enumerate}
\end{remark}
\vspace{1mm}

In the full DF scheme, since the relay decodes the codeword $X_1^n$ completely,
there is no limit in the amount of information that can be sent as common
message. However, in the partial DF scheme, we are introducing the variable
$U_1$ between $X_1$ and $V_1$, effectively prohibiting $V_1=X_1$. Therefore, the
structure of the codebook imposes that the relay should be in a better condition
to decode the common message $V_1^n$ than the second destination. If that is not
the case, we should employ the CF scheme presented in the following section.

\subsection{Compress-and-Forward}
\label{ssec:IN-CF}

In this scheme, the relay does not decode any message and it only sends a
compressed version of its observation. The destinations jointly decode this
information with their message and the common layer of the interference.
Transmission takes place in $B+L$ time blocks, similarly
to~\cite{wu_optimal_2013,Behboodi-piantanida2013}, and during the last $L$
blocks, the relay repeats its message to assure a correct decoding at both
destinations.

Let $\mathcal{P}_4$ be the set of PDs that factor as
\begin{equation}
p(q) p(v_1 x_1\vert q) p(v_2 x_2\vert q) p(x_3\vert q) p(\hat{y}_3\vert x_3 y_3
q), \label{eq:IB-CF-pdf}
\end{equation}
and let us define the following set of expressions
\begin{subequations}\label{eq:IB-CF-join-def1}
\begin{align}
I_{k1} &\triangleq \min\{ \IC{X_k}{Y_k \hat{Y}_3}{V_k V_j X_3 Q}, \nonumber\\
&\qquad\qquad \IC{X_k X_3}{Y_k}{V_k V_j Q} - I_k \}, \label{eq:IB-CF-joint_1} \\
I_{k2} &\triangleq \min\{ \IC{X_k}{Y_k \hat{Y}_3}{V_j X_3 Q}, \nonumber\\
&\qquad\qquad \IC{X_k X_3}{Y_k}{V_j Q} - I_k \},\\
I_{k3} &\triangleq \min\{ \IC{X_k V_j}{Y_k \hat{Y}_3}{V_k X_3 Q}, \nonumber\\
&\qquad\qquad \IC{X_k V_j X_3}{Y_k}{V_k Q} - I_k \},\\
I_{k4} &\triangleq \min\{ \IC{X_k V_j}{Y_k \hat{Y}_3}{X_3 Q}, \nonumber\\
&\qquad\qquad \IC{X_k V_j X_3}{Y_k}{Q} - I_k \}
\end{align}
\end{subequations}
where $I_k \triangleq \IC{\hat{Y}_3}{Y_3}{X_k V_j X_3 Y_k Q}$ and
\begin{subequations}\label{eq:IB-CF-join-def2}
\begin{align}
I_{k1}' &\triangleq \IC{X_k}{Y_k}{V_k V_j Q},\\
I_{k2}' &\triangleq \IC{X_k}{Y_k}{V_j Q},\\
I_{k3}' &\triangleq \IC{X_k V_j}{Y_k}{V_k Q},\\
I_{k4}' &\triangleq \IC{X_k V_j}{Y_k}{Q}.
\end{align}
\end{subequations}

\begin{theorem}[CF scheme] \label{th:IB-CF}
Given a specific $P_4 \in \mathcal{P}_4$, let $\mathcal{R}_{\textrm{CF}_0}(P_4)$
be the region of nonnegative rate pairs $(R_1,R_2)$ that satisfy%
\begin{subequations}\label{eq:IB-CF-joint}
\begin{align}
R_k &\leq I_{k2},\\
R_k +R_j &\leq \min\{ I_{k1}+I_{j4}, I_{k3}+I_{j3} \},\\
2R_k +R_j &\leq I_{k1}+I_{k4}+I_{j3},
\end{align}
\end{subequations}
and $\mathcal{R}_{\textrm{CF}_k}(P_4)$ defined by
\begin{subequations}\label{eq:IB-CF2-joint}
\begin{align}
R_k &\leq I_{k2},\\
R_j &\leq I_{j2}',\\
R_k +R_j &\leq \min\{ I_{k1}+I_{j4}', I_{k4}+I_{j1}', I_{k3}+I_{j3}' \},\\
2R_k +R_j &\leq I_{k1}+I_{k4}+I_{j3}',\\
R_k +2R_j &\leq I_{k3}+I_{j1}'+I_{j4}'.
\end{align}
\end{subequations}
Then, an achievable region for the IRC is defined by the union of
$\mathcal{R}_{\textrm{CF}_0}(P_4) \cup \mathcal{R}_{\textrm{CF}_1}(P_4) \cup
\mathcal{R}_{\textrm{CF}_2}(P_4)$ over all joint distributions
$P_4\in\mathcal{P}_4$, as defined in \eqref{eq:IB-CF-pdf}.
\end{theorem}
\begin{IEEEproof}
Since the relay does not decode any message, the codewords $V_k^n$ and $X_k^n$
carry the common and full message of the present block, respectively. The
variable $X_3$ is independent of the sources' signals and is used to reconstruct
the relay's observation $Y_3$.

Each expression $I_{ki}$ resembles the CF inner bound for the relay channel, and
when the relay is ignored it reduces to the expression $I_{ki}'$.
The region $\mathcal{R}_{\textrm{CF}_0}$~\eqref{eq:IB-CF-joint} is obtained when
both destinations decode the compression index, whereas in region
$\mathcal{R}_{\textrm{CF}_k}$~\eqref{eq:IB-CF2-joint} only destination~$k$
decodes it.

Since the compression index is sent with block-Markov coding, each destination
needs to assure the correct decoding of it in each block, which results in
additional bounds not shown here. However, the union
$\mathcal{R}_{\textrm{CF}_0} \cup \mathcal{R}_{\textrm{CF}_1} \cup
\mathcal{R}_{\textrm{CF}_2}$ after the maximization over all joint PDs provides
that these bounds are redundant.
See Appendix~\ref{sec:AP-Proof-CF} for details.
\end{IEEEproof}

\vspace{1mm}
\begin{remark}
The relay only generates one compression index that is decodable by both
destinations, i.e., the compression rate is determined by the worst channel. It
is possible, however, to improve the performance with successive refinement that
is not used here because of its complexity. As we shall see in the next section,
two layers of successive refinement are not needed as far as the constant gap is
concerned.
\end{remark}

\vspace{1mm}
\begin{remark}
If both users ignore the compression index, this strategy reduces to the
Han-Kobayashi scheme, a special case of $\mathcal{R}_{\textrm{CF}_0}$.
Additionally, $\mathcal{R}_{\textrm{CF}_0}$ is equal to the extension of
NNC~\cite[Thm. 1]{kang_new_2013} for one relay, i.e., $N=1$.
\end{remark}

\vspace{1mm}
\begin{remark}
The region $\mathcal{R}_{\textrm{CF}_0}$ contains both the CF and GCF schemes
presented in~\cite[Thm. 1 and 2]{tian_gaussian_2011}.
It is easy to see that the bounds on the partial rates of the first
scheme~\cite[(5)--(8)]{tian_gaussian_2011} are below~\eqref{eq:IB-CF-join-def1}
if we relax the constraint~\cite[(9)]{tian_gaussian_2011} to $\I{X_3}{Y_j}
\geq \IC{Y_3}{\hat{Y}_3}{X_3 Y_j}$ with $j\in \{1,2\}$.
Additionally, relaxing $R_0$ in~\cite[Thm. 2]{tian_gaussian_2011}, shows that
GCF$_1$ is equal to $\mathcal{R}_{\textrm{CF}_0}$ with $V_1=V_2=\emptyset$ and
GCF$_2$ is equal to $\mathcal{R}_{\textrm{CF}_0}$ with $V_1=X_1$ and
$V_2=X_2$.
Therefore, the capacity results~\cite[Thm. 4 and 5]{tian_gaussian_2011} are
achieved by the proposed CF scheme.
\end{remark}

\section{Constant Gap Results and Discussion}
\label{sec:Gap}

In this section, we evaluate the gap between the achievable regions and the
outer bound in the Gaussian case (Fig.~\ref{fig:IN-Gaussian_model}). Then, we
identify the strategies that achieve the best constant gap to the capacity
region for any SNR value. This is summarized in Table~\ref{tab:CG-optimal},
while the value of the gap for each strategy is shown in
Table~\ref{tab:CG-value}.

\begin{table}[!t]
{%
\newcommand{\mc}[3]{\multicolumn{#1}{#2}{#3}}
\renewcommand{\arraystretch}{1.2} 
\begin{center}
\begin{tabular}{|c|c|c|}
\cline{2-3}
\mc{1}{c|}{}            & $\SNR{31}<\SNR{21}$ & $\SNR{31}\geq \SNR{21}$\\
\hline
$\SNR{31}<\SNR{11}$     & CF                  & partial DF\\
\hline
$\SNR{31}\geq \SNR{11}$ & \mc{2}{c|}{full DF}\\
\hline
\end{tabular}
\end{center}
}%
\caption{SNR regimes and corresponding best constant-gap strategies.
\label{tab:CG-optimal}}
\end{table}

\begin{figure*}[!t]%
\centering
\subfloat[Maximum attainable sum-rate]{\label{fig:CG-plots_a}
\pgfplotsset{every axis legend/.append style={
at={(0.02,0.98)},
anchor=north west}}
\pgfplotsset{every axis/.append style={
font=\footnotesize,
thick,
mark repeat={11},
tick style={line width=1pt}}}
\begin{tikzpicture}
\begin{axis}[
width=.48\textwidth,
xlabel={\SNR{31} [dB]},
ylabel={Max. sum-rate [bit]},
ylabel style={yshift=-15pt},
xmin=-15, xmax=25,
ymin=2.4, ymax=6.75,
grid=major,
legend entries={OB, DF, CF, HK},
]
\addplot[blue, mark=triangle]
  table[x=rho, y=outer] {sumrate_s31.dat};
\addplot[red, mark=*]
  table[x=rho, y=p_df] {sumrate_s31.dat};
\addplot[cyan, mark=square]
  table[x=rho, y=cf] {sumrate_s31.dat};
\addplot[magenta, mark=diamond]
  table[x=rho, y=hk] {sumrate_s31.dat};
\addplot[mark=none, dashed] coordinates {(20,2.4) (20,6.75)};
\addplot[mark=none, dashed] coordinates {(8,2.4) (8,6.75)};
\addplot[mark=none, dashed] coordinates {(-12,2.4) (-12,6.75)};
\node at (axis cs:22.5,2.67) {\scriptsize f-DF};
\node at (axis cs:14,2.63) {\scriptsize p-DF};
\draw[<-] (axis cs:8,2.65) -- (axis cs:11.5, 2.65);
\draw[->] (axis cs:16.5,2.65) -- (axis cs:20, 2.65);
\node at (axis cs:-2,2.65) {\scriptsize CF};
\draw[<-] (axis cs:-12,2.65) -- (axis cs:-4, 2.65);
\draw[->] (axis cs:0,2.65) -- (axis cs:8, 2.65);
\node at (axis cs:-13.5,2.66) {\scriptsize HK};
\end{axis}
\end{tikzpicture}
}%
%
\hfil
\subfloat[Gap per dimension]{\label{fig:CG-plots_b}
\pgfplotsset{every axis legend/.append style={
at={(0.02,0.98)},
anchor=north west}}
\pgfplotsset{every axis/.append style={
font=\footnotesize,
thick,
mark repeat={11},
tick style={line width=1pt}}}
\begin{tikzpicture}
\begin{axis}[
width=.48\textwidth,
xlabel={\SNR{31} [dB]},
ylabel={Gap per dimension [bit]},
ylabel style={yshift=-10pt},
xmin=-15, xmax=25,
ymin=0.27, ymax=1.6,
grid=major,
legend entries={DF, CF},
]
\addplot[red, mark=*]
  table[x=rho, y=p_df] {gap_s31.dat};
\addplot[cyan, mark=square]
  table[x=rho, y=cf] {gap_s31.dat};
\addplot[mark=none, dashed] coordinates {(20,0) (20,2)};
\addplot[mark=none, dashed] coordinates {(8,0) (8,2)};
\addplot[mark=none, dashed] coordinates {(-12,0) (-12,2)};
\node at (axis cs:22.5,.355) {\scriptsize f-DF};
\node at (axis cs:14,.345) {\scriptsize p-DF};
\draw[<-] (axis cs:8,.35) -- (axis cs:11.5, .35);
\draw[->] (axis cs:16.5,.35) -- (axis cs:20, .35);
\node at (axis cs:-2,.35) {\scriptsize CF};
\draw[<-] (axis cs:-12,.35) -- (axis cs:-4, .35);
\draw[->] (axis cs:0,.35) -- (axis cs:8, .35);
\node at (axis cs:-13.5,.35) {\scriptsize HK};
\addplot[mark=none, dashed] coordinates {(20,1) (25,1)};
\addplot[mark=none, dashed] coordinates {(8,1.5) (20,1.5)};
\addplot[mark=none, dashed] coordinates {(-10,1.32) (8,1.32)};
\end{axis}
\end{tikzpicture}
}
\caption{Performance analysis for the Gaussian IRC
(Fig.~\ref{fig:IN-Gaussian_model}) with the following fixed SNRs:
$\SNR{11}=\SNR{22}=20$dB, $\SNR{12}=\SNR{21}=8$dB, $\SNR{13}=\SNR{23}=20$dB.
\label{fig:CG-plots}}%
\end{figure*}
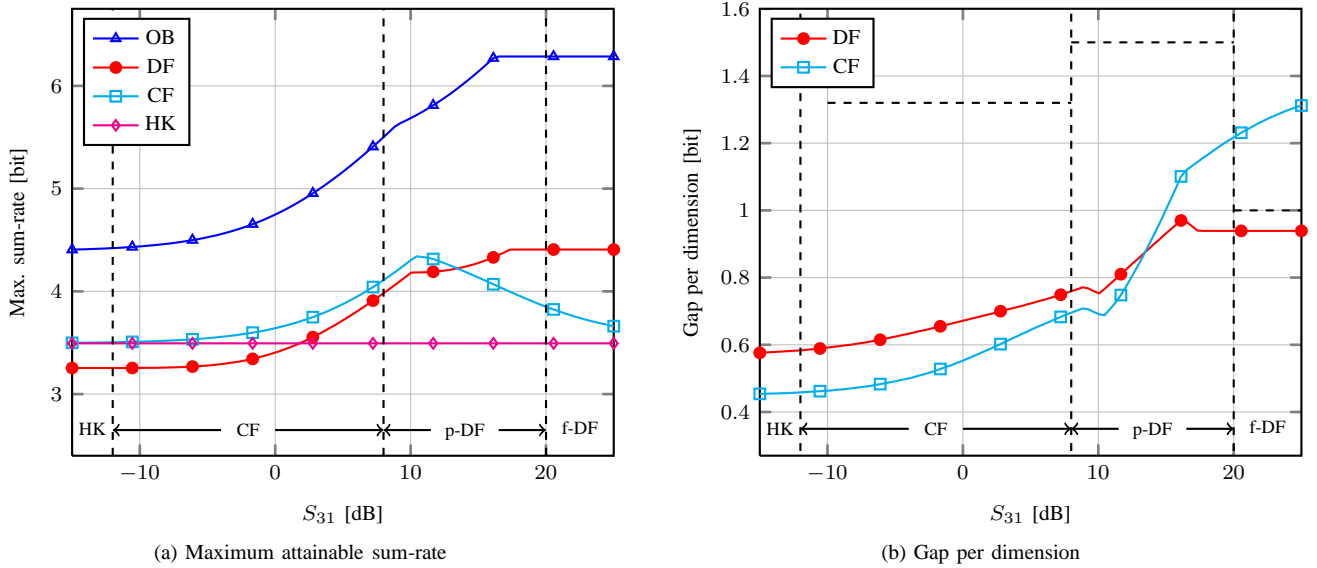

\subsection{DF Scheme Achieves Capacity to Within \texorpdfstring{$1.5$}{1.5} 
Bits}

Table~\ref{tab:CG-value} shows two different constant-gap values for this
scheme, $1.5$ bits being the larger. The difference comes from the choice
of input PD used in the inner bound as we see next.

When the relay is close to the source, i.e., when \SNR{31} is high enough, the
relay is able to decode the entire message without penalizing the rate $R_1$.
Therefore, as mentioned in Section~\ref{ssec:IN-DF}, the input PD verifies
$U_1=X_1$ and the inner bound is found in Corollary~\ref{th:IB-DF}.

\vspace{1mm}
\begin{proposition}
\label{th:CG-DF}
If $\SNR{31} \geq \SNR{11}$, the full DF scheme presented in
Corollary~\ref{th:IB-DF} achieves capacity to within $1$ bit.
\end{proposition}
\begin{IEEEproof}
The mentioned constant gap is quite conservative in the majority of cases since
it arises from choosing a fixed input PD for the inner bound (which reduces the
achievable rate) and using the loose outer bound from Corollary~\ref{cl:OB-AG}.
See Appendix~\ref{sec:AP-Proof-CG-DF} for details.
\end{IEEEproof}

\begin{remark}
The capacity result in~\cite[Thm. 3]{maric_relaying_2012} is contained in this
regime. This capacity result, which is valid for general memoryless channels,
relies on three conditions, namely,
\begin{enumerate}
 \item the relay can only observe one source signal;
 \item the IRC is \emph{physically degraded}, i.e., $(X_1 X_2) \mkv (X_3 Y_3)
\mkv (Y_1 Y_2)$; and,
 \item the IRC is under the \emph{strong interference} regime, i.e., $\IC{X_k
X_3}{Y_k}{X_j} \leq \IC{X_k X_3}{Y_j}{X_j}$.
\end{enumerate}
The IRC model~\eqref{eq:PD-general_pdf} used in this work only verifies the
first condition. However, if we further assume that the conditions of
physically degradedness and strong interference hold, the full DF scheme
presented in Corollary~\ref{th:IB-DF} also achieves capacity (see
Remark~\ref{rm:IB-DF-cap}). As we see next, the lack of these two assumptions
imposes the $1$-bit gap.

First, our Gaussian model~\eqref{eq:PD-Gaussian_case} does not admit any kind of
degradedness, however, if $\SNR{31} \geq \SNR{11}$, we can bound the
corresponding term by $0.5$ bits, as in~\eqref{eq:AP-CG-DF-g1},
$$
\IC{X_1}{Y_1}{X_2 X_3 Y_3 Q} = \C{ \frac{\SNR{11}}{ 1 +\SNR{31}} } 
 \leq \frac{1}{2}.
$$

Second, the strong interference condition renders the rate-splitting useless,
since both encoders send only common messages, and allows the development of a
tighter outer bound, similar to the IC with strong interference~\cite[Remark
6.9]{gamal_network_2011}. Without common messages, not only the binning term
$I_b$ disappears but also the simplifications made in
Appendix~\ref{sec:AP-Proof-CG-DF}, namely the choice of
auxiliaries~\eqref{eq:AP-CG-DF-aux} and the uncorrelation between $X_1$ and
$X_3$, can be dropped.
For example, as seen in Appendix~\ref{sec:AP-Proof-CG-DF}, the choice of
auxiliaries~\eqref{eq:AP-CG-DF-aux} inflicts half a bit of gap
in~\eqref{eq:AP-CG-DF-g2} and~\eqref{eq:AP-CG-DF-g3}, while another half a bit
of gap is due to the uncorrelation between $X_1$ and $X_3$
in~\eqref{eq:AP-CG-DF-g2} and due to the binning term $I_b$
in~\eqref{eq:AP-CG-DF-g3}.

Therefore, the $1$-bit gap the full DF scheme presents in contrast to the
capacity-achieving scheme of~\cite{maric_relaying_2012} comes from the last two
conditions, which are not assumed by our model.
\end{remark}

\begin{table}[!t]
{%
\newcommand{\mc}[3]{\multicolumn{#1}{#2}{#3}}
\newcommand{\mr}[3]{\multirow{#1}{#2}{#3}}
\begin{center}
\begin{tabular}{cccc}
\toprule
\mc{2}{c}{SNR regime}                                      & CF     & DF \\
\midrule
\mr{2}{*}{$\SNR{31}<\SNR{21}$}    & $\SNR{31}<\SNR{11}$    & $1.32$ & -- \\
\cmidrule(r){2-4}
                                  & $\SNR{31}\geq\SNR{11}$ & $1.32$ & $1$ \\
\midrule
\mr{2}{*}{$\SNR{31}\geq\SNR{21}$} & $\SNR{31}\geq\SNR{11}$ & --     & $1$ \\
\cmidrule(r){2-4}
                                  & $\SNR{31}<\SNR{11}$    & --     & $1.5$ \\
\bottomrule
\end{tabular}
\end{center}
}%
\caption{Maximum gap in bits of each scheme for each SNR regime.
\label{tab:CG-value}} 
\end{table}

\vspace*{1mm}

If the source-to-relay link is not good enough for the relay to decode the
entire message, the relay should decode it partially, i.e., $U_1\neq X_1$.
However, due to the structure of the codebook, the relay should still be able to
decode the common message.

\vspace{1mm}
\begin{proposition}
\label{th:CG-pDF}
If $\SNR{31} \geq \SNR{21}$, the partial DF scheme presented in
Theorem~\ref{th:IB-pDF} achieves capacity to within $1.5$ bits.
\end{proposition}
\begin{IEEEproof}
Similarly to the proof of Proposition~\ref{th:CG-DF}, we reduce the inner bound
by fixing the input PD and enlarge the outer bound by choosing a subset of
bounds from it. 
See Appendix~\ref{sec:AP-Proof-CG-pDF} for details.
\end{IEEEproof}

\vspace{1mm}
\begin{remark}
The gap between the original expression in the inner bound, \IC{V_1
U_1}{Y_3}{X_3 Q}, and the one used to compact the region, \IC{U_1}{Y_3}{X_3 Q},
is $0.5$ bit at most with the choice of auxiliaries~\eqref{eq:AP-CG-DF-aux}
and~\eqref{eq:AP-CG-pDF-aux} used in Appendix~\ref{sec:AP-Proof-CG-pDF}. This is
the cause of the larger gap for the partial DF scheme.
\end{remark}

\vspace{1mm}
\begin{remark}
If $\SNR{31} \geq \SNR{11}$ and $\SNR{31} \geq \SNR{21}$ the DF scheme, full or
partial, achieves a constant gap to capacity. Nonetheless, this regime appears
in Table~\ref{tab:CG-optimal} as ``full DF'' since its gap is smaller.
\end{remark}

\subsection{CF Scheme Achieves Capacity to Within \texorpdfstring{$1.32$}{1.32} 
Bits}

The CF scheme does not impose any condition on the sources' codebook structure,
nonetheless, a constant gap could only be found in the regime $\SNR{31} \leq
\SNR{21}$.

\vspace{1mm}
\begin{proposition}
\label{th:CG-CF}
If $\SNR{31} \leq \SNR{21}$ the CF scheme presented in Theorem~\ref{th:IB-CF}
achieves capacity to within $1.32$ bits.
\end{proposition}
\begin{IEEEproof}
The proof follows similar steps as the previous ones.
See Appendix~\ref{sec:AP-Proof-CG-CF} for details.
\end{IEEEproof}

\subsection{Limited Relaying Benefit}

It sounds reasonable that for a really low SNR in the source-to-relay link, the
use of relaying has limited benefit. In this case, it might be preferable, due
to complexity, to shut the relay down and fall back to the much simpler
Han-Kobayashi scheme for the IC.

\vspace{1mm}
\begin{proposition}
\label{th:CG-NoRelay}
If $\SNR{31} \leq \SNR{11}/(1 +\SNR{12})$ and $\SNR{31} \leq \SNR{21}/(1
+\SNR{22})$, the Han-Kobayashi scheme (without relay) achieves the capacity of
the IS-IRC within $1$ bit, i.e., relaying does not improve the achievable rate
in more than $1$ bit.
\end{proposition}
\begin{IEEEproof}
See Appendix~\ref{sec:AP-Proof-CG-NoRelay}.
\end{IEEEproof}
\vspace{1mm}

The two conditions over the source-to-relay link presented above can be
interpreted as follows. In the first case, $\SNR{31} \leq \SNR{11}/(1
+\SNR{12})$ implies that, by treating the interference from source~$2$ as noise,
destination~$1$ can still have a better observation on source~$1$'s signal than
the relay does. Therefore, the relay's observation cannot help much for
destination~$1$ to decode its own signal.

On the other hand, $\SNR{31} \leq \SNR{21}/(1 +\SNR{22})$ implies that, by
treating its own signal as noise, destination~$2$ can still have a better
observation on source~$1$'s signal than the relay does. Therefore, the relay's
observation cannot help much for destination~$2$ to learn/decode the
interference from source~$1$.

\subsection{Numerical Example}

To illustrate the regimes described before, we plot the maximum attainable
sum-rate for the outer bound and each inner bound in Fig.~\ref{fig:CG-plots_a}.
Additionally, we delimit each regime with vertical dashed lines and we add the
Han-Kobayashi scheme as a means of comparison. The SNR of each link in the
channel remains fixed while we vary the SNR of the source-to-relay link
\SNR{31}.

All the inner bounds present in the figure are the simplified versions used in
the computation of the gap, i.e., there is no maximization of the PDs employed
in them. The curve labeled DF is the maximum achievable rate attained by either
the simplified inner bound of Proposition~\ref{th:CG-DF} or~\ref{th:CG-pDF}; the
reader should refer to the appropriate appendix for details. The HK inner bound
is not optimized either since we use the auxiliaries proposed
in~\cite{telatar_bounds_2007}, but this is needed to make a fair comparison with
our schemes. Moreover, Corollary~\ref{cl:OB-AG} is the outer bound used in here.


We see that when the source-to-relay link is strong DF outperforms CF, namely in
the regime labeled ``f-DF'', i.e., when $\SNR{31}\geq\SNR{11}$. As the quality
of this link degrades, CF achieves higher rates and eventually surpasses DF,
mainly in the `CF'' regime, i.e., when $\SNR{31}<\SNR{21}$. Below certain
threshold in the quality of the source-to-relay link, the DF scheme even
achieves lower rates than the HK scheme. The cause of this might lie in the
numerous simplifications made to the scheme. However, due to the many
auxiliaries present in the scheme, we did not carry out an extensive
optimization of the scheme to prove this conjecture. Finally, when the
source-to-relay link is really weak, CF performs as good as the Han-Kobayashi
scheme.

Another way of analyzing these curves is by looking at the gap per dimension,
as in Fig.~\ref{fig:CG-plots_b}. Here, the maximum theoretical gap in each
regime is represented by horizontal dashed lines, and we see that they hold.

\section{Summary and Concluding Remarks}
\label{sec:Summary}

We derived a novel outer bound and two inner bounds for a class of IRCs where 
the relay can only observe one of the sources. These bounds allowed us to 
identify the main SNR regimes of interest, and for them, we found the adequate 
relaying strategies that achieve capacity of the Gaussian IRC to within a 
constant gap regardless of the channel parameters.

While the proposed inner and outer bounds suggest the existence of different SNR 
regimes for the Gaussian IRC, in which different coding strategies are needed to 
achieve a constant gap to capacity, whether there exists a single coding scheme 
that achieves the constant gap in all SNR regimes is still an open question. In 
other words, there may be ways to improve the outer bound, the inner bounds, or 
both, which remains an interesting future work.

Additionally, the general IRC where the relay observe both sources is not an 
straightforward extension of our work. The central difficulty lies in the way of 
modeling the interference signals used in the injective semideterministic model 
and hence the derivation of an adequate outer bound. Since in the general IRC 
$X_3$ can be arbitrarily correlated to both $X_1$ and $X_2$, the interference 
signal $S_k$ is no longer independent of the input $X_j$, with $(k,j) \in 
\left\{ (1,2), (2,1) \right\}$. This, in turn, forbids us of single-letterizing 
the outer bound the way we did. A new technique to derive outer bounds for this 
problem is therefore needed, which also remains as future work.

\appendices
\section{Strongly Typical Sequences and Delta-Convention}
\label{app:typical}

Following~\cite{csiszar1982information}, we use in this paper \emph{strongly
typical sets} and the so-called \emph{Delta-Convention}. 
Some useful facts are recalled here.
Let $X$ and $Y$ be random variables on some finite sets $\cX$ and $\cY$,
respectively. We denote by $p_{X,Y}$ (resp. $p_{Y|X}$, and $p_X$) the joint
probability distribution of $(X,Y)$ (resp. conditional distribution of $Y$ given
$X$, and marginal distribution of $X$). 

\vspace{1mm}
\begin{definition}[Number of occurrences]
$\,$ For any sequence $x^n\in\cX^n$ and any symbol $a\in\cX$, notation
$N(a|x^n)$ stands for the number of occurrences of $a$ in $x^n$.
\end{definition}

\vspace{1mm}
\begin{definition}[Typical sequence]
A sequence $x^n\in\cX^n$ is called \emph{(strongly) $\delta$-typical} w.r.t.\
$X$ (or simply \emph{typical} if the context is clear) if
\[
\abs{\frac1n N(a|x^n) - p_X(a)} \leq \delta \ \text{ for each } a\in\cX \ ,
\]
and $N(a|x^n)=0$ for each $a\in\cX$ such that $p_X(a)=0$.
The set of all such sequences is denoted by $\typ{n}{X}$.
\end{definition}

\vspace{1mm}
\begin{definition}[Conditionally typical sequence]
$\,$ Let $x^n\in\cX^n$. A sequence $y^n\in\cY^n$ is called \emph{(strongly)
$\delta$-typical (w.r.t. $Y$) given $x^n$}  if
\begin{multline*}
\abs{\frac1n N(a,b|x^n,y^n) - \frac1n N(a|x^n)p_{Y|X}(b|a)} \leq \delta \\ 
\text{for each } a\in\cX, b\in\cY \ ,
\end{multline*}
and, $N(a,b|x^n,y^n)=0$ for each $a\in\cX$, $b\in\cY$ such that
$p_{Y|X}(b|a)=0$.
The set of all such sequences is denoted by $\typ{n}{Y|x^n}$.
\end{definition}

\newtheorem{DeltaConvention}{Delta-Convention~\cite{
csiszar1982information}}
\renewcommand{\theDeltaConvention}{\unskip}
\vspace{1mm}
\begin{DeltaConvention}
For any sets $\cX$, $\cY$, there exists a sequence $\{\delta_n\}_{n\in\bN^*}$
such that the lemmas stated below hold.\footnote{%
	As a matter of fact, $\delta_n\to0$ and $\sqrt{n}\,\delta_n\to\infty$ as
$n\to\infty$.}
From now on, typical sequences are understood with $\delta=\delta_n$. 
Typical sets are still denoted by $\typ{n}{\cdot}$.
\end{DeltaConvention}

\vspace{1mm}
\begin{lemma}[{\hspace{1sp}\cite[Lemma~1.2.12]{csiszar1982information}}]
There exists a sequence $\eta_n\toas{n\to\infty}0$ such that
\[
p_X(\typ{n}{X}) \geq 1 - \eta_n \ .
\]
\end{lemma}

\vspace{1mm}
\begin{lemma}[{\hspace{1sp}\cite[Lemma~1.2.13]{csiszar1982information}}]
\label{lem:cardTyp}
There exists a sequence $\eta_n\toas{n\to\infty}0$ such that, for each
$x^n\in\typ{n}{X}$,
\begin{IEEEeqnarray*}{c}
\abs{\frac1n \log \norm{\typ{n}{X}} - H(X)} \leq \eta_n 		\ ,\\
\abs{\frac1n \log \norm{\typ{n}{Y|x^n}} - H(Y|X)} \leq \eta_n	\ .
\end{IEEEeqnarray*}
\end{lemma}

\vspace{1mm}
\begin{lemma}[Asymptotic equipartition property]
\label{lem:AEP}
$\!$There exists a sequence $\eta_n\toas{n\to\infty}0$ such that, for each
$x^n\in\typ{n}{X}$ and each $y^n\in\typ{n}{Y|x^n}$,
\begin{IEEEeqnarray*}{c}
\abs{-\frac1n \log p_X(x^n) - H(X)} \leq \eta_n 			
\ ,\\
\abs{-\frac1n \log p_{Y|X}(y^n|x^n) - H(Y|X)} \leq \eta_n		\ .
\end{IEEEeqnarray*}
\end{lemma}

\vspace{1mm}
\begin{lemma}[Joint typicality lemma~\cite{gamal_network_2011}]
\label{lem:jointTypicality}
There exists a sequence $\eta_n\toas{n\to\infty}0$ such that
\begin{multline*}
\abs{-\frac1n \log p_Y(\typ{n}{Y|x^n}) - I(X;Y)} \leq \eta_n \\
\text{for each }x^n\in\typ{n}{X}\ .
\end{multline*}
\end{lemma}

\begin{IEEEproof}
\begin{IEEEeqnarray*}{rCl}
p_Y(\typ{n}{Y|x^n})
	&=&					
\sum_{y^n\in\typ{n}{Y|x^n}} p_Y(y^n)				\\
	&\stackrel{(a)}{\leq}&	\norm{\typ{n}{Y|x^n}}\,2^{-n[H(Y)-\alpha_n]}	
\\
	&\stackrel{(b)}{\leq}&	2^{n[H(Y|X)+\beta_n]}\,2^{-n[H(Y)-\alpha_n]}
\\
	&=&					
2^{-n[I(X;Y)-\beta_n-\alpha_n]}					\ ,
\end{IEEEeqnarray*}
where
\begin{itemize}
\item step~$(a)$ follows from the fact that $\typ{n}{Y|x^n}\subset\typ{n}{Y}$
and Lemma~\ref{lem:AEP}, for some sequence $\alpha_n\toas{n\to\infty}0$,
\item step~$(b)$ from Lemma~\ref{lem:cardTyp}, for some sequence
$\beta_n\toas{n\to\infty}0$.
\end{itemize}
The reverse inequality $p_Y(\typ{n}{Y|x^n})\geq 2^{-n[I(X;Y)+\beta_n+\alpha_n]}$
can be proved following similar argument. 
\end{IEEEproof}

\section{Proof of Theorem~\ref{th:OB-IS_IRC} (IS-IRC Outer Bound)}
\label{sec:AP-Proof_OB}

The proof follows by using a similar approach to that developed
in~\cite{telatar_bounds_2007} and it was partially presented
in~\cite{bassi_allerton_2013,bassi_isit_2014}. As explained before, the inputs
$X_1$ and $X_3$ are arbitrarily correlated and they are independent of $X_2$.
Since we are not considering noise correlation in the outputs, the interference
signals $\ve{S_1}$ and $S_2$ are therefore independent.

First, let us recall that the inputs $X_1^n$ and $X_2^n$ are functions of the
messages $W_1$ and $W_2$, each one independent of the other, and the relay's
input is a deterministic function of its past observations, i.e., $X_{3i} =
\phi_i\left(Y_3^{i-1}\right)$, $i\in[1:n]$. Then, we add two new random
variables $\ve{V_1^{}}^{\!n}$ and $V_2^n$, which are obtained by passing
$X_1^n$, $X_2^n$ and $X_3^n$ through the memoryless channel $p_{\ve{S_1} \vert
X_1 X_3}p_{S_2\vert X_2}$.

A multi-letter outer bound on each rate can be derived using Fano's inequality,
i.e.,
$$ n(R_k - \epsilon_n) \leq \I{X_k^n}{Y_k^n}, $$
where $\epsilon_n$ denotes
a sequence such that $\epsilon_n\rightarrow 0$ as $n\rightarrow\infty$.
Therefore, we present different derivations of \I{X_k^n}{Y_k^n} in the sequel.
We first see that
\begin{subequations}\label{eq:AP-Proof_OB-1}
\begin{align}
\I{X_1^n}{Y_1^n} &\leq \I{X_1^n X_3^n}{Y_1^n} \nonumber\\
 &= \H{Y_1^n} - \HC{Y_1^n}{X_1^n X_3^n} \nonumber\\
 &= \H{Y_1^n} - \HC{S_2^n}{X_1^n X_3^n} \label{eq:AP-Proof_OB-1a}\\
 &= \H{Y_1^n} - \boxed{\H{S_2^n}}, \label{eq:AP-Proof_OB-1b}
\end{align}
\end{subequations}
where~\eqref{eq:AP-Proof_OB-1a} follows from the IS model; and
in~\eqref{eq:AP-Proof_OB-1b} we take into account that the interference signal
$S_2^n$ is independent of the inputs $(X_1^n X_3^n)$.
We can provide the interference $X_2^n$,
\begin{equation}
\I{X_1^n}{Y_1^n} \leq \IC{X_1^n X_3^n}{Y_1^n}{X_2^n}, \label{eq:AP-Proof_OB-2}
\end{equation}
where~\eqref{eq:AP-Proof_OB-2} follows from the fact that $X_2^n$ is
independent of $(X_1^n X_3^n)$.
Also, we can augment the bound with the auxiliary $V_1^n$,
\begin{subequations}\label{eq:AP-Proof_OB-3}
\begin{align}
\MoveEqLeft[1]
\I{X_1^n}{Y_1^n} \leq \I{X_1^n X_3^n}{Y_1^n V_1^n} \nonumber\\
 &= \I{X_1^n X_3^n}{V_1^n} +\IC{X_1^n X_3^n}{Y_1^n}{V_1^n} \nonumber\\
 &= \H{V_1^n} -\HC{V_1^n}{X_1^n X_3^n} +\HC{Y_1^n}{V_1^n} -\HC{Y_1^n}{X_1^n
X_3^n} \label{eq:AP-Proof_OB-3a}\\
 &= \boxed{\H{S_1^n}} -\HC{Y_2^n}{X_1^n X_2^n X_3^n} +\HC{Y_1^n}{V_1^n}
-\boxed{\H{S_2^n}}, \label{eq:AP-Proof_OB-3b}
\end{align}
\end{subequations}
where in the fourth term of~\eqref{eq:AP-Proof_OB-3a} we use the Markov
chain $V_1^n \mkv (X_1^n X_3^n) \mkv (\cdots)$; and~\eqref{eq:AP-Proof_OB-3b}
is due to the channel property and the fact that interchanging $V_1$ and
$S_1$ does not change the entropies in question, i.e., $\H{V_1^n} = \H{S_1^n}$
and $\HC{V_1^n}{X_1^n X_3^n} = \HC{S_1^n}{X_1^n X_3^n} = \HC{S_1^n}{X_1^n X_2^n
X_3^n} = \HC{Y_2^n}{X_1^n X_2^n X_3^n}$.
We repeat the same procedure with the auxiliary $\ve{V_1^n}$,
\begin{subequations}\label{eq:AP-Proof_OB-4}
\begin{align}
\MoveEqLeft[1]
\I{X_1^n}{Y_1^n} \leq \I{X_1^n X_3^n}{Y_1^n \ve{V_1^n}} \nonumber\\
 &= \I{X_1^n X_3^n}{\ve{V_1^n}} +\IC{X_1^n X_3^n}{Y_1^n}{\ve{V_1^n}} 
\displaybreak[2]\nonumber\\
 &= \H{\ve{V_1^n}} -\HC{\ve{V_1^n}}{X_1^n X_3^n} +\HC{Y_1^n}{\ve{V_1^n}}
-\HC{Y_1^n}{X_1^n X_3^n} \displaybreak[2]\label{eq:AP-Proof_OB-4a}\\
 &= \boxed{\H{\ve{S_1^n}}} -\!\HC{Y_2^n Y_3^n}{X_1^n X_2^n X_3^n}\!
+\!\HC{Y_1^n}{\ve{V_1^n}}\!-\!\boxed{\H{S_2^n}}, \label{eq:AP-Proof_OB-4b}
\end{align}
\end{subequations}
where in~\eqref{eq:AP-Proof_OB-4a} we use the Markov chain $\ve{V_1^n} \mkv
(X_1^n X_3^n) \mkv (\cdots)$; and in~\eqref{eq:AP-Proof_OB-4b} we again
interchange $\ve{V_1}$ and $\ve{S_1}$, i.e., $\H{V_1^n} = \H{S_1^n}$
and $\HC{\ve{V_1^n}}{X_1^n X_3^n} = \HC{\ve{S_1^n}}{X_1^n X_3^n} =
\HC{\ve{S_1^n}}{X_1^n X_2^n X_3^n} = \HC{Y_2^n Y_3^n}{X_1^n X_2^n X_3^n}$.
We can now increase the bound with both $X_2^n$ and $V_1^n$,
\begin{subequations}\label{eq:AP-Proof_OB-5}
\begin{align}
\MoveEqLeft[1]
\I{X_1^n}{Y_1^n} \leq \IC{X_1^n X_3^n}{Y_1^n V_1^n}{X_2^n} \nonumber\\
 &= \IC{X_1^n X_3^n}{V_1^n}{X_2^n} +\IC{X_1^n X_3^n}{Y_1^n}{V_1^n X_2^n}
\nonumber\\
 &= \HC{V_1^n}{X_2^n} -\HC{V_1^n}{X_1^n X_3^n} +\IC{X_1^n X_3^n}{Y_1^n}{V_1^n
X_2^n} \label{eq:AP-Proof_OB-5b}\\
 &= \boxed{\H{S_1^n}} -\HC{Y_2^n}{X_1^n X_2^n X_3^n} +\IC{X_1^n
X_3^n}{Y_1^n}{V_1^n X_2^n}, \label{eq:AP-Proof_OB-5c}
\end{align}
\end{subequations}
where the key steps in~\eqref{eq:AP-Proof_OB-5b} and~\eqref{eq:AP-Proof_OB-5c}
are the same as in~\eqref{eq:AP-Proof_OB-3a} and~\eqref{eq:AP-Proof_OB-3b}.
Similarly, we can derive
\begin{align}
\I{X_1^n}{Y_1^n} &\leq \boxed{\H{\ve{S_1^n}}} -\HC{Y_2^n Y_3^n}{X_1^n X_2^n
X_3^n} \nonumber\\
 &\quad +\IC{X_1^n X_3^n}{Y_1^n}{\ve{V_1^n} X_2^n}. \label{eq:AP-Proof_OB-6}
\end{align}

In an analogous way as~\eqref{eq:AP-Proof_OB-1}, \eqref{eq:AP-Proof_OB-2},
\eqref{eq:AP-Proof_OB-3}, and \eqref{eq:AP-Proof_OB-5}, we derive similar
bounds for the rate $R_2$,
\begin{align}
\I{X_2^n}{Y_2^n} &\leq \H{Y_2^n} -\boxed{\H{S_1^n}} \label{eq:AP-Proof_OB-7},\\
\I{X_2^n}{Y_2^n} &\leq \IC{X_2^n}{Y_2^n}{X_1^n X_3^n},
\label{eq:AP-Proof_OB-8}\\
\I{X_2^n}{Y_2^n} &\leq \boxed{\H{S_2^n}} -\HC{Y_1^n}{X_1^n X_2^n X_3^n} 
\nonumber\\
 &\quad +\HC{Y_2^n}{V_2^n} -\boxed{\H{S_1^n}}, \label{eq:AP-Proof_OB-9}\\
\I{X_2^n}{Y_2^n} &\leq \boxed{\H{S_2^n}} -\HC{Y_1^n}{X_1^n X_2^n X_3^n} 
\nonumber\\
 &\quad +\IC{X_2^n}{Y_2^n}{X_1^n V_2^n X_3^n}. \label{eq:AP-Proof_OB-10}
\end{align}
Additionally, if we add the sequence $Y_3^n$ next to $Y_2^n$ in the first steps
of the derivation of~\eqref{eq:AP-Proof_OB-7} and~\eqref{eq:AP-Proof_OB-9}, we
obtain
\begin{align}
\I{X_2^n}{Y_2^n} &\leq \H{Y_2^n Y_3^n} -\boxed{\H{\ve{S_1^n}}},
\label{eq:AP-Proof_OB-11}\\
\I{X_2^n}{Y_2^n} &\leq \boxed{\H{S_2^n}} -\HC{Y_1^n}{X_1^n X_2^n X_3^n} 
\nonumber\\
 &\quad +\HC{Y_2^n Y_3^n}{V_2^n} -\boxed{\H{\ve{S_1^n}}}. 
\label{eq:AP-Proof_OB-12}
\end{align}

The use of Fano's inequality and all the possible linear combinations of the
expressions~\eqref{eq:AP-Proof_OB-1}--\eqref{eq:AP-Proof_OB-12} where the boxed
terms get canceled gives rise to multi-letter bounds that can be
single-letterized, as summarized in Table~\ref{tab:AP-Proof_bounds}. For
instance, \eqref{eq:AP-Proof_OB-2} and~\eqref{eq:AP-Proof_OB-8} allow us to find
bounds on the single rates, whereas the addition of~\eqref{eq:AP-Proof_OB-5}
and~\eqref{eq:AP-Proof_OB-7} gives us the sum-rate~\eqref{eq:OB-IS_IRC4},
\begin{subequations}
\begin{align}
\MoveEqLeft[1]
n(R_1 +R_2 -\epsilon_n') \leq \I{X_1^n}{Y_1^n} +\I{X_2^n}{Y_2^n} \nonumber\\
 &\leq \IC{X_1^n X_3^n}{Y_1^n}{V_1^n X_2^n} +\I{X_1^n X_2^n X_3^n}{Y_2^n}
\label{eq:AP-Proof_OB-13a}\\
 &\leq \sum_{i=1}^n \IC{X_{1i} X_{3i}}{Y_{1i}}{V_{1i} X_{2i}} +\I{X_{1i} X_{2i}
X_{3i}}{Y_{2i}} \!\! \label{eq:AP-Proof_OB-13b}\\
 &= n [\, \IC{X_1 X_3}{Y_1}{V_1 X_2 Q} +\IC{X_1 X_2 X_3}{Y_2}{Q} ],
\label{eq:AP-Proof_OB-13c}
\end{align}
\end{subequations}
where~\eqref{eq:AP-Proof_OB-13a} follows from the addition
of~\eqref{eq:AP-Proof_OB-5c} and~\eqref{eq:AP-Proof_OB-7};
\eqref{eq:AP-Proof_OB-13b} is due to the chain rule of the mutual information,
the fact that removing conditioning increases the entropy, and the Markov chain
$(Y_{1i} Y_{2i}) \mkv (X_{1i} X_{2i} X_{3i}) \mkv (\cdots)$;
and~\eqref{eq:AP-Proof_OB-13c} follows from the addition of the time-sharing
variable $Q$ uniformly distributed in $[1:n]$.

\begin{table}[t]
\begin{center}
\begin{tabular}{rcl}
\toprule
$R_1$ & \eqref{eq:OB-IS_IRC1} & \eqref{eq:AP-Proof_OB-2}*\\
 & \eqref{eq:OB-IS_IRC2} & \eqref{eq:AP-Proof_OB-2}\\
\midrule
$R_2$ & \eqref{eq:OB-IS_IRC3} & \eqref{eq:AP-Proof_OB-8}\\
\midrule
$R_1+R_2$ & \eqref{eq:OB-IS_IRC4} & \eqref{eq:AP-Proof_OB-5}
+\eqref{eq:AP-Proof_OB-7}\\
 & \eqref{eq:OB-IS_IRC5} & \eqref{eq:AP-Proof_OB-3} +\eqref{eq:AP-Proof_OB-9}\\
 & \eqref{eq:OB-IS_IRC6} & \eqref{eq:AP-Proof_OB-1} +\eqref{eq:AP-Proof_OB-10}\\
 & \eqref{eq:OB-IS_IRC7} & \eqref{eq:AP-Proof_OB-5}*+\eqref{eq:AP-Proof_OB-7}\\
 & \eqref{eq:OB-IS_IRC8} & \eqref{eq:AP-Proof_OB-3}*+\eqref{eq:AP-Proof_OB-9}\\
 & \eqref{eq:OB-IS_IRC9} & \eqref{eq:AP-Proof_OB-1}*+\eqref{eq:AP-Proof_OB-10}\\
 & \eqref{eq:OB-IS_IRC10} &\eqref{eq:AP-Proof_OB-6}*+\eqref{eq:AP-Proof_OB-11}\\
 & \eqref{eq:OB-IS_IRC11} &\eqref{eq:AP-Proof_OB-4}*+\eqref{eq:AP-Proof_OB-12}\\
\midrule
$2R_1+R_2$ & \eqref{eq:OB-IS_IRC12} & \eqref{eq:AP-Proof_OB-5}
+\eqref{eq:AP-Proof_OB-1} +\eqref{eq:AP-Proof_OB-9}\\
 & \eqref{eq:OB-IS_IRC13} & \eqref{eq:AP-Proof_OB-5} +\eqref{eq:AP-Proof_OB-1}*%
+\eqref{eq:AP-Proof_OB-9}\\
 & \eqref{eq:OB-IS_IRC14} & \eqref{eq:AP-Proof_OB-5}*+\eqref{eq:AP-Proof_OB-1}
+\eqref{eq:AP-Proof_OB-9}\\
 & \eqref{eq:OB-IS_IRC15} & \eqref{eq:AP-Proof_OB-5}*+\eqref{eq:AP-Proof_OB-1}*%
+\eqref{eq:AP-Proof_OB-9}\\
 & \eqref{eq:OB-IS_IRC16} & \eqref{eq:AP-Proof_OB-6}*+\eqref{eq:AP-Proof_OB-1}
+\eqref{eq:AP-Proof_OB-12}\\
 & \eqref{eq:OB-IS_IRC17} & \eqref{eq:AP-Proof_OB-6}*+\eqref{eq:AP-Proof_OB-1}*%
+\eqref{eq:AP-Proof_OB-12}\\
\midrule
$R_1+2R_2$ & \eqref{eq:OB-IS_IRC18} & \eqref{eq:AP-Proof_OB-3}
+\eqref{eq:AP-Proof_OB-10} +\eqref{eq:AP-Proof_OB-7}\\
 & \eqref{eq:OB-IS_IRC19} & \eqref{eq:AP-Proof_OB-3}*+\eqref{eq:AP-Proof_OB-10}
+\eqref{eq:AP-Proof_OB-7}\\
 & \eqref{eq:OB-IS_IRC20} & \eqref{eq:AP-Proof_OB-4}*+\eqref{eq:AP-Proof_OB-10}
+\eqref{eq:AP-Proof_OB-11}\\
\bottomrule
\end{tabular}
\end{center}
 \caption{Combination of multi-letter outer bounds. Terms with * need the
addition of $Y_3^n$.\label{tab:AP-Proof_bounds}}
\end{table}

In this way, we obtain all the bounds in~\eqref{eq:OB-IS_IRC} except for the
ones with the pair $(Y_1 Y_3)$. For them, we need to add the sequence $Y_3^n$
next to $Y_1^n$ before applying the chain rule in the mutual information. These
terms are denoted with * in Table~\ref{tab:AP-Proof_bounds}. For example,
continuing from~\eqref{eq:AP-Proof_OB-13a} we obtain the
bound~\eqref{eq:OB-IS_IRC7},
\begin{subequations}
\begin{align}
\MoveEqLeft[0]
n(R_1 +R_2 -\epsilon_n') \nonumber\\
 &\leq \IC{X_1^n X_3^n}{Y_1^n Y_3^n}{V_1^n X_2^n} +\I{X_1^n X_2^n X_3^n}{Y_2^n} 
\nonumber\\
 &\leq \sum_{i=1}^n \IC{X_{1i}}{Y_{1i} Y_{3i}}{V_{1i} X_{2i} X_{3i}} +\I{X_{1i}
X_{2i} X_{3i}}{Y_{2i}} \!\!\! \label{eq:AP-Proof_OB-14a}\\
 &= n [\, \IC{X_1}{Y_1 Y_3}{V_1 X_2 X_3 Q} +\IC{X_1 X_2 X_3}{Y_2}{Q} ]
\label{eq:AP-Proof_OB-14b}
\end{align}
\end{subequations}
where~\eqref{eq:AP-Proof_OB-14a} follows from the fact that $X_{3i}$ is a
function of $Y_3^{i-1}$.

\section{Proof of Corollary~\ref{cl:OB-AG}}
\label{sec:AP-Proof-OB-decorr}

The expression of the bounds~\eqref{eq:OB-IS_IRC1}--\eqref{eq:OB-IS_IRC3} in
the Gaussian case is
\begin{align}
R_1 &\leq \C{(1-\rho^2)(\SNR{11} +\SNR{31})}, \label{eq:AP-decorr_b1}\\
R_1 &\leq \C{\SNR{11} +\SNR{13} +2\rho \sqrt{\SNR{11} \SNR{13}}},
\label{eq:AP-decorr_b2}\\
R_2 &\leq \C{\SNR{22}}, \label{eq:AP-decorr_b3}
\end{align}
where we assume the channel coefficients $h_{11}$ and $h_{13}$ have the same
sign, otherwise, the analysis is the same by inverting the sign in $\rho$. For
any $|\rho|\leq 1$, we can upper bound the previous terms as follows
\begin{align}
R_1 &\leq \C{\SNR{11} +\SNR{31}}, \\
R_1 &\leq \C{\SNR{11} +\SNR{13}} + \frac{1}{2}, \displaybreak[2]\\
R_2 &\leq \C{\SNR{22}},
\end{align}
which, in turn, gives us~\eqref{eq:OB-Gaussian1}--\eqref{eq:OB-Gaussian3}.

All the other bounds behave similarly. If both $X_1$ and $X_3$ appear in the
conditioning part of a mutual information, it does not depend on $\rho$,
like~\eqref{eq:AP-decorr_b3}. If only $X_3$ appears in the conditioning, it
depends on $(1-\rho^2)$, like~\eqref{eq:AP-decorr_b1}. Otherwise, it
depends on $2\rho \sqrt{(\,\cdot\,)}$, like~\eqref{eq:AP-decorr_b2}. In the
first two situations, the expressions are maximized with its value at $\rho=0$,
whereas, the last one has its maximum at $\rho=1$.

The bounds containing $V_1$ in the conditioning part, but not $X_3$,
e.g.~\eqref{eq:OB-IS_IRC4}, present a more complicated behavior and it is not
clear which value of $\rho$ maximizes the bound. We analyze the
sum-rate~\eqref{eq:OB-IS_IRC4} in the sequel.

Let us first define
\begin{align*}
  \pmb{H} &= \begin{bmatrix}
    h_{11} & h_{13} \\ h_{21} & h_{23}
  \end{bmatrix}, 
  \\
  \pmb{Q} &= \begin{bmatrix}
    1 & \rho \\ \rho & 1
  \end{bmatrix}  
  = \underbrace{\begin{bmatrix}
    \frac{1}{\sqrt{2}} &  \frac{1}{\sqrt{2}} \\
    \frac{1}{\sqrt{2}} & -\frac{1}{\sqrt{2}}
  \end{bmatrix} }_{\pmb{U}}
  \underbrace{\begin{bmatrix}
    1+\rho & 0 \\ 0 & 1-\rho
  \end{bmatrix} }_{\pmb{\Lambda}}
  \underbrace{
\begin{bmatrix}
    \frac{1}{\sqrt{2}} &  \frac{1}{\sqrt{2}} \\
    \frac{1}{\sqrt{2}} & -\frac{1}{\sqrt{2}}
  \end{bmatrix} }_{\pmb{U}^T},
\end{align*}%
where we have normalized the sources' power and noise power. We are interested
in 
\begin{align*}
D_0 &\triangleq \det( \pmb{I} + \pmb{H} \pmb{H}^T ) \\ &=
 \det(\pmb{I} + \pmb{H} \pmb{U}  \pmb{U}^T \pmb{H}^T) =
 \det(\pmb{I} + \pmb{G} \pmb{G}^T), \\
D &\triangleq \det( \pmb{I} + \pmb{H} \pmb{Q} \pmb{H}^T ) \\ &=
 \det(\pmb{I} + \pmb{H} \pmb{U} \pmb{\Lambda} \pmb{U}^T \pmb{H}^T) =
 \det(\pmb{I} + \pmb{G} \pmb{\Lambda} \pmb{G}^T)
\end{align*}%
where we define $\pmb{G} \triangleq \pmb{H} \pmb{U} = [g_{ij}]_{i,j=1,2}$. For
convenience, we also define the normalized matrix $\pmb{V}$ such that
\begin{align*}
  \pmb{G} &= \begin{bmatrix} \sqrt{G_1} & 0 \\ 0 & \sqrt{G_2} \end{bmatrix}
\pmb{V}, \quad G_i \triangleq { g_{i1}^2 + g_{i2}^2 }, \ i=1,2
\end{align*}%
where $v_{ij} \triangleq g_{ij}/\sqrt{G_i}$. Note that $v_{i1}^2 +
v_{i2}^2 = 1$, $i=1,2$. We let $V_{ij} \triangleq v_{ij}^2$ hereafter. 

Then, we can rewrite 
\begin{align*}
D_0 &= 1 +G_1 +G_2 +G_1 G_2 \underbrace{\det(\pmb{V} \pmb{V}^T)}_{\gamma} \\
D &= 1 + G_1 (1+\underbrace{(V_{11} - V_{12})}_{\alpha_1}\rho)
 +G_2 (1+\underbrace{(V_{21} -V_{22})}_{\alpha_2} \rho) \\
&\quad + G_1 G_2 \gamma (1-\rho^2) 
\end{align*}%
where $\gamma\in[0,1]$ and $\alpha_1,\alpha_2 \in [-1,1]$. In fact,
$\gamma$ can be presented as a function of $\alpha_1$ and $\alpha_2$%
\begin{subequations}
\begin{align}
\gamma &= (v_{11}v_{22} - v_{21}v_{12})^2 \\
 &\ge \bigl(\sqrt{V_{11}V_{22}} - \sqrt{V_{21}V_{12}}\bigr)^2  \\
 &= \frac{1-\alpha_1 \alpha_2}{2} - \frac{1}{2}
\sqrt{(1-\alpha_1^2)(1-\alpha_2^2)} \triangleq \gamma_*.
\label{eq:AP-decorr_gamma}
\end{align}
\end{subequations}

Given the sum-rate~\eqref{eq:OB-IS_IRC4},
\begin{align*}
\MoveEqLeft[0.25]
R_1 + R_2 \leq \IC{X_1 X_3}{Y_1}{V_1 X_2} \!+\!\I{X_1 X_2 X_3}{Y_2} \nonumber\\
 &= \IC{X_1 X_3}{Y_1 V_1}{X_2} \!-\!\IC{X_1 X_3}{V_1}{X_2} \!+\!
 \I{X_1 X_2X_3}{Y_2},
\end{align*}
the ultimate goal is to quantify the maximum gap between the value of this
bound with and without correlation in the inputs $(X_1 X_3)$. In other words, we
shall obtain an upper bound on
\begin{equation}
\frac{D}{D_0} \frac{1+G_2}{1+G_2 (1+\alpha_2\rho)} \frac{1+G_2 (1+\alpha_2\rho)
+\SNR{22}}{1 +G_2 +\SNR{22}}. \label{eq:AP-decorr_bound}
\end{equation}
If $\SNR{22} \rightarrow 0$, the expression~\eqref{eq:AP-decorr_bound} tends
to $D/D_0$, and since the eigenvalues of $\pmb{\Lambda}$ are less or equal than
$2$, it can be easily upper-bounded,
\begin{equation*}
\frac{D}{D_0} = \frac{\det(\pmb{I} + \pmb{G} \pmb{\Lambda}
\pmb{G}^T)}{\det(\pmb{I} + \pmb{G} \pmb{G}^T)} \leq \frac{\det(\pmb{I} +
2 \pmb{G} \pmb{G}^T)}{\det(\pmb{I} + \pmb{G} \pmb{G}^T)} \leq 2.
\end{equation*}

On the other hand, if $\SNR{22} \rightarrow \infty$,~\eqref{eq:AP-decorr_bound} 
becomes
\begin{align*}
\frac{D}{D_0} \frac{1+G_2}{1+G_2 (1+\alpha_2\rho)} &=
 \frac{ 1 + G_1 \displaystyle \frac{1+\alpha_1\rho + G_2 \gamma
(1-\rho^2)}{1+G_2 (1+\alpha_2 \rho)} }{
  1 + G_1 \displaystyle \frac{1+G_2 \gamma}{1+G_2} } \\
  &= \frac{ 1 + G_1 A }{ 1 + G_1 B }.
\end{align*}%
We observe that this function is upper-bounded by $1$ when $A\le B$, while it is
otherwise upper-bounded by $A/B$. Therefore, it suffices to find an upper bound
on $A/B$ that can be rewritten as%
\begin{align}
\frac{A}{B} &= \frac{ (1 +\! \alpha_1 \rho) +\! G_2 \gamma (1-\!\rho^2) +\!
  G_2 (1+\!\alpha_1 \rho) +\! G_2^2 \gamma (1-\!\rho^2) }{ (1+G_2\gamma)(1+
  G_2 (1+\alpha_2 \rho)) } \nonumber\\
&= (1 + \alpha_1 \rho) \frac{ 1+G_2 }{ 1+ G_2 (1+\alpha_2 \rho) }
 \frac{1 +G_2 \displaystyle \frac{\gamma (1-\rho^2)}{1 + \alpha_1 \rho} }{
1 +G_2\gamma}. \label{eq:AP-decorr_bound_AB}
\end{align}
Without loss of generality, we assume that $\rho\ge0$. The case when $\rho<0$
follows straightforwardly by simply changing both signs of $\alpha_1$ and
$\alpha_2$. In the following, we shall show that
\begin{align*}
 \frac{A}{B} &\le 2.
\end{align*}%
First, from~\eqref{eq:AP-decorr_bound_AB}, we derive a trivial upper bound
\begin{subequations}
\begin{align}
\frac{A}{B} &\le (1\!+\!\alpha_1 \rho)
 \max\!\left\{ 1, \,\frac{1}{1\!+\!\alpha_2\rho}  \right\}
 \max\!\left\{ 1, \,\frac{1-\rho^2}{1\!+\!\alpha_1\rho} \right\}\!
 \label{eq:AP-decorr_bound_AB1}\\ 
&= \max\left\{ 1 -\rho^2, \,1 +\alpha_1\rho, \,\frac{1-\rho^2}{1+\alpha_2\rho},
\,\frac{1 + \alpha_1 \rho}{1+\alpha_2\rho} \right\},
 \label{eq:AP-decorr_bound_AB2}
\end{align}
\end{subequations}%
where both maximizations in~\eqref{eq:AP-decorr_bound_AB1} come from the
monotonicity of $\frac{1 +G_2x}{1 +G_2y}$ w.r.t. $G_2$ and that it is bounded by
the extreme values for $G_2=0$ and $G_2\rightarrow\infty$. Note that only the
last term in~\eqref{eq:AP-decorr_bound_AB2} is not always upper-bounded by $2$.
In the following, we focus on the case $\frac{1-\rho^2}{1+\alpha_1 \rho} < 1$,
i.e., $\alpha_1 > -\rho$, since the opposite would imply that the last term
in~\eqref{eq:AP-decorr_bound_AB2} is upper-bounded by the third term. In this
case~$(\alpha_1 > -\rho)$, the third term in \eqref{eq:AP-decorr_bound_AB}, and
thus $A/B$, is decreasing with $\gamma$. Therefore, the worst case in which
$A/B$ is maximized is when $\gamma$ achieves $\gamma_*$. It suffices to show
that
\begin{align*}
\sup_{G_2 \ge 0}  \frac{1\!+\!\alpha_1 \rho\!+\!G_2  \left( {1\!+\!\alpha_1
\rho} \!+\!\gamma_* ({1\!-\!\rho^2}) \right)\!+\!G_2^2 \gamma_* ({1\!-\!\rho^2})
}{ (1+ G_2 \gamma_*)(1 + G_2 (1+\alpha_2 \rho)) } \le 2,
\end{align*}%
$\forall\, {(\alpha_1,\alpha_2,\rho)\in\mathcal{A}}$ where we define the set
$\mathcal{A}$
\begin{align*}
\mathcal{A} \triangleq \{\alpha_1, \alpha_2 \in (-1,1),\ \rho \in (0,1)\ : \
\alpha_1 > \alpha_2,\ \alpha_1 > -\rho\}. 
\end{align*}%
We observe that for each point at the boundary of the set $\mathcal{A}$, the
objective function is upper-bounded by $2$. Note that, in the denominator,
$\gamma_* > 0$ since $\alpha_1 \ne \alpha_2$, and $1+\alpha_2\rho > 0$ since
$\rho<1$. Therefore, the objective function is the ratio between two quadratic
functions in the form $(a_0+a_1 G_2 + a_2 G_2^2)/((1 +b_1 G_2 )(1 +b_2 G_2 ))$
with $a_0, a_1, a_2 \ge 0$ and $b_1, b_2 > 0$, that are continuous functions of
$(\alpha_1, \alpha_2, \rho)$.  Let us first assume that $b_1\ne b_2$. It is
readily shown that
\begin{align}
f(G_2) &= \frac{a_0+a_1 G_2 + a_2 G_2^2}{(1 + b_1 G_2)(1 + b_2 G_2)}
 \label{eq:AP-decorr_obj_func} \\
 &= c_0 + \frac{c_1}{1+b_1 G_2} + \frac{c_2}{1+b_2 G_2}, \quad \forall\, G_2 
\end{align}%
where $(c_0,c_1,c_2)$ is a continuous function of $\left\{ a_i \right\}$ and
$\left\{ b_i \right\}$. Then, we differentiate the function $f(G_2)$
\begin{align*}
f'(G_2) = -\frac{b_1 c_1}{(1+b_1 G_2)^2} -\frac{b_2 c_2}{(1+b_2 G_2)^2}.
\end{align*}%
It is clear that there is at most one solution in $[0,\infty]$ such that
$f'(G_2) = 0$. If such a solution does not exist, then $f'(G_2)$ is either
strictly positive or strictly negative in $[0,\infty]$. In this case, both
extreme values $f(0)$ and $f(\infty)$ are upper-bounded by $2$
from~\eqref{eq:AP-decorr_bound_AB}. If such a solution does exist, it is in
the following form  
\begin{align}
G_2^* &= \frac{\beta -1}{b_1 -b_2 \beta}, \quad \beta \triangleq
\sqrt{-\frac{b_1 c_1}{b_2 c_2}}, \quad \frac{c_1}{c_2} < 0.
\label{eq:AP-decorr_xopt} 
\end{align}%
Note that the function $f$ defined in~\eqref{eq:AP-decorr_obj_func},
alternatively denoted as $f_{b_1,b_2}$, converges pointwise to $f_{b,b}$ when
$b_1,b_2\to b$, $\forall\,b>0$, and that $f'_{b_1,b_2}$ converges uniformly to
$f_{b,b}'$. Therefore, the solution~\eqref{eq:AP-decorr_xopt} holds even when
$b_1 = b_2$ by taking the limit.
Finally, let us define a set $\mathcal{B}$ of $(\alpha_1, \alpha_2, \rho)$ such
that $c_1 / c_2 < 0$ and $G_2^* \ge 0$. It remains to show that  
\begin{align}
\sup_{(\alpha_1, \alpha_2, \rho) \in \mathcal{A} \cap \mathcal{B}} f(G_2^*)
\le 2. \label{eq:AP-decorr_num_opt}
\end{align}%
Since $\mathcal{A} \cap \mathcal{B}$ is a bounded set and the objective
function is continuous in $(\alpha_1, \alpha_2, \rho)$ in $\mathcal{A} \cap
\mathcal{B}$, we can perform numerical optimization and obtain the value $2$,
which confirms the claim in~\eqref{eq:AP-decorr_num_opt}.

Similar steps can be performed in every other bound containing $V_1$ in the
conditioning, which concludes the proof.

\section{Proof of Theorem~\ref{th:IB-pDF} (Partial DF Scheme)}
\label{sec:AP-Proof-pDF}

Each source transmits $B$ messages during $B+1$ time blocks, each of them of
length $n$. The messages are sent using block-Markov coding and the destinations
employ backward decoding to retrieve them.

The second source splits its message $\tilde{m}_2$ into a common message $m_2$
and a private one $w_2$, with partial rates $R_{20}$ and $R_{22}$, respectively,
such that $R_{2}=R_{20}+R_{22}$. On the other hand, the first source splits its
message $\tilde{m}_1$ into three parts: $(m_1,w_1',w_1'')$. The relay decodes
and retransmits the common message and a part of the private one, i.e.,
$(m_1,w_1')$, whereas the other part is only decoded by the final destination.
The rate of the first user is therefore the sum of these three partial rates:
$R_{1}=R_{10}+R_{11}'+R_{11}''$.

\begin{table*}[t]
\begin{center}
\renewcommand{\arraystretch}{1.2} 
\begin{tabular}{|l|l|l|l|l|}
\hline
$b=1$ & $b=2$ & $\ldots$ & $b=B$ & $b=B+1$ \\
\hline
\hline
$v^n_3(1)$ & $v^n_3(t_{11})$ & $\ldots$ & $v^n_3(t_{1(B-1)})$ & $v^n_3(t_{1B})$
\\
\hline
$x^n_3(1,1)$ & $x^n_3(t_{11},w_{11}')$ & $\ldots$ & 
$x^n_3(t_{1(B-1)},w_{1(B-1)}')$ & $x^n_3(t_{1B},w_{1B}')$ \\
\hline
\hline
$v^n_1(1,t_{11})$ & $v^n_1(t_{11},t_{12})$ & $\ldots$ & 
$v^n_1(t_{1(B-1)},t_{1B})$ & $v^n_1(t_{1B},1)$ \\
\hline
$x^n_1(1,t_{11},1,w_{11}',w_{11}'')$ & 
$x^n_1(t_{11},t_{12},w_{11}',w_{12}',w_{12}'')$ & $\ldots$ & 
$x^n_1(t_{1(B-1)},t_{1B},w_{1(B-1)}',w_{1B}',w_{1B}'')$ & 
$x^n_1(t_{1B},1,w_{1B}',1,1)$ \\
\hline
\hline
$v^n_2(1)$ & $v^n_2(m_{21})$ & $\ldots$ & $v^n_2(m_{2(B-1)})$ & $v^n_2(m_{2B})$
\\
\hline
$x^n_2(1,1)$ & $x^n_2(m_{21},w_{21})$ &$\ldots$ &$x^n_2(m_{2(B-1)},w_{2(B-1)})$
& $x^n_2(m_{2B},w_{2B})$\\
\hline
\end{tabular}
\end{center}
\caption{Codewords in the proposed partial DF scheme for the IRC.}
\label{tab:AP-pDF-codewords}
\end{table*}

\subsection{Code Generation}
\begin{enumerate}
 \item Generate the time-sharing sequence $q^n$ where each element is
independent and identically distributed~(i.i.d.) according to the PD
$$ p(q^n) = \prod_{i=1}^n p_Q(q_i). $$
 \item For each sequence $q^n$, generate $2^{nT_{10}}$ conditionally
independent sequences $v_3^n(t_0)$, where $t_0 \in \left[ 1: 2^{nT_{10}}
\right]$, and distributed according to the conditional PD
$$ p(v_3^n\vert q^n) = \prod_{i=1}^n p_{V_3\vert Q}(v_{3i}\vert q_i). $$
 \item For each $v_3^n(t_0)$, generate $2^{nR_{11}'}$ conditionally independent
sequences $x_3^n(t_0, r_0)$, where $r_0 \in \big[ 1: 2^{nR_{11}'} \big]$, and
distributed according to the conditional PD
$$ p(x_3^n\vert v_3^n(t_0), q^n) = \prod_{i=1}^n p_{X_3\vert V_3 Q}(x_{3i}\vert
v_{3i}(t_0), q_i). $$
 \item For each $v_3^n(t_0)$, generate $2^{nT_{10}}$ conditionally independent
sequences $v_1^n(t_0, t_1)$, where $t_1 \in \left[ 1: 2^{nT_{10}} \right]$, and
distributed according to the conditional PD
$$ p(v_1^n\vert v_3^n(t_0), q^n) = \prod_{i=1}^n p_{V_1\vert V_3 Q}(v_{1i}\vert
v_{3i}(t_0), q_i). $$
 \item Partition the set $\left[ 1: 2^{nT_{10}} \right]$ into $2^{nR_{10}}$
cells and label them $\mathcal{T}(m_1)$, where $m_1 \in \left[ 1: 2^{nR_{10}} 
\right]$.
 \item For every pair $(x_3^n(t_0, r_0),v_1^n(t_0, t_1))$, generate 
$2^{nR_{11}'}$ conditionally independent sequences $u_1^n(t_{0}, t_{1}, r_{0}, 
r_{1})$, where $r_{1} \in \big[ 1: 2^{nR_{11}'} \big]$, and distributed 
according to the conditional PD
\begin{multline*}
p(u_1^n\vert v_1^n(t_0, t_1), x_3^n(t_0, r_0), v_3^n(t_0), q^n) = \\
\prod_{i=1}^n p(u_{1i} \vert v_{1i}(t_0, t_1), x_{3i}(t_0, r_0), v_{3i}(t_0),
q_i).
\end{multline*}
 \item For each $u_1^n(t_{0}, t_{1}, r_{0}, r_{1})$, generate 
$2^{nR_{11}''}$ conditionally independent sequences $x_1^n(t_{0}, t_{1}, r_{0}, 
r_{1},r_{2})$, where $r_2 \in \big[ 1: 2^{nR_{11}''} \big]$, and distributed 
according to the conditional PD
\begin{multline*}
\!\!\!\!
p(x_1^n\vert u_1^n(\cdot), v_1^n(t_0, t_1), x_3^n(t_0, r_0), v_3^n(t_0), q^n)
= \\
\prod_{i=1}^n p(x_{1i}\vert u_{1i}(\cdot), v_{1i}(t_0, t_1), x_{3i}(t_0, r_0),
v_{3i}(t_0), q_i).
\end{multline*}
 \item For each sequence $q^n$, generate $2^{nR_{20}}$ conditionally independent
sequences $v_2^n(s_0)$, where $s_0 \in \left[ 1: 2^{nR_{20}} \right]$, and 
distributed according to the conditional PD
$$ p(v_2^n\vert q^n) = \prod_{i=1}^n p_{V_2\vert Q}(v_{2i}\vert q_i). $$
 \item For each $v_2^n(s_0)$, generate $2^{nR_{22}}$ conditionally independent
sequences $x_2^n(s_0, s_1)$, where $s_1 \in \left[ 1: 2^{nR_{22}} \right]$, and
distributed according to the conditional PD
$$ p(x_2^n\vert v_2^n(s_0), q^n) = \prod_{i=1}^n p_{X_2\vert V_2 Q}(x_{2i}\vert
v_{2i}(s_0), q_i). $$
\end{enumerate}

\subsection{Encoding Part}
Encoding in block $b$ proceeds as follows,
\begin{enumerate}
 \item The relay already knows the indices $(t_{1(b-1)}, w_{1(b-1)}')$ from
decoding step~$1$ in the previous block, thus it transmits $x_3^n(t_{1(b-1)},
w_{1(b-1)}')$. For block $b=1$, it transmits the dummy message $x_3^n(1,1)$.
 \item Encoder 1 wants to transmit $\tilde{m}_{1b}=(m_{1b}, w_{1b}', w_{1b}'')$,
thus, it searches for an index $t_{1b}\in \mathcal{T}(m_{1b})$ such that $\big(
v_1^n(t_{1(b-1)}, t_{1b}), x_3^n(t_{1(b-1)}, w_{1(b-1)}'), v_3^n(t_{1(b-1)}),
q^n \big) \allowbreak \in\typc{n}{V_1 X_3 V_3 Q}$. The success of this step
requires that
\begin{equation}
T_{10}-R_{10} > I_b +\delta', \label{eq:IB-DF-bincond} 
\end{equation}
where $\delta'>0$ is an arbitrarily small constant and $I_b \triangleq
\IC{X_3}{V_1}{V_3 Q}$.
It then transmits the codeword $x_1^n(t_{1(b-1)}, t_{1b}, w_{1(b-1)}', 
w_{1b}', w_{1b}'')$. The source sends the dummy messages
$\tilde{m}_{10}=(1,1,1)$ and $\tilde{m}_{1(B+1)} =(1,1,1)$ known to all users at
the beginning and at the end of the transmission.
 \item Encoder 2 sends its message $\tilde{m}_{2(b-1)}=(m_{2(b-1)}, \allowbreak
w_{2(b-1)})$ through the codeword $x_2^n(m_{2(b-1)}, w_{2(b-1)})$. During block 
$b=1$, it sends the dummy message $x_2^n(1,1)$.
 \end{enumerate}
See Table~\ref{tab:AP-pDF-codewords} for references.


\subsection{Decoding Part}
\begin{enumerate}
 \item Let $\delta>\delta'$. At the end of block $b\in [1:B]$ and assuming its
past message estimates are correct, the relay looks for the unique pair of
indices $(t_{1b}, w_{1b}') \equiv (i,j)$ such that%
\begin{multline*}
\!\!\!\!\!\!\!\!\!\!\!\!\!\!\!\!\!
\big( v_3^n(t_{1(b-1)}), x_3^n(t_{1(b-1)}, w_{1(b-1)}'), v_1^n(t_{1(b-1)},i),
y_{3b}^n, q^n\!\!, \\
u_1^n(t_{1(b-1)},i,w_{1(b-1)}',j) \big) \in\typ{n}{V_3 X_3 V_1 U_1 Y_{3} Q}.
\!\!\!\!
\end{multline*}
The probability of error becomes arbitrarily small if%
\begin{subequations}\label{eq:IB-pDF-dec_bound1}
\begin{align}
R_{11}' &< \IC{U_1}{Y_3}{V_1 X_3 Q} -\delta,\\
T_{10}+R_{11}' &< \IC{V_1 U_1}{Y_3}{X_3 Q} +I_b -\delta.
\end{align}
\end{subequations}
 \item Starting at the end of block $B+1$ and assuming its past message 
estimates are correct, destination~1 looks for the indices $(t_{1(b-1)}, 
w_{1(b-1)}', w_{1b}'', m_{2(b-1)}) \equiv (i,j,k,l)$ backwardly such that
\begin{multline*}
\!\!\!\!\!\!\!\!\!\!\!\!\!\!\!\!\!
\big( v_3^n(i), v_1^n(i,t_{1b}), x_3^n(i,j), u_1^n(i,t_{1b},j,w_{1b}'),
v_2^n(l), y_{1b}^n, q^n\!\!, \\
x_1^n(i,t_{1b},j,w_{1b}',k) \big) \in\typ{n}{V_3 V_1 X_3 U_1 X_1 V_2 Y_{1} Q}.
\!\!\!\!\!
\end{multline*}
The probability of error becomes arbitrarily small if%
\begin{subequations}\label{eq:IB-pDF-dec_bound2}
\begin{align}
R_{11}''                                   \!&< \IC{X_1}{Y_1}{V_1 U_1 V_2 X_3 Q}
-\delta,\\
R_{11}'\!+\!R_{11}''                       \!&< \IC{X_1 X_3}{Y_1}{V_1 V_2 V_3 Q}
\!+\!I_b \!-\!\delta,\!\!\!\\
\!\!\!\!\!\!\!\!
T_{10}\!+\!R_{11}'\!+\!R_{11}''            \!&< \IC{X_1 X_3}{Y_1}{V_2 Q} +I_b
-\delta,\\
R_{11}''\!+\!R_{20}                        \!&< \IC{X_1 V_2}{Y_1}{V_1 U_1 X_3 Q}
-\delta,\\
\!\!\!\!\!\!\!\!
R_{11}'\!+\!R_{11}''\!+\!R_{20}            \!&< \IC{X_1 V_2 X_3}{Y_1}{V_1 V_3 Q}
\!+\!I_b \!-\!\delta,\!\!\!\\
\!\!\!\!\!\!\!\!\!\!\!\!\!\!\!\!\!\!
T_{10}\!+\!R_{11}'\!+\!R_{11}''\!+\!R_{20} \!&< \IC{X_1 V_2 X_3}{Y_1}{Q} +I_b
-\delta.
\end{align}
\end{subequations}
 \item Destination~2 performs similarly, thus, it looks for
the indices $(t_{1(b-1)}, m_{2(b-1)}, w_{2(b-1)}) $ $\equiv
(i,k,l)$ backwardly such that%
\begin{multline*}
\big( v_3^n(i), v_1^n(i,t_{1b}), v_2^n(k), x_2^n(k,l), y_{2b}^n, q^n \big) \\
\in\typ{n}{V_3 V_1 V_2 X_2 Y_{2} Q}.
\end{multline*}
The probability of error becomes arbitrarily small if %
\begin{subequations}\label{eq:IB-pDF-dec_bound3}
\begin{align}
R_{22} &< \IC{X_2}{Y_2}{V_1 V_2 V_3 Q} -\delta,\\
R_{20}+R_{22} &< \IC{X_2}{Y_2}{V_1 V_3 Q} -\delta,\\
T_{10}+R_{22} &< \IC{V_1 X_2 V_3}{Y_2}{V_2 Q} -\delta,\\
T_{10}+R_{20}+R_{22} &< \IC{V_1 X_2 V_3}{Y_2}{Q} -\delta,\\
T_{10} &< \IC{V_1 V_3}{Y_2}{X_2 Q} -\delta. \label{eq:IB-pDF-dec_bound3-5}
\end{align}
\end{subequations}
\end{enumerate}

\vspace{1mm}
\begin{remark}
If at this point we replace $U_1$ with $X_1$, the region boils down to the one
attained by the full DF scheme (Corollary~\ref{th:IB-DF}). See
Appendix~\ref{sec:AP-Proof-3}.
\end{remark}

\vspace{1mm}
\begin{remark}
The bound \eqref{eq:IB-pDF-dec_bound3-5} represents the perfect decoding of the
common layer of interference. This bound is needed, however, because of the
block-Markov coding technique and the assumption that the index $t_{1b}$ present
in $v_1^n(\cdot)$ is correct. Nonetheless, this term only appears in some of the
additional bounds shown below and it does not affect the final region
$\mathcal{R}_{\textrm{p-DF}}$.
\end{remark}
\vspace{1mm}

After running Fourier-Motzkin elimination (FME) to the set
\eqref{eq:IB-DF-bincond}--\eqref{eq:IB-pDF-dec_bound3} and letting $n
\rightarrow\infty$, we obtain the region
$\mathcal{R}_{\textrm{p-DF}}(P_2)$~\eqref{eq:IB-pDF-A} with the term \IC{V_1
U_1}{Y_3}{X_3 Q} instead of \IC{U_1}{Y_3}{X_3 Q} in~\eqref{eq:IB-pDF-A1},
\eqref{eq:IB-pDF-A11}, and~\eqref{eq:IB-pDF-A13}, plus four additional bounds
\begin{subequations}\label{eq:IB-pDF-dec_bound4}
\begin{align}
R_1 &< \IC{X_1 X_3}{Y_1}{V_1 V_2 V_3 Q} + \IC{V_1 V_3}{Y_2}{X_2 Q},
\label{eq:IB-pDF-dec_bound4-1}\\
R_1 &< \IC{U_1}{Y_3}{V_1 X_3 Q} + \IC{X_1}{Y_1}{V_1 U_1 V_2 X_3 Q} \nonumber\\
&\quad + \IC{V_1 V_3}{Y_2}{X_2 Q} -I_b, \label{eq:IB-pDF-dec_bound4-2}\\
R_2 &< \IC{X_1 V_2}{Y_1}{V_1 U_1 X_3 Q} + \IC{X_2}{Y_2}{V_1 V_2 V_3 Q},
\label{eq:IB-pDF-dec_bound4-3}\\
R_2 &< \IC{X_1 V_2}{Y_1}{V_1 U_1 X_3 Q} + \IC{V_1 X_2 V_3 }{Y_2}{V_2 Q} -I_b.
\label{eq:IB-pDF-dec_bound4-4}
\end{align}
\end{subequations}
These bounds on the single rates arise from the decoding of the common message
of the interference at the interfered receiver. It is reasonable to assume that
the maximizing PD will render these bounds \emph{inactive}, i.e., if the single
rates are penalized due to the large amount of common information, another PD
with \emph{less} common information will increase the achievable rate.

In order to eliminate the bounds~\eqref{eq:IB-pDF-dec_bound4} --a necessary
condition to later compare to the outer bound-- we proceed in a similar way
as~\cite[Lemma 2]{chong_han-kobayashi_2008}. First, let us define, for a given
PD $p \in \mathcal{P}_2$, the region $\mathcal{R}_{\textrm{p-DF}}^o(p)$ as the
\emph{original} region after FME, i.e., the region
$\mathcal{R}_{\textrm{p-DF}}(p)$~\eqref{eq:IB-pDF-A} with the term
\IC{V_1 U_1}{Y_3}{X_3 Q} instead of \IC{U_1}{Y_3}{X_3 Q} plus the four
bounds~\eqref{eq:IB-pDF-dec_bound4}.

Second, we define $\mathcal{R}_{\textrm{p-DF}}^{c_1}(p)$ as the region
$\mathcal{R}_{\textrm{p-DF}}^o(p)$ without bounds~\eqref{eq:IB-pDF-dec_bound4-3}
and~\eqref{eq:IB-pDF-dec_bound4-4}. For this reason, it is easy to see that
$\mathcal{R}_{\textrm{p-DF}}^o(p) \subseteq
\mathcal{R}_{\textrm{p-DF}}^{c_1}(p)$. On the other hand, when
either~\eqref{eq:IB-pDF-dec_bound4-3} or~\eqref{eq:IB-pDF-dec_bound4-4} is
active in $\mathcal{R}_{\textrm{p-DF}}^o(p)$, then
$\mathcal{R}_{\textrm{p-DF}}^o(p^{**})$ with $p^{**} = \sum_{v_2} p$ attains
higher rates than $\mathcal{R}_{\textrm{p-DF}}^{c_1}(p)$. The PD $p^{**}$ is the
marginal of $p$ w.r.t. $V_2$, therefore, effectively eliminating the common
message from the second source. In summary,
$\mathcal{R}_{\textrm{p-DF}}^{c_1}(p) \subseteq \mathcal{R}_{\textrm{p-DF}}^o(p)
\cup \mathcal{R}_{\textrm{p-DF}}^o(p^{**})$. After maximizing over all joint
PDs, we obtain $\mathcal{R}_{\textrm{p-DF}}^{c_1} =
\mathcal{R}_{\textrm{p-DF}}^o$, thus~\eqref{eq:IB-pDF-dec_bound4-3}
and~\eqref{eq:IB-pDF-dec_bound4-4} are redundant.

Third, we reduce the achievable region~$\mathcal{R}_{\textrm{p-DF}}^{c_1}(p)$ by
replacing the terms \IC{V_1 U_1}{Y_3}{X_3 Q} with \IC{U_1}{Y_3}{X_3 Q}, let us
call this new reduced region~$\mathcal{R}_{\textrm{p-DF}}^{c_2}(p)$. We define
the region~$\mathcal{R}_{\textrm{p-DF}}(p)$ based
on~$\mathcal{R}_{\textrm{p-DF}}^{c_2}(p)$ and eliminate the
bounds~\eqref{eq:IB-pDF-dec_bound4-1} and~\eqref{eq:IB-pDF-dec_bound4-2} from
it. After this, it is easy to prove that both
$\mathcal{R}_{\textrm{p-DF}}^{c_2}(p) \subseteq \mathcal{R}_{\textrm{p-DF}}(p)$
and $\mathcal{R}_{\textrm{p-DF}}(p) \subseteq
\mathcal{R}_{\textrm{p-DF}}^{c_2}(p) \cup
\mathcal{R}_{\textrm{p-DF}}^{c_2}(p^*)$, with $p^* = \sum_{v_1 v_3} p$, hold.
Therefore, after the maximization, we obtain $\mathcal{R}_{\textrm{p-DF}} =
\mathcal{R}_{\textrm{p-DF}}^{c_2}$.

It is worth mentioning that the region
$\mathcal{R}_{\textrm{p-DF}}$~\eqref{eq:IB-pDF-A} is not the optimal one for
partial DF because of the aforementioned reduction, i.e.
$\mathcal{R}_{\textrm{p-DF}} = \mathcal{R}_{\textrm{p-DF}}^{c_2} \subseteq
\mathcal{R}_{\textrm{p-DF}}^{c_1} = \mathcal{R}_{\textrm{p-DF}}^o$. However, as
we see later, this loss does not prevent us from obtaining a constant-gap
result.

\begin{table*}
\begin{center}
\renewcommand{\arraystretch}{1.2} 
\begin{tabular}{|l|l|l|l|l|l|l|}
\hline
$b=1$ & $b=2$ & $\ldots$ & $b=B$ & $b=B+1$ & $\ldots$ & $b=B+L$ \\
\hline
$v^n_1(m_{11})$ & $v^n_1(m_{12})$ & $\ldots$ & $v^n_1(m_{1B})$ & $v^n_1(1)$ &
$\ldots$ & $v^n_1(1)$ \\
\hline
$x^n_1(m_{11},w_{11})$ & $x^n_1(m_{12},w_{12})$ & $\ldots$ &
$x^n_1(m_{1B},w_{1B})$ & $x^n_1(1,1)$ & $\ldots$ & $x^n_1(1,1)$ \\
\hline
\hline
$v^n_2(m_{21})$ & $v^n_2(m_{22})$ & $\ldots$ & $v^n_2(m_{2B})$ & $v^n_2(1)$ &
$\ldots$ & $v^n_2(1)$ \\
\hline
$x^n_2(m_{21},w_{21})$ & $x^n_2(m_{22},w_{22})$ & $\ldots$ &
$x^n_2(m_{2B},w_{2B})$ & $x^n_2(1,1)$ & $\ldots$ & $x^n_2(1,1)$ \\
\hline
\hline
$\hat{y}^n_3(1,l_1)$ & $\hat{y}^n_3(l_1,l_2)$ & $\ldots$ &
$\hat{y}^n_3(l_{B-1},l_B)$ & $\emptyset$ & $\ldots$ & $\emptyset$ \\
\hline
$x^n_3(1)$ & $x^n_3(l_1)$ & $\ldots$ & $x^n_3(l_{B-1})$ & $x^n_3(l_B)$ &
$\ldots$ & $x^n_3(l_B)$ \\
\hline
\end{tabular}
\end{center}
\caption{Codewords in the proposed CF scheme for the IRC. 
\label{tab:AP-CF-codewords}}
\end{table*}

\section{Proof of Corollary~\ref{th:IB-DF} (Full DF Scheme)}
\label{sec:AP-Proof-3}

Since $U_1 = X_1$, the first source does not split its private message in two,
i.e., $R_{11}'' = 0$ and $R_{1} = R_{10} + R_{11}'$. The codebook generation,
encoding and decoding is carried out as in the partial DF scheme.

After running Fourier-Motzkin elimination to the set
\eqref{eq:IB-DF-bincond}--\eqref{eq:IB-pDF-dec_bound3} and letting $n
\rightarrow\infty$, we obtain the region
$\mathcal{R}_{\textrm{f-DF}}(P_3)$~\eqref{eq:IB-DF-A}, plus three additional
bounds
\begin{subequations}\label{eq:IB-DF-dec_bound4}
\begin{flalign}
\!\!\!
R_1 \!&<\! \IC{X_1 X_3}{Y_1}{V_1 V_2 V_3 Q} + \IC{V_1 V_3}{Y_2}{X_2 Q},
\label{eq:IB-DF-dec_bound4-1}\\
\!\!\!
R_1 \!&<\! \IC{X_1}{Y_3}{V_1 X_3 Q} + \IC{V_1 V_3}{Y_2}{X_2 Q} -I_b,
\label{eq:IB-DF-dec_bound4-2}\\
\!\!\!
R_2 \!&<\! \IC{X_1 V_2 X_3}{Y_1}{V_1 V_3 Q} \!+\! \IC{X_2}{Y_2}{V_1 V_2 V_3 Q}
\!+\!I_b.\!\! \label{eq:IB-DF-dec_bound4-3}
\end{flalign}
\end{subequations}
As in the partial DF scheme, these bounds are redundant when maximized over
all possible PDs. Let us define $\mathcal{R}_{\textrm{f-DF}}^o(P_3)$ as the
\emph{original} region after FME. Then, it is clear that for a given PD $p \in
\mathcal{P}_3$, $\mathcal{R}_{\textrm{f-DF}}^o(p) \subseteq
\mathcal{R}_{\textrm{f-DF}}(p)$, because of the presence
of~\eqref{eq:IB-DF-dec_bound4}.

When either~\eqref{eq:IB-DF-dec_bound4-1} or~\eqref{eq:IB-DF-dec_bound4-2} is
active in $\mathcal{R}_{\textrm{f-DF}}^o(p)$, then
$\mathcal{R}_{\textrm{f-DF}}^o(p^*)$ with $p^* = \sum_{v_1 v_3} p$ attains
higher rates than $\mathcal{R}_{\textrm{f-DF}}(p)$.
Similarly, when~\eqref{eq:IB-DF-dec_bound4-3} is active,
$\mathcal{R}_{\textrm{f-DF}}^o(p^{**})$ with $p^{**} = \sum_{v_2} p$ outperforms
$\mathcal{R}_{\textrm{f-DF}}(p)$.
Succinctly, $\mathcal{R}_{\textrm{f-DF}}(p) \subseteq
\mathcal{R}_{\textrm{f-DF}}^o(p) \cup \mathcal{R}_{\textrm{f-DF}}^o(p^*) \cup
\mathcal{R}_{\textrm{f-DF}}^o(p^{**})$.

Therefore, after maximizing over all possible PDs,
$\mathcal{R}_{\textrm{f-DF}} = \mathcal{R}_{\textrm{f-DF}}^o$, which
renders~\eqref{eq:IB-DF-dec_bound4} redundant.

\section{Proof of Theorem~\ref{th:IB-CF} (CF Scheme)}
\label{sec:AP-Proof-CF}

As before, each source $k \in \{1,2\}$ splits its message $\tilde{m}_k$ into a
common message $m_k$ and a private one $w_k$, each with partial rate $R_{k0}$
and $R_{kk}$, respectively, such that $R_{k}=R_{k0}+R_{kk}$. But now, each
source transmits $B$ messages during $B+L$ time blocks, each of them of length
$n$. During these additional $L$ time blocks, the relay repeats the same
compression index to ensure a correct decoding at each
destination~\cite{wu_optimal_2013,Behboodi-piantanida2013}.

\subsection{Code Generation}
\begin{enumerate}
 \item Generate the time-sharing sequence $q^n$ where each element is
independent and identically distributed~(i.i.d.) according to the PD
$$ p(q^n) = \prod_{i=1}^n p_Q(q_i). $$
 \item For each source $k \in \{1,2\}$ and the sequence $q^n$, generate
$2^{nR_{k0}}$ conditionally independent sequences $v_k^n(m_k)$, where
$m_{k} \in \left[ 1: 2^{nR_{k0}} \right]$, and distributed according to the
conditional PD
$$ p(v_k^n \vert q^n) = \prod_{i=1}^n p_{V_k\vert Q}(v_{ki}\vert q_i ). $$
 \item For each source $k \in \{1,2\}$ and for each $v_k^n(m_{k})$,
generate $2^{nR_{kk}}$ conditionally independent sequences $x_k^n(m_k,
w_k)$, where $w_k \in \left[ 1: 2^{nR_{kk}} \right]$, and distributed
according to the conditional PD
$$ p(x_k^n \vert v_k^n(m_{k}), q^n ) = \prod_{i=1}^n p_{X_k\vert V_k
Q}(x_{ki}\vert v_{ki}(m_{k}), q_i ). $$
 \item For the sequence $q^n$, generate $2^{n\hat{R}}$ conditionally
independent sequences $x_3^n(l_1)$, where $l_1 \in \bigl[ 1: 2^{n\hat{R}}
\bigr]$ for $\hat{R} = \IC{\hat{Y}_3}{Y_3}{X_3 Q} + \delta'$, and
distributed according to the conditional PD
$$ p(x_3^n \vert q^n) = \prod_{i=1}^n p_{X_3\vert Q}(x_{3i} \vert q_i). $$
 \item For the sequence $q^n$ and each $x_3^n(l_1)$, generate
$2^{n\hat{R}}$ conditionally independent sequences $\hat{y}_3^n(l_1,l_2)$,
where $l_2 \in \bigl[ 1: 2^{n\hat{R}} \bigr]$, and distributed according to the
conditional PD
$$ p(\hat{y}_3^n\vert x_3^n(l_1), q^n) = \prod_{i=1}^n p_{\hat{Y}_3\vert X_3
Q}(\hat{y}_{3i} \vert x_{3i}(l_1), q_i). $$
\end{enumerate}

\subsection{Encoding Part}
Encoding in block $b$ proceeds as follows,
\begin{enumerate}
 \item Each source $k \in \{1,2\}$ uses its present message $\tilde{m}_{kb}$ to
choose the codeword it transmits, $x_k^n( m_{kb}, w_{kb} )$ for blocks $b\in
[1:B]$. During blocks $b \in [B+1: B+L]$, the sources send the dummy message
$\tilde{m}_{kb}=1$ known to all users.
 \item At the end of block $b\in[1:B]$, the relay looks for at least one index
$l_b$, with $l_0=1$ s.t. $ \big( x_3^n(l_{b-1}), \allowbreak
\hat{y}_{3}^n(l_{b-1}, l_b), y_{3b}^n, q^n \big) \in\typc{n}{X_3 \hat{Y}_3 Y_3
Q}$. The probability of finding such $l_b$ goes to one as $n$ approaches
infinity. It then transmits $x_3^n(l_b)$ in the next time block. Moreover, for
blocks $b\in [B+1:B+L]$, the last compression index $l_B$ is repeated.
\end{enumerate}
See Table~\ref{tab:AP-CF-codewords} for references.


\subsection{Decoding Part}
\begin{enumerate}
 \item Destination~1 decodes the compression index in two steps. First, it looks
for the unique index $l_B\equiv l$ such that, $\forall\ b \in [B+1:B+L]$,
\begin{multline*}
\big( v_1^n(1), x_{1}^n(1,1), v_{2}^n(1), {x}_{3}^n(l), {y}_{1b}^n, q^n \big) \\
\in\typ{n}{V_1 X_1 V_2 X_3 Y_1  Q}.
\end{multline*}
For a finite but sufficiently large $L$, the probability of incorrectly decoding
$l_B$ can be made arbitrarily small.
 \item After finding $l_B$, destination~1 looks for the indices $(m_{1b},
w_{1b}, m_{2b}, l_{b-1}) \equiv (i,j,k,l)$ for $b \in [1:B]$ such that 
\begin{multline*}
\big( v_1^n(i), x_1^n(i,j), v_2^n(k), x_3^n(l), \hat{y}_3^n(l,l_b), 
y_{1b}^n, q^n \big) \\ \in\typ{n}{V_1 X_1 V_2 X_3 \hat{Y}_3 Y_1 Q}.
\end{multline*}
The probability of error can be made arbitrarily small 
provided that,
\begin{subequations}\label{eq:AP-CF-dec_bound1}
\begin{align}
R_{11} &< I_{11} -\delta, \\
R_{10} +R_{11} &< I_{12} -\delta, \\
R_{20} +R_{11} &< I_{13} -\delta, \label{eq:AP-CF-b3}\displaybreak[2]\\
\!\!\!\!
R_{10} +R_{11} +R_{20} &< I_{14} -\delta, \label{eq:AP-CF-b4}\displaybreak[2]\\
R_{20} &< \IC{V_2 X_3}{Y_1}{X_1 Q} - I_1 -\delta, \label{eq:AP-CF-b5}\\
I_1 &< \IC{X_3}{Y_1}{X_1 V_2 Q} -\delta \label{eq:AP-CF-b6}
\end{align}
\end{subequations}
where $I_1 \triangleq \IC{\hat{Y}_3}{Y_3}{X_1 V_2 X_3 Y_1 Q} +\delta'$ and
\begin{align*}
\!\!\!\!\!\!\!\!\!\!\!\!\!\!
I_{11} \!&\triangleq\! \min\{ \IC{X_1}{Y_1 \hat{Y}_3}{V_1 V_2 X_3 Q}, \IC{X_1
X_3}{Y_1}{V_1 V_2 Q} \!-\! I_1 \}\\
\!\!\!\!\!\!\!\!\!\!\!\!\!\!
I_{12} \!&\triangleq\! \min\{ \IC{X_1}{Y_1 \hat{Y}_3}{V_2 X_3 Q}, \IC{X_1
X_3}{Y_1}{V_2 Q} - I_1 \} \displaybreak[2]\\
\!\!\!\!\!\!\!\!\!\!\!\!\!\!
I_{13} \!&\triangleq\! \min\{ \IC{X_1 V_2}{Y_1 \hat{Y}_3}{V_1 X_3 Q}, \IC{X_1
V_2 X_3}{Y_1}{V_1 Q} \!-\! I_1 \}\\
\!\!\!\!\!\!\!\!\!\!\!\!\!\!
I_{14} \!&\triangleq\! \min\{ \IC{X_1 V_2}{Y_1 \hat{Y}_3}{X_3 Q}, \IC{X_1 V_2
X_3}{Y_1}{Q} - I_1 \}.
\end{align*}
 \item If destination~1 ignores the compression index, it looks for the indices
$(m_{1b}, w_{1b}, m_{2b}) \equiv (i,j,k)$ for $b \in [1:B]$ such that
$$ \left( v_1^n(i), x_1^n(i,j), v_2^n(k), y_{1b}^n, q^n \right)  \in\typ{n}{V_1 X_1 V_2 Y_1 Q}.$$
The probability of error can be made arbitrarily small provided that,
\begin{subequations}\label{eq:AP-CF-dec_bound2}
\begin{align}
R_{11} &< \IC{X_1}{Y_1}{V_1 V_2 Q} -\delta,\\
R_{10} +R_{11} &< \IC{X_1}{Y_1}{V_2 Q} -\delta,\\
R_{20} +R_{11} &< \IC{X_1 V_2}{Y_1}{V_1 Q} -\delta,\\
R_{10} +R_{11} +R_{20} &< \IC{X_1 V_2}{Y_1}{Q} -\delta.
\end{align}
\end{subequations}
 \item Destination~2 performs similarly, and all the above inequalities hold by
swapping the indices $1$ and $2$.
\end{enumerate}

It is noteworthy that the bound in the rate of the interfering common
message~\eqref{eq:AP-CF-b5}, i.e., $R_{j0} \leq \IC{V_j X_3}{Y_k}{X_k Q} - I_k$,
is a by-product of the CF scheme. Although the error in decoding the index of
the interfering common message is normally not taken into account in the IC,
this bound is needed in order to assure that the compression index $l_b$ is the
right one at time $b$. Nonetheless, both the bound~\eqref{eq:AP-CF-b5}
and~\eqref{eq:AP-CF-b6} are redundant as we see next.

When~\eqref{eq:AP-CF-b5} does not hold, \eqref{eq:AP-CF-b3}
and~\eqref{eq:AP-CF-b4} become:
\begin{subequations}
\begin{align}
R_{11} &< \IC{X_1 V_2 X_3}{Y_1}{V_1 Q} -\IC{V_2 X_3}{Y_1}{X_1 Q}  \nonumber\\
 &= \IC{X_1}{Y_1}{V_1 Q},\\
R_{10} +R_{11} &< \IC{X_1 V_2 X_3}{Y_1}{Q} -\IC{V_2 X_3}{Y_1}{X_1 Q} \nonumber\\
 &= \IC{X_1}{Y_1}{Q}.
\end{align}
\end{subequations}
This is included in the region~\eqref{eq:AP-CF-dec_bound2} for the special case 
$V_2=\emptyset$.

Moreover, if~\eqref{eq:AP-CF-b6} does not hold, the first five bounds
of~\eqref{eq:AP-CF-dec_bound1} become:
\begin{subequations}
\begin{align}
R_{11} &< \IC{X_1 X_3}{Y_1}{V_1 V_2 Q} - I_1 \nonumber\\
 &< \IC{X_1 X_3}{Y_1}{V_1 V_2 Q} -\IC{X_3}{Y_1}{X_1 V_2 Q} \nonumber\\
 &= \IC{X_1}{Y_1}{V_1 V_2 Q}, \\
R_{10} \!+\!R_{11} &< \IC{X_1}{Y_1}{V_2 Q}, \\
R_{20} \!+\!R_{11} &< \IC{X_1 V_2}{Y_1}{V_1 Q}, \displaybreak[2]\\
R_{10} \!+\!R_{11} \!+\!R_{20} &< \IC{X_1 V_2}{Y_1}{V_1 Q}, \\
R_{20} &< \IC{V_2}{Y_1}{X_1 Q}.
\end{align}
\end{subequations}
This region is also included in~\eqref{eq:AP-CF-dec_bound2}. Therefore, when
either condition~\eqref{eq:AP-CF-b5} or~\eqref{eq:AP-CF-b6} does not hold for a
given distribution, the region~\eqref{eq:AP-CF-dec_bound1} is included
inside~\eqref{eq:AP-CF-dec_bound2}, i.e., destination~1 should ignore the relay
to achieve higher rates. Since the final region is the union over all possible
PDs of~\eqref{eq:AP-CF-dec_bound1} and~\eqref{eq:AP-CF-dec_bound2} for both
users, we can drop~\eqref{eq:AP-CF-b5} and~\eqref{eq:AP-CF-b6} because they do
not affect the final region after the maximization. This result can be seen as
an extension of~\cite{Behboodi-piantanida2013}.

Before running Fourier-Motzkin elimination to this system, we shall make same
clarifications. First, let us define $\mathcal{R}_{\textrm{CF}_3}(P_4)$ as the
region obtained with the distribution $P_4$ when both users ignore the
compression index, i.e., the Han-Kobayashi inner bound. The regions
$\mathcal{R}_{\textrm{CF}_1}(P_4)$ and $\mathcal{R}_{\textrm{CF}_2}(P_4)$ are
the ones obtained when only the first or second user decodes the relay's
message, respectively. $\mathcal{R}_{\textrm{CF}_0}(P_4)$ corresponds to the
region when both users decode the compression index.

Second, even though the expressions $I_{ki}$ look rather complex, there exists 
an ordering between them analogous to $I_{ki}'$ that allows us to reduce the 
number of bounds. In other words, the following inequalities hold,
\begin{equation}
I_{k1} \leq I_{k2} \leq I_{k4} \textnormal{ and } I_{k1} \leq I_{k3} \leq
I_{k4}. \label{eq:AP-P4-relation}
\end{equation}
To check this, take each term of $I_{11}$ and $I_{12}$ separately
\begin{subequations}
\begin{align}
I_{11} &\leq \IC{X_1}{Y_1 \hat{Y}_3}{V_1 V_2 X_3 Q} \nonumber\\
 &= \HC{Y_1 \hat{Y}_3}{V_1 V_2 X_3 Q} - \HC{Y_1 \hat{Y}_3}{X_1 V_2 X_3 Q},
\label{eq:AP-P4-terms1}\\
I_{11} &\leq \IC{X_1 X_3}{Y_1}{V_1 V_2 Q} - I_1 \nonumber\\
 &= \HC{Y_1}{V_1 V_2 Q} - \HC{Y_1}{X_1 V_2 X_3 Q} - I_1,
\label{eq:AP-P4-terms2}\displaybreak[2]\\
I_{12} &\leq \IC{X_1}{Y_1 \hat{Y}_3}{V_2 X_3 Q} \nonumber\\
  &= \HC{Y_1 \hat{Y}_3}{V_2 X_3 Q} - \HC{Y_1 \hat{Y}_3}{X_1 V_2 X_3 Q},
\label{eq:AP-P4-terms3}\displaybreak[2]\\
I_{12} &\leq \IC{X_1 X_3}{Y_1}{V_2 Q} - I_1 \nonumber\\
 &= \HC{Y_1}{V_2 Q} - \HC{Y_1}{X_1 V_2 X_3 Q} - I_1. \label{eq:AP-P4-terms4}
\end{align}
\end{subequations}
Since conditioning reduces entropy, we have that~\eqref{eq:AP-P4-terms1}
$\leq$~\eqref{eq:AP-P4-terms3} and~\eqref{eq:AP-P4-terms2}
$\leq$~\eqref{eq:AP-P4-terms4}, which leads to $I_{11} \leq I_{12}$. The same
reasoning applies for the other $I_{ki}$ in~\eqref{eq:AP-P4-relation}.


\subsubsection{Final Region
\texorpdfstring{$\mathcal{R}_{\textrm{CF}_3}$}{RCF3}}

After running FME to the system composed by~\eqref{eq:AP-CF-dec_bound2} and its
symmetric one for the second user, and letting $n \rightarrow\infty$, we obtain 
the region $\mathcal{R}_{\textrm{CF}_3}^o(p)$:
\begin{align*}
R_k &\leq \min\{ I_{k2}', I_{k1}'+I_{j3}' \},\\
R_k +R_j &\leq \min\{ I_{k1}'+I_{j4}', I_{k3}'+I_{j3}' \}, \\
2R_k +R_j &\leq I_{k1}'+I_{k4}'+I_{j3}'.
\end{align*}
This region has two redundant bounds as shown
in~\cite{chong_han-kobayashi_2008}:
\begin{subequations}\label{eq:AP-CF_3-red}
\begin{align}
R_1 &\leq \IC{X_1}{Y_1}{V_1 V_2 Q} + \IC{V_1 X_2}{Y_2}{V_2 Q},
\label{eq:AP-CF_3-red1}\\
R_2 &\leq \IC{X_1 V_2}{Y_1}{V_1 Q} + \IC{X_2}{Y_2}{V_1 V_2 Q}.
\label{eq:AP-CF_3-red2}
\end{align}
\end{subequations}
If we define $\mathcal{R}_{\textrm{CF}_3}^c(p)$ as the \emph{compact} version of
the \emph{original} region $\mathcal{R}_{\textrm{CF}_3}^o(p)$, i.e., without the
two redundant bounds, we can readily see that $\mathcal{R}_{\textrm{CF}_3}^o(p)
\subseteq \mathcal{R}_{\textrm{CF}_3}^c(p)$ for a given distribution $p \in
\mathcal{P}_4$ since $\mathcal{R}_{\textrm{CF}_3}^c(p)$ has fewer bounds.

If a pair of rates $(R_1,R_2)$ belongs to $\mathcal{R}_{\textrm{CF}_3}^c(p)$ 
but not to $\mathcal{R}_{\textrm{CF}_3}^o(p)$, it is
because~\eqref{eq:AP-CF_3-red} does not hold. Let us first assume that
$$ R_1 > \IC{X_1}{Y_1}{V_1 V_2 Q} + \IC{V_1 X_2}{Y_2}{V_2 Q}. $$
With this condition, $\mathcal{R}_{\textrm{CF}_3}^c(p)$ becomes:
\begin{align*}
R_1 &\leq \IC{X_1}{Y_1}{V_2 Q}, \\
R_2 &\leq \IC{V_2}{Y_2}{Q}, \\
R_1+R_2 &\leq \IC{X_1 V_2}{Y_1}{Q},
\end{align*}
together with some additional bounds. We may compare this region with 
$\mathcal{R}_{\textrm{CF}_3}^o(p^*)$, where $p^*=\sum_{v_1}p$,
\begin{align*}
R_1 &\leq \IC{X_1}{Y_1}{V_2 Q}, \\
R_2 &\leq \IC{X_2}{Y_2}{Q}, \\
R_1+R_2 &\leq \IC{X_1 V_2}{Y_1}{Q} +\IC{X_2}{Y_2}{V_2 Q}.
\end{align*}
It is clear that, when~\eqref{eq:AP-CF_3-red1} is violated,
$\mathcal{R}_{\textrm{CF}_3}^c(p) \subseteq \mathcal{R}_{\textrm{CF}_3}^o(p^*)$.

Similarly, if~\eqref{eq:AP-CF_3-red2} does not hold, we see that
$\mathcal{R}_{\textrm{CF}_3}^c(p) \subseteq
\mathcal{R}_{\textrm{CF}_3}^o(p^{**})$, where $p^{**}=\sum_{v_2}p$. Therefore,
in the general case, 
$$ \mathcal{R}_{\textrm{CF}_3}^c(p) \subseteq \mathcal{R}_{\textrm{CF}_3}^o(p)
\cup \mathcal{R}_{\textrm{CF}_3}^o(p^{*}) \cup
\mathcal{R}_{\textrm{CF}_3}^o(p^{**}). $$
Since we have already shown that $\mathcal{R}_{\textrm{CF}_3}^o(p) \subseteq
\mathcal{R}_{\textrm{CF}_3}^c(p)$, when maximizing over all joint PDs, we have
that $\mathcal{R}_{\textrm{CF}_3}^o = \mathcal{R}_{\textrm{CF}_3}^c$.

\subsubsection{Final Regions
\texorpdfstring{$\mathcal{R}_{\textrm{CF}_1}$}{RCF1} and
\texorpdfstring{$\mathcal{R}_{\textrm{CF}_2}$}{RCF2}}

Now, we go to $\mathcal{R}_{\textrm{CF}_1}^o(p)$, where only the first user 
decodes the compression index. In this case, the region that is obtained after 
running FME is:
\begin{align*}
R_1 &\leq \min\{ I_{12}, I_{11}+I_{23}' \}, \\
R_2 &\leq \min\{ I_{22}', I_{13}+I_{21}' \}, \\
R_1 +R_2 &\leq \min\{ I_{11}+I_{24}', I_{14}+I_{21}', I_{13}+I_{23}' \}, \\
2R_1+ R_2 &\leq I_{11}+I_{14}+I_{23}', \\
R_1 +2R_2 &\leq I_{13}+I_{21}'+I_{24}'.
\end{align*}
Here, we have another two redundant bounds:
\begin{subequations}
\begin{align}
R_1 &\leq I_{11} + \IC{V_1 X_2}{Y_2}{V_2 Q}, \label{eq:AP-CF_1-red1}\\
R_2 &\leq I_{13} + \IC{X_2}{Y_2}{V_1 V_2 Q}. \label{eq:AP-CF_1-red2}
\end{align}
\end{subequations}
Once again, for a given distribution $p \in \mathcal{P}_4$, we define 
$\mathcal{R}_{\textrm{CF}_1}^o(p)$ as the original region with all the bounds 
and $\mathcal{R}_{\textrm{CF}_1}^c(p)$ as the compact one without the redundant 
bounds. Since $\mathcal{R}_{\textrm{CF}_1}^c(p)$ has fewer bounds, we can 
readily see that $\mathcal{R}_{\textrm{CF}_1}^o(p) \subseteq 
\mathcal{R}_{\textrm{CF}_1}^c(p)$.

If~\eqref{eq:AP-CF_1-red1} does not hold, $\mathcal{R}_{\textrm{CF}_1}^c(p)$ 
becomes:
\begin{align*}
R_1 &\leq I_{12}, \\
R_2 &\leq \IC{V_2}{Y_2}{Q}, \\
R_1+R_2 &\leq I_{14},
\end{align*}
together with some additional bounds. We may compare this region with 
$\mathcal{R}_{\textrm{CF}_1}^o(p^*)$, where $p^*=\sum_{v_1}p$,
\begin{align*}
R_1 &\leq I_{12}, \\
R_2 &\leq \IC{X_2}{Y_2}{Q}, \\
R_1+R_2 &\leq I_{14} +\IC{X_2}{Y_2}{V_2 Q}.
\end{align*}
As we see, when~\eqref{eq:AP-CF_1-red1} is violated,
$\mathcal{R}_{\textrm{CF}_1}^c(p) \subseteq \mathcal{R}_{\textrm{CF}_1}^o(p^*)$.

Since this region is not symmetric, we also need to see what happens
when~\eqref{eq:AP-CF_1-red2} does not hold. In this case, 
$\mathcal{R}_{\textrm{CF}_1}^c(p)$ becomes:
\begin{subequations}\label{eq:AP-CF_1-stp1}
\begin{align}
R_1 &\leq I_{14}-I_{13}, \label{eq:AP-CF_1-stp1a}\\
R_2 &\leq \IC{X_2}{Y_2}{V_1 Q}, \\
R_1+R_2 &\leq \IC{V_1 X_2}{Y_2}{Q},
\end{align}
\end{subequations}
together with some additional bounds. Now, let us take $p^{**}=\sum_{v_2}p$ and 
calculate $\mathcal{R}_{\textrm{CF}_1}^o(p^{**})$:
\begin{subequations}\label{eq:AP-CF_1-stp2}
\begin{align}
R_1 &\leq I_{14}^*, \\
R_2 &\leq \IC{X_2}{Y_2}{V_1 Q}, \displaybreak[2]\\
R_2 &\leq I_{13}^* + \IC{X_2}{Y_2}{V_1 Q}, \displaybreak[2]\\
R_1+R_2 &\leq I_{13}^* +\IC{V_1 X_2}{Y_2}{Q}
\end{align}
\end{subequations}
where
\begin{align*}
I_{13}^* &\triangleq \min\{ \IC{X_1}{Y_1 \hat{Y}_3}{V_1 X_3 Q}, \\
&\qquad\qquad \IC{X_1 X_3}{Y_1}{V_1 Q} - \IC{Y_3}{\hat{Y}_3}{X_1 X_3 Y_1 Q}
\},\\
I_{14}^* &\triangleq \min\{ \IC{X_1}{Y_1 \hat{Y}_3}{X_3 Q}, \\
&\qquad\qquad \IC{X_1 X_3}{Y_1}{Q} - \IC{Y_3}{\hat{Y}_3}{X_1 X_3 Y_1 Q} \}.
\end{align*}
We shall recall that the PD $p$ is such that the rates $R_1$ and $R_2$ are
nonnegative in $\mathcal{R}_{\textrm{CF}_1}^c(p)$. However, this does not mean
that $I_{13}^*$ or $I_{14}^*$ should be positive since they depend on $p^{**}$.
If any of the two expressions is negative, $\mathcal{R}_{\textrm{CF}_1}^c(p)
\nsubseteq \mathcal{R}_{\textrm{CF}_1}^o(p^{**})$, which is not what we are
looking for. We first assume that both quantities are positive.

Let us define with a subscript $a$ and $b$ the first and second term of the 
minimums in the expressions $I_{ki}$, respectively. Then, if $I_{13}=I_{13a}$, 
the first rate in $\mathcal{R}_{\textrm{CF}_1}^c(p)$ becomes:
\begin{subequations}\label{eq:AP-CF_1-stp3}
\begin{align}
R_1 &\leq I_{14a}-I_{13a} = \IC{V_1}{Y_1 \hat{Y}_3}{X_3 Q} \leq  I_{14a}^*, \\
R_1 &\leq I_{14b}-I_{13a} \nonumber\\
 &= \IC{X_1 V_2 X_3}{Y_1}{Q} -\IC{Y_3}{\hat{Y}_3}{X_1 V_2 X_3 Y_1 Q} \nonumber\\
 &\quad - \IC{X_1 V_2}{Y_1 \hat{Y}_3}{V_1 X_3 Q}\nonumber\\
 &= \IC{X_1 V_2 X_3}{Y_1}{Q} - \IC{Y_3}{\hat{Y}_3}{X_1 V_2 X_3 Y_1 Q}
\nonumber\\
 &\quad - \IC{X_1 V_2}{Y_1}{V_1 X_3 Q}- \IC{X_1 V_2}{\hat{Y}_3}{V_1 X_3 Y_1
Q}\nonumber\\
 &= \IC{V_1 X_3}{Y_1}{Q} - \IC{X_1 V_2 Y_3}{\hat{Y}_3}{V_1 X_3 Y_1 Q}\nonumber\\
 &= \IC{V_1 X_3}{Y_1}{Q} - \IC{Y_3}{\hat{Y}_3}{V_1 X_3 Y_1 Q} \leq  I_{14b}^*
\label{eq:AP-CF_1-stp3b}
\end{align}
\end{subequations}
where in the last step we take into account that $\hat{Y}_3 \mkv (X_3 Y_3 Q) 
\mkv (X_1 V_2)$. On the other hand, if $I_{13}=I_{13b}$, the first rate in 
$\mathcal{R}_{\textrm{CF}_1}^c(p)$ becomes:
\begin{equation}
R_1 \leq I_{14b}-I_{13b} = \IC{V_1}{Y_1}{Q} \leq  I_{14a}^*.
\label{eq:AP-CF_1-stp4}
\end{equation}
Also, in $\mathcal{R}_{\textrm{CF}_1}^o(p^{**})$:
\begin{align}
R_1 &\leq I_{14b}^* = \IC{X_1 X_3}{Y_1}{Q} - \IC{Y_3}{\hat{Y}_3}{X_1 X_3 Y_1 Q} 
\nonumber\\
 &= \IC{V_1}{Y_1}{Q} + I_{13b}^*. \label{eq:AP-CF_1-stp5}
\end{align}
If we assume that $I_{13}^*\geq 0$, \eqref{eq:AP-CF_1-stp3b}
and~\eqref{eq:AP-CF_1-stp5} assure us that $I_{14}^*\geq 0$.
Putting~\eqref{eq:AP-CF_1-stp1} through~\eqref{eq:AP-CF_1-stp5} together, we
have shown that $\mathcal{R}_{\textrm{CF}_1}^c(p) \subseteq
\mathcal{R}_{\textrm{CF}_1}^o(p^{**})$. However, if $I_{13}^*<0$ we shall
consider the case where the first user also ignores the compression index, i.e.
$\mathcal{R}_{\textrm{CF}_3}^o(p^{**})$,
\begin{subequations}\label{eq:AP-CF_1-stp6}
\begin{align}
R_1 &\leq \IC{X_1}{Y_1}{Q}, \\
R_2 &\leq \IC{X_2}{Y_2}{V_1 Q}, \\
R_1+R_2 &\leq \IC{X_1}{Y_1}{V_1 Q} +\IC{V_1 X_2}{Y_2}{Q}.
\end{align}
\end{subequations}
The region in~\eqref{eq:AP-CF_1-stp1} looks smaller
than~\eqref{eq:AP-CF_1-stp6}, with the exception of the rate $R_1$ that we
analyze in the sequel. If $I_{13}=I_{13a}$, in~\eqref{eq:AP-CF_1-stp1a} we have
that,
\begin{subequations}\label{eq:AP-CF_1-stp7}
\begin{align}
R_1 &\leq I_{14}-I_{13} = \min\{ I_{14a}, I_{14b} \} - I_{13a} \leq I_{14b} -
I_{13a} \nonumber\\
 &= \IC{V_1 X_3}{Y_1}{Q} - \IC{Y_3}{\hat{Y}_3}{V_1 X_3 Y_1 Q}
\displaybreak[2]\label{eq:AP-CF_1-stp7a}\\
 &= \IC{V_1 X_3}{Y_1}{Q} - \IC{X_1}{\hat{Y}_3}{V_1 X_3 Y_1 Q} \nonumber\\
&\quad - \IC{Y_3}{\hat{Y}_3}{X_1 X_3 Y_1 Q} \label{eq:AP-CF_1-stp7b}
\displaybreak[2]\\
 &< \IC{V_1 X_3}{Y_1}{Q} - \IC{X_1}{\hat{Y}_3}{V_1 X_3 Y_1 Q} \nonumber\\
&\quad - \IC{X_1 X_3}{Y_1}{V_1 Q} \label{eq:AP-CF_1-stp7c} \displaybreak[2]\\
 &\leq \IC{X_1 X_3}{Y_1}{Q} - \IC{X_1}{\hat{Y}_3}{V_1 X_3 Y_1 Q} \nonumber\\
&\quad - \IC{X_1 X_3}{Y_1}{V_1 Q} \nonumber\\
 &= \IC{V_1}{Y_1}{Q} - \IC{X_1}{\hat{Y}_3}{V_1 X_3 Y_1 Q} \nonumber\\
 &\leq \IC{V_1}{Y_1}{Q},
\end{align}
\end{subequations}
where~\eqref{eq:AP-CF_1-stp7a} comes from~\eqref{eq:AP-CF_1-stp3b}, 
\eqref{eq:AP-CF_1-stp7b} is due to the Markov chain $\hat{Y}_3 \mkv (X_3Y_3Q) 
\mkv X_1$, and~\eqref{eq:AP-CF_1-stp7c} is due to the assumption $I_{13}^*<0$, 
i.e. $\IC{X_1 X_3}{Y_1}{V_1 Q} < \IC{Y_3}{\hat{Y}_3}{X_1 X_3 Y_1 Q}$.

On the other hand, if $I_{13}=I_{13b}$, we have already shown
in~\eqref{eq:AP-CF_1-stp4} that $R_1 \leq \IC{V_1}{Y_1}{Q}$. Therefore, if
$I_{13}^*<0$, the region $\mathcal{R}_{\textrm{CF}_3}^o(p^{**})$ is larger than
$\mathcal{R}_{\textrm{CF}_1}^c(p)$ when $R_2 > I_{13} + \IC{X_2}{Y_2}{V_1 V_2
Q}$. To sum up, in the general case, $$ \mathcal{R}_{\textrm{CF}_1}^c(p)
\subseteq \mathcal{R}_{\textrm{CF}_1}^o(p) \cup
\mathcal{R}_{\textrm{CF}_1}^o(p^{*}) \cup \mathcal{R}_{\textrm{CF}_1}^o(p^{**})
\cup \mathcal{R}_{\textrm{CF}_3}^o(p^{**}), $$ and since
$\mathcal{R}_{\textrm{CF}_1}^o(p) \subseteq \mathcal{R}_{\textrm{CF}_1}^c(p)$,
if we maximize over all joint possible joint distributions we obtain
$\mathcal{R}_{\textrm{CF}_1}^c \cup \mathcal{R}_{\textrm{CF}_3}^c =
\mathcal{R}_{\textrm{CF}_1}^o \cup \mathcal{R}_{\textrm{CF}_3}^o$.

The symmetric region $\mathcal{R}_{\textrm{CF}_2}^o(p)$ where only the second
user decodes the compression index behaves similarly. We can redo the whole
proof by simply swapping the subindices $1$ and $2$. Consequently, if we
maximize over all joint possible joint distributions we have that
$\mathcal{R}_{\textrm{CF}_2}^c \cup \mathcal{R}_{\textrm{CF}_3}^c =
\mathcal{R}_{\textrm{CF}_2}^o \cup \mathcal{R}_{\textrm{CF}_3}^o$.

\subsubsection{Final Region
\texorpdfstring{$\mathcal{R}_{\textrm{CF}_0}$}{RCF0}}

Finally, when both users decode the compression index, the region we obtain 
after running FME is,
\begin{align*}
R_k &\leq \min\{ I_{k2}, I_{k1}+I_{j3} \},\\
R_k +R_j &\leq \min\{ I_{k1}+I_{j4}, I_{k3}+I_{j3} \}, \\
2R_k +R_j &\leq I_{k1}+I_{k4}+I_{j3}
\end{align*}
where the redundant terms are
\begin{align*}
R_1 &\leq I_{11} + I_{23},\\
R_2 &\leq I_{13} + I_{21}.
\end{align*}
We omit the complete proof for this region since it follows the same steps as
the previous ones. The conclusion here is that the region
$\mathcal{R}_{\textrm{CF}_0}^c(p)$, the one without the redundant terms, is
larger than $\mathcal{R}_{\textrm{CF}_0}^o(p)$, and also,
\begin{multline*}
\mathcal{R}_{\textrm{CF}_0}^c(p) \subseteq \mathcal{R}_{\textrm{CF}_0}^o(p) \cup
\mathcal{R}_{\textrm{CF}_0}^o(p^{*}) \cup \mathcal{R}_{\textrm{CF}_1}^o(p^{*})
\cup \\
\mathcal{R}_{\textrm{CF}_0}^o(p^{**})\cup \mathcal{R}_{\textrm{CF}_2}^o(p^{**}).
\end{multline*}
Therefore, if we maximize over all possible joint distributions we have
$$ \mathcal{R}_{\textrm{CF}_0}^c \cup \mathcal{R}_{\textrm{CF}_1}^c \cup 
\mathcal{R}_{\textrm{CF}_2}^c \cup \mathcal{R}_{\textrm{CF}_3}^c = 
\mathcal{R}_{\textrm{CF}_0}^o \cup \mathcal{R}_{\textrm{CF}_1}^o \cup 
\mathcal{R}_{\textrm{CF}_2}^o \cup \mathcal{R}_{\textrm{CF}_3}^o. $$
Since the region $\mathcal{R}_{\textrm{CF}_3}$ is a special case of
$\mathcal{R}_{\textrm{CF}_0}$ in the maximization, we can eliminate it. The
final region without redundant terms is~\eqref{eq:IB-CF-joint} when both
destinations decode the compression index, and the
region~\eqref{eq:IB-CF2-joint} when one of them ignores it.

\section{Proof of Proposition~\ref{th:CG-DF} (Full DF Constant Gap)}
\label{sec:AP-Proof-CG-DF}

The comparison between the full DF inner bound~\eqref{eq:IB-DF-A} and the outer 
bound is complex mainly due to the different PDs in each bound and the presence
of the binning terms. However, as we see next, we can propose some
simplifications to help us calculate the difference between the bounds.

First, let us assume the following set of auxiliary random variables,
\begin{subequations}\label{eq:AP-CG-DF-aux}
\begin{align}
V_1 &= h_{21} X_1 + h_{23} X_3 + Z_2',\\
V_2 &= h_{12} X_2 + Z_1',\\
V_3 &= \frac{h_{23}}{\sqrt{1+\SNR{21}}} X_3 + Z_2''
\end{align}
\end{subequations}
where $\SNR{21}\triangleq |h_{21}|^2 P_1 / N_2$, and $Z_k'$ and $Z_k''$ are
independent copies of $Z_k$. This choice fulfills the Markov chains
in~\eqref{eq:IB-DF-pdf}. Nonetheless, since it is a particular choice of
variables, the region might be smaller than the optimal one.

Second, let us assume that $X_1$ and $X_3$ are independent. Then, the binning 
term becomes upper-bounded regardless of the channel coefficients,
$$ I_b = \C{ \frac{\SNR{23}}{1 +\SNR{21} +\SNR{23}} } \leq \frac{1}{2}
\textnormal{ bit.} $$
We can reduce the achievable region~\eqref{eq:IB-DF-A} if we add $-I_b$
to~\eqref{eq:IB-DF-A3} and~\eqref{eq:IB-DF-A9} which render~\eqref{eq:IB-DF-A4}
and~\eqref{eq:IB-DF-A8} redundant. We further shrink the region by replacing
$-I_b$ with $-\frac{1}{2}$ which gives us,
\begin{subequations}\label{eq:AP-CG-DF-A}
\begin{align}
R_1 &\leq \IC{X_1}{Y_3}{X_3 Q}\\
R_1 &\leq \IC{X_1 X_3}{Y_1}{V_2 Q}\\
R_2 &\leq \IC{X_2}{Y_2}{V_1 V_3 Q} -\frac{1}{2}\\
R_1 \!+\!R_2 &\leq \IC{X_1 X_3}{Y_1}{V_1 V_2 V_3 Q} +\IC{V_1 X_2 V_3}{Y_2}{Q}\\
R_1 \!+\!R_2 &\leq \IC{X_1}{Y_3}{V_1 X_3 Q} + \IC{V_1 X_2 V_3}{Y_2}{Q} 
-\frac{1}{2}\\
R_1 \!+\!R_2 &\leq \IC{X_1 V_2 X_3}{Y_1}{V_1 V_3 Q} + \IC{V_1 X_2 V_3}{Y_2}{V_2
Q}\\
R_1 \!+\!R_2 &\leq \IC{X_1 V_2 X_3}{Y_1}{Q} + \IC{X_2}{Y_2}{V_1 V_2 V_3 Q} 
-\frac{1}{2} \displaybreak[2]\\
2 R_1 \!+\!R_2 &\leq \IC{X_1 X_3}{Y_1}{V_1 V_2 V_3 Q} + \IC{X_1 V_2 X_3}{Y_1}{Q}
\nonumber\\
&\quad + \IC{V_1 X_2 V_3}{Y_2}{V_2 Q}\\
2 R_1 \!+\!R_2 &\leq \IC{X_1}{Y_3}{V_1 X_3 Q} + \IC{X_1 V_2 X_3}{Y_1}{Q}
\nonumber\\
&\quad + \IC{V_1 X_2 V_3}{Y_2}{V_2 Q} -\frac{1}{2}\\
R_1 \!+\!2 R_2 &\leq \IC{X_1 V_2 X_3}{Y_1}{V_1 V_3 Q} + \IC{X_2}{Y_2}{V_1 V_2
V_3 Q} \nonumber\\
&\quad + \IC{V_1 X_2 V_3}{Y_2}{Q}
\end{align}
\end{subequations}
These bounds look similar to the following subset of the outer
bound~\eqref{eq:OB-IS_IRC}: \eqref{eq:OB-IS_IRC1}--\eqref{eq:OB-IS_IRC7}, 
\eqref{eq:OB-IS_IRC12}, \eqref{eq:OB-IS_IRC14}, and~\eqref{eq:OB-IS_IRC18},
which allows us to compare them. However, as the PDs present in the inner
and outer bounds are different, we compare the expression of each bound in the
Gaussian case since they only depend on the SNRs of the links.

The reduced region~\eqref{eq:AP-CG-DF-A} for the Gaussian case is,
\begin{subequations}\label{eq:AP-CG-DF-B}
\begin{align}
R_1 &\leq \C{ \SNR{31} } \label{eq:AP-CG-DF-B1} \\
R_1 &\leq \C{ G_2(\SNR{11} +\SNR{13}) } \label{eq:AP-CG-DF-B2} \\
R_2 &\leq \C{ G_1\SNR{22} } -\frac{1}{2} \label{eq:AP-CG-DF-B3} \\
R_1 \!+\!R_2 &\leq \C{ G_2 \frac{\SNR{11} +\SNR{13} +\delta +
\SNR{11}\SNR{23}/(1+\SNR{21})}{ 1 +\SNR{21} +2\SNR{23} } } \nonumber\\
&\quad +\C{ \SNR{21} +\SNR{22} +\SNR{23} } +\frac{1}{2}\log_2 G_1 \\
R_1 \!+\!R_2 &\leq \C{ \frac{ \SNR{31} }{ 1 +\SNR{21} } } + \C{ \SNR{21}
+\SNR{22} +\SNR{23} } \nonumber\\
&\quad + \frac{1}{2}\log_2 G_1 -\frac{1}{2} \displaybreak[2]\\
R_1 \!+\!R_2 &\leq \C{ \SNR{12} +\frac{\SNR{11} +\SNR{13} +\delta
+\SNR{11}\SNR{23}/(1+\SNR{21})}{ 1 +\SNR{21} +2\SNR{23} } } \nonumber\\
 &\quad + \C{ \SNR{21} \!+\SNR{23} \!+\frac{\SNR{22}}{1 \!+\SNR{12}}} \!+\!
\frac{1}{2}\log_2 G_1G_2\\
R_1 \!+\!R_2 &\leq \C{ \SNR{11} +\SNR{12} +\SNR{13} } + \C{ G_1
\frac{\SNR{22}}{1 +\SNR{12}} } \nonumber\\
&\quad + \frac{1}{2}\log_2 G_2 -\frac{1}{2} \displaybreak[2]\\
2 R_1 \!+\!R_2 &\leq \C{ G_2 \frac{\SNR{11} +\SNR{13} +\delta
+\SNR{11}\SNR{23}/(1+\SNR{21})}{ 1 +\SNR{21} +2\SNR{23} } } \nonumber\\
&\quad + \C{ \SNR{11} +\SNR{12} +\SNR{13} } + \frac{1}{2}\log_2 G_1G_2
\nonumber\\
 &\quad + \C{ \SNR{21} +\SNR{23} +\frac{\SNR{22}}{1 +\SNR{12}}} \\
2 R_1 \!+\!R_2 &\leq \C{ \frac{ \SNR{31} }{ 1 +\SNR{21} } } + \C{ \SNR{11} 
+\SNR{12} +\SNR{13} } -\frac{1}{2} \nonumber\\
&\quad + \C{ \SNR{21} \!+\SNR{23} \!+\frac{\SNR{22}}{1 \!+\SNR{12}}} \!+\!
\frac{1}{2}\log_2 G_1G_2 \\
R_1 \!+\!2 R_2 &\leq \C{ \SNR{12} +\frac{\SNR{11} +\SNR{13} +\delta
+\SNR{11}\SNR{23}/(1+\SNR{21})}{ 1 +\SNR{21} +2\SNR{23} } } \nonumber\\
 &\quad + \C{ G_1 \frac{\SNR{22}}{1 +\SNR{12}} } + \C{ \SNR{21} +\SNR{22}
+\SNR{23} } \nonumber\\
 &\quad + \frac{1}{2}\log_2 G_1G_2,
\end{align}
\end{subequations}
where
\begin{align*}
\delta &\triangleq \left( \sqrt{\SNR{11} \SNR{23}} \pm \sqrt{\SNR{13}
\SNR{21}} \right)^2,\\
G_1 &\triangleq \frac{1 +2\SNR{21} +2\SNR{23} +\SNR{21}^2 +2\SNR{21}\SNR{23}}{1 
+3\SNR{21} +3\SNR{23} +2\SNR{21}^2 +4\SNR{21}\SNR{23}}, \\
G_2 &\triangleq \frac{1 +\SNR{12}}{1 +2\SNR{12}}.
\end{align*}

To illustrate the procedure for bounding the gap, we show the single-rate gaps
in the sequel. Consider,
\begin{align}
\Delta_{R_1} &= \eqref{eq:OB-Gaussian1} -\eqref{eq:AP-CG-DF-B1} \nonumber\\
 &= \C{ \SNR{11} + \SNR{31} } -\C{ \SNR{31} } \nonumber\\
 &= \C{ \frac{\SNR{11}}{ 1 +\SNR{31}} } 
 \leq \frac{1}{2}, \label{eq:AP-CG-DF-g1}
\end{align}
where the last inequality is due to $\SNR{31}\geq\SNR{11}$, otherwise, the gap
would be unbounded. Additionally,
\begin{align}
\Delta_{R_1} &= \eqref{eq:OB-Gaussian2} -\eqref{eq:AP-CG-DF-B2} \nonumber\\
 &= \C{ \SNR{11} +\SNR{13} } +\frac{1}{2} -\C{ G_2(\SNR{11} +\SNR{13}) }
\nonumber\\
 &\leq \frac{1}{2} -\frac{1}{2}\log_2 G_2 
 \leq 1, \label{eq:AP-CG-DF-g2}
\end{align}
where the last two inequalities are due to $\frac{1}{2}\leq G_2\leq 1$. For
$R_2$ we have,
\begin{align}
\Delta_{R_2} &= \eqref{eq:OB-Gaussian3} -\eqref{eq:AP-CG-DF-B3} \nonumber\\
 &= \C{ \SNR{22} } -\C{ G_1\SNR{22} } +\frac{1}{2} \nonumber\\
 &\leq \frac{1}{2} -\frac{1}{2}\log_2 G_1 
 \leq 1, \label{eq:AP-CG-DF-g3}
\end{align}
where the last two inequalities are due to $\frac{1}{2}\leq G_1\leq 1$. In
summary, if we compare the appropriate pair of bounds and we assume
$\SNR{31}\geq \SNR{11}$, we obtain the following gaps
\begin{align*}
\Delta_{R_1} &\leq \frac{1}{2}, & 
\Delta_{R_1 +R_2} &\leq 2, \\ 
\Delta_{R_1} &\leq 1, & 
\Delta_{R_1 +R_2} &\leq 2, \\ 
\Delta_{R_2} &\leq 1, & 
\Delta_{2 R_1 +R_2} &\leq 3, \displaybreak[2]\\ 
\Delta_{R_1 +R_2} &\leq 2, & 
\Delta_{2 R_1 +R_2} &\leq 3, \\ 
\Delta_{R_1 +R_2} &\leq 2, & 
\Delta_{R_1 +2 R_2} &\leq \frac{5}{2}. 
\end{align*}
Therefore, the gap between the outer bound and the full DF inner bound, when  
$\SNR{31}\geq \SNR{11}$, is $1$ bit per real dimension at most.

\section{Proof of Proposition~\ref{th:CG-pDF} (Partial DF Constant Gap)}
\label{sec:AP-Proof-CG-pDF}

The analysis of the gap for the partial DF scheme follows similar steps as for 
the full DF scheme. We enlarge the set of auxiliary random variables used in 
Appendix~\ref{sec:AP-Proof-CG-DF} with
\begin{equation}
U_1 = h_{31} X_1 + Z_3'. \label{eq:AP-CG-pDF-aux}
\end{equation}
Then, we reduce the achievable region using the assumptions of independence 
between $X_1$ and $X_3$ and the upper bound in the binning term, which gives us,
\begin{subequations}\label{eq:AP-CG-pDF-A}
\begin{align}
R_1 &\leq \IC{U_1}{Y_3}{X_3 Q} + \IC{X_1}{Y_1}{V_1 U_1 V_2 X_3 Q},
\label{eq:AP-CG-pDF-A1}\\
R_1 &\leq \IC{X_1 X_3}{Y_1}{V_2 Q}, \displaybreak[2]\\
R_2 &\leq \IC{X_2}{Y_2}{V_1 V_3 Q} -\frac{1}{2},\\
R_1 \!+\!R_2 &\leq \IC{X_1 X_3}{Y_1}{V_1 V_2 V_3 Q} \!+\! \IC{V_1 X_2
V_3}{Y_2}{Q}, \!\!\!\\
R_1 \!+\!R_2 &\leq \IC{U_1}{Y_3}{V_1 X_3 Q} + \IC{X_1}{Y_1}{V_1 U_1 V_2 X_3 Q}
\nonumber\\
 &\quad + \IC{V_1 X_2 V_3}{Y_2}{Q} -\frac{1}{2}, \label{eq:AP-CG-pDF-A5}\\
R_1 \!+\!R_2 &\leq \IC{X_1 V_2 X_3}{Y_1}{V_1 V_3 Q} \!+\! \IC{V_1 X_2
V_3}{Y_2}{V_2 Q}, \\
R_1 \!+\!R_2 &\leq \IC{U_1}{Y_3}{V_1 X_3 Q} + \IC{X_1 V_2}{Y_1}{V_1 U_1 X_3 Q}
\nonumber\\
 &\quad + \IC{V_1 X_2 V_3}{Y_2}{V_2 Q} -\frac{1}{2}, \label{eq:AP-CG-pDF-A7}\\
R_1 \!+\!R_2 &\leq \IC{X_1 V_2 X_3}{Y_1}{Q} + \IC{X_2}{Y_2}{V_1 V_2 V_3 Q} 
-\frac{1}{2}, \displaybreak[2]\\
R_1 \!+\!R_2 &\leq \IC{U_1}{Y_3}{X_3 Q} + \IC{X_1 V_2}{Y_1}{V_1 U_1 X_3 Q}
\nonumber\\
 &\quad + \IC{X_2}{Y_2}{V_1 V_2 V_3 Q}, \label{eq:AP-CG-pDF-A9}\\
\!\!\!\!
2 R_1 \!+\!R_2 &\leq \IC{X_1 X_3}{Y_1}{V_1 V_2 V_3 Q} + \IC{X_1 V_2 X_3}{Y_1}{Q}
\nonumber\\
 &\quad + \IC{V_1 X_2 V_3}{Y_2}{V_2 Q}, \displaybreak[2]\\
\!\!\!\!
2 R_1 \!+\!R_2 &\leq \IC{X_1 X_3}{Y_1}{V_1 V_2 V_3 Q} \!+\! \IC{X_1
V_2}{Y_1}{V_1 U_1 X_3 Q} \nonumber\\
 &\quad + \IC{U_1}{Y_3}{X_3 Q} + \IC{V_1 X_2 V_3}{Y_2}{V_2 Q},
\label{eq:AP-CG-pDF-A11} \displaybreak[2]\\
\!\!\!\!
2 R_1 \!+\!R_2 &\leq \IC{U_1}{Y_3}{V_1 X_3 Q} + \IC{X_1}{Y_1}{V_1 U_1 V_2 X_3 Q}
-\!\frac{1}{2} \nonumber\\
 &\quad +\! \IC{X_1 V_2 X_3}{Y_1}{Q} \!+\! \IC{V_1 X_2 V_3}{Y_2}{V_2 Q},
\label{eq:AP-CG-pDF-A12} \displaybreak[2]\\
\!\!\!\!
R_1 \!+\!2 R_2 &\leq \IC{X_1 V_2 X_3}{Y_1}{V_1 V_3 Q} + \IC{X_2}{Y_2}{V_1 V_2
V_3 Q} \nonumber\\
 &\quad + \IC{V_1 X_2 V_3}{Y_2}{Q}, \displaybreak[2]\\
\!\!\!\!
R_1 \!+\!2 R_2 &\leq \IC{U_1}{Y_3}{V_1 X_3 Q} + \IC{X_1 V_2}{Y_1}{V_1 U_1 X_3 Q}
-\!\frac{1}{2} \nonumber\\
 &\quad +\! \IC{X_2}{Y_2}{V_1 V_2 V_3 Q} \!+\! \IC{V_1 X_2 V_3}{Y_2}{Q}.
\label{eq:AP-CG-pDF-A14}
\end{align}
\end{subequations}
We can compare these bounds with a larger subset of the outer bound 
\eqref{eq:OB-IS_IRC}: \eqref{eq:OB-IS_IRC1}--\eqref{eq:OB-IS_IRC9}, 
\eqref{eq:OB-IS_IRC12}--\eqref{eq:OB-IS_IRC14}, and 
\eqref{eq:OB-IS_IRC18}--\eqref{eq:OB-IS_IRC19}.

Half of the bounds in~\eqref{eq:AP-CG-pDF-A} are the same as
in~\eqref{eq:AP-CG-DF-A}, while the other half --composed by the bounds 
\eqref{eq:AP-CG-pDF-A1}, \eqref{eq:AP-CG-pDF-A5}, \eqref{eq:AP-CG-pDF-A7}, 
\eqref{eq:AP-CG-pDF-A9}, \eqref{eq:AP-CG-pDF-A11}, \eqref{eq:AP-CG-pDF-A12},
and~\eqref{eq:AP-CG-pDF-A14}-- have the following new terms:
\begin{align*}
\IC{U_1}{Y_3}{X_3 Q} &= \C{ \SNR{31} } +\frac{1}{2}\log_2 G_{31}, \\
\IC{U_1}{Y_3}{V_1 X_3 Q} &= \C{ \frac{\SNR{31}}{1 +\SNR{21}} } 
+\frac{1}{2}\log_2 G_{32}, \\
\IC{X_1}{Y_1}{V_1 U_1 V_2 X_3 Q} &= \C{ G_2 \frac{\SNR{11}}{1 +\SNR{21} 
+\SNR{31}} },\\
\IC{X_1 V_2}{Y_1}{V_1 U_1 X_3 Q} &= \C{ \SNR{12} +\frac{\SNR{11}}{1 +\SNR{21} 
+\SNR{31}} } \nonumber\\
 &\quad + \frac{1}{2}\log_2 G_2 
\end{align*}
where
\begin{equation*}
G_{31} \triangleq \frac{1 +\SNR{31}}{1 +2\SNR{31}},\ \textnormal{ and }\
G_{32} \triangleq \frac{1 +\SNR{21} +\SNR{31}}{1 +\SNR{21} +2\SNR{31}}.
\end{equation*}

Let us analyze only one of the gaps that change,
\begin{align}
\Delta_{R_1} &= \eqref{eq:OB-Gaussian1} -\eqref{eq:AP-CG-pDF-A1} 
 = \C{ \SNR{11} + \SNR{31} } -\C{ \SNR{31} } \nonumber\\
&\quad -\frac{1}{2}\log_2 G_{31} -\C{ G_2 \frac{\SNR{11}}{1 +\SNR{21} +\SNR{31}}
} \nonumber\\
 &\leq \C{ \frac{\SNR{21}}{ 1 +\SNR{31}} } -\frac{1}{2}\log_2 G_{31}G_2
 \,\leq\, \frac{3}{2}, \label{eq:AP-CG-pDF-g1}
\end{align}
where the last inequality is due to $\SNR{31}\geq\SNR{21}$, otherwise, the gap
would be unbounded.

The gap between each pair of bounds in the inner and outer bound is,
\begin{align*}
\Delta_{R_1} &\leq \frac{3}{2}, & 
\Delta_{R_1 +R_2} &\leq 2, \\ 
\Delta_{R_1} &\leq 1, & 
\Delta_{R_1 +R_2} &\leq 2, \\ 
\Delta_{R_2} &\leq 1, & 
\Delta_{2 R_1 +R_2} &\leq 3, \displaybreak[2]\\ 
\Delta_{R_1 +R_2} &\leq 2, & 
\Delta_{2 R_1 +R_2} &\leq \frac{7}{2}, \displaybreak[2]\\ 
%
\Delta_{R_1 +R_2} &\leq \frac{5}{2}, & 
\Delta_{2 R_1 +R_2} &\leq \frac{7}{2}, \displaybreak[2]\\ 
%
\Delta_{R_1 +R_2} &\leq 2, & 
\Delta_{R_1 +2 R_2} &\leq \frac{5}{2}, \displaybreak[2]\\ 
\Delta_{R_1 +R_2} &\leq \frac{5}{2}, & 
\Delta_{R_1 +2 R_2} &\leq 3. 
\end{align*}
In the previous calculations we assumed that $\SNR{31} \geq \SNR{21}$.
Therefore, under this condition, the gap between the outer bound and the partial
DF inner bound is $1.5$ bits per real dimension at most.

\section{Proof of Proposition~\ref{th:CG-CF} (CF Constant Gap)}
\label{sec:AP-Proof-CG-CF}

In this section, we show the constant gap result for the CF inner bound. As 
with the previous two schemes, we propose some simplifications to help in the 
analysis which, at the same time, reduce the region. First, we only take the
region $\mathcal{R}_{\textrm{CF}_0}$~\eqref{eq:IB-CF-joint} into account. This
means that we force both end users to decode the compression index when we have
already stated in the proof of the scheme that sometimes is better to ignore
this message.

Second, the compressed channel observation of the relay is obtained by adding 
an independent Gaussian noise $Z\sim \mathcal{N}(0,N)$ to its channel output,
$$ \hat{Y}_3 = Y_3 + Z. $$

Third, the random variables used in the scheme have the following structure. 
Given the independent random variables $V_1$, $V_2$, $X_1'$, and $X_2'$, all 
distributed according to $\mathcal{N}(0,1)$, we construct $X_1$ and $X_2$ as 
follows:
\begin{align*}
X_1 &= \sqrt{\alpha_1 P_1}V_1 + \sqrt{\bar{\alpha}_1 P_1}X_1', \\
X_2 &= \sqrt{\alpha_2 P_2}V_2 + \sqrt{\bar{\alpha}_2 P_2}X_2'
\end{align*}
where $\alpha_i \in [0,1]$ and $\bar{\alpha}_i \triangleq 1- \alpha_i$.
Furthermore, inspired by~\cite{etkin_gaussian_2008} and taking into account the
presence of the relay's compressed channel output, we choose the fixed power
split strategy
\begin{align*}
\bar{\alpha}_1\left( 1 +\SNR{21} +\frac{\SNR{31}}{1+N} \right) &= 1, \\
\bar{\alpha}_2\left( 1 +\SNR{12} \right) &= 1.
\end{align*}

\begin{figure*}[!b]
\normalsize
\vspace*{-5pt}
\hrulefill
\begin{align*}
I_{11} &= \min\left\{ %
\frac{1}{2}\log_2\left\{ \frac{(1+N)(1 +\bar{\alpha}_1\SNR{11} 
+\bar{\alpha}_2\SNR{12}) + \bar{\alpha}_1\SNR{31}(1 
+\bar{\alpha}_2\SNR{12})}{(1+N)(1+\bar{\alpha}_2\SNR{12})} \right\}, 
\frac{1}{2}\log_2\left\{ \frac{N(1 +\bar{\alpha}_1\SNR{11} 
+\bar{\alpha}_2\SNR{12} +\SNR{13})}{(1+N) (1+\bar{\alpha}_2\SNR{12})} \right\} 
\right\}, \\
I_{12} &= \min\left\{ %
\frac{1}{2}\log_2\left\{ \frac{(1+N)(1 +\SNR{11} +\bar{\alpha}_2\SNR{12}) + 
\SNR{31}(1 +\bar{\alpha}_2\SNR{12})}{(1+N)(1+\bar{\alpha}_2\SNR{12})} \right\}, 
 \frac{1}{2}\log_2\left\{ \frac{N(1 +\SNR{11} +\bar{\alpha}_2\SNR{12}
+\SNR{13})}{(1+N) (1+\bar{\alpha}_2\SNR{12})} \right\} \right\}, \\
I_{13} &= \min\left\{ %
\frac{1}{2}\log_2\left\{ \frac{(1+N)(1 +\bar{\alpha}_1\SNR{11} +\SNR{12}) + 
\bar{\alpha}_1\SNR{31}(1 +\SNR{12})}{(1+N)(1+\bar{\alpha}_2\SNR{12})} \right\}, 
 \frac{1}{2}\log_2\left\{ \frac{N(1 +\bar{\alpha}_1\SNR{11} 
+\SNR{12} +\SNR{13})}{(1+N) (1+\bar{\alpha}_2\SNR{12})} \right\} \right\}, \\
I_{14} &= \min\left\{ %
\frac{1}{2}\log_2\left\{ \frac{(1+N)(1 +\SNR{11} +\SNR{12}) + \SNR{31}(1 
+\SNR{12})}{(1+N)(1+\bar{\alpha}_2\SNR{12})} \right\}, 
 \frac{1}{2}\log_2\left\{ \frac{N(1 +\SNR{11} +\SNR{12} +\SNR{13})}{(1+N)
(1+\bar{\alpha}_2\SNR{12})} \right\} \right\}, \\
I_{21} &= \min\left\{ %
\frac{1}{2}\log_2\left\{ \frac{(1+N)(1 +\bar{\alpha}_1\SNR{21} 
+\bar{\alpha}_2\SNR{22}) + \bar{\alpha}_1\SNR{31}(1
+\bar{\alpha}_2\SNR{22})}{(1+N)(1 
+\bar{\alpha}_1\SNR{21}) +\bar{\alpha}_1\SNR{31}} \right\}, 
 \frac{1}{2}\log_2\left\{ \frac{N(1
+\bar{\alpha}_1\SNR{21} 
+\bar{\alpha}_2\SNR{22} +\SNR{23})}{(1+N)(1 +\bar{\alpha}_1\SNR{21}) 
+\bar{\alpha}_1\SNR{31}} \right\} \right\}, \\
I_{22} &= \min\left\{ %
\frac{1}{2}\log_2\left\{ \frac{(1+N)(1 +\bar{\alpha}_1\SNR{21} +\SNR{22}) + 
\bar{\alpha}_1\SNR{31}(1 +\SNR{22})}{(1+N)(1 +\bar{\alpha}_1\SNR{21}) 
+\bar{\alpha}_1\SNR{31}} \right\}, 
 \frac{1}{2}\log_2\left\{ \frac{N(1 +\bar{\alpha}_1\SNR{21} 
+\SNR{22} +\SNR{23})}{(1+N)(1 +\bar{\alpha}_1\SNR{21}) +\bar{\alpha}_1\SNR{31}} 
\right\} \right\}, \\
I_{23} &= \min\left\{ %
\frac{1}{2}\log_2\left\{ \frac{(1+N)(1 +\SNR{21} +\bar{\alpha}_2\SNR{22}) + 
\SNR{31}(1 +\bar{\alpha}_2\SNR{22})}{(1+N)(1 +\bar{\alpha}_1\SNR{21}) 
+\bar{\alpha}_1\SNR{31}} \right\},
 \frac{1}{2}\log_2\left\{ \frac{N(1 +\SNR{21} +\bar{\alpha}_2\SNR{22}
+\SNR{23})}{(1+N)(1 +\bar{\alpha}_1\SNR{21}) +\bar{\alpha}_1\SNR{31}} \right\}
\right\},\displaybreak[2]\\
I_{24} &= \min\left\{ %
\frac{1}{2}\log_2\left\{ \frac{(1+N)(1 +\SNR{21} +\SNR{22}) + \SNR{31}(1 
+\SNR{22})}{(1+N)(1+\bar{\alpha}_1\SNR{21})+\bar{\alpha}_1\SNR{31}} \right\}, 
 \frac{1}{2}\log_2\left\{ \frac{N(1 +\SNR{21} +\SNR{22} 
+\SNR{23})}{(1+N)(1+\bar{\alpha}_1\SNR{21})+\bar{\alpha}_1\SNR{31}} \right\} 
\right\}.
\end{align*}
\end{figure*}

The expression of the bounds~\eqref{eq:IB-CF-join-def1} in the Gaussian case,
where we have assumed $N_3=1$ for simplicity, can be found at the bottom of
next page.

We start by calculating the gap for the single rate $R_1\leq I_{12a}$ with the 
bound~\eqref{eq:OB-IS_IRC1} from the outer bound:
\begin{subequations}
\begin{align}
\Delta_{R_1} &= \IC{X_1}{Y_1 Y_3}{X_2 X_3 Q} - \IC{X_1}{Y_1 \hat{Y}_3}{V_2 X_3 
Q} \nonumber\\
 &\leq \frac{1}{2} \log_2 \{ 1 +\SNR{11} +\SNR{31} \} \nonumber\\
&\quad -\frac{1}{2}\log_2\left\{ \frac{(1+N)(1 +\SNR{11}/2) + \SNR{31}}{1+N}
\right\} \label{eq:AP-CG-CF-A1}\\
 &= \frac{1}{2} \log_2\left\{ 1+ \frac{ (1+N)\SNR{11}/2 +N\SNR{31}}{(1+N)(1 
+\SNR{11}/2) + \SNR{31}} \right\} \displaybreak[2]\nonumber\\
 &\leq %
  \begin{cases}
  \frac{1}{2} + \C{ \frac{N}{1+N} } &\textnormal{if } \SNR{31}<\SNR{11}\\
  \log_2 \frac{3}{2} + \C{ N } &\textnormal{if } \SNR{31}\geq\SNR{11}
  \end{cases} \label{eq:AP-CG-CF-A2}
\end{align}
\end{subequations}
where in~\eqref{eq:AP-CG-CF-A1} we have reduced the expression of the inner
bound by adding $(1+N)\bar{\alpha}_2$ in the denominator and then, we apply the
fixed power split strategy; and~\eqref{eq:AP-CG-CF-A2} is obtained by
eliminating either $(1+N)(1 +\SNR{11}/2)$ or $\SNR{31}$ from the denominator and
taking into account that $\SNR{31}\lessgtr\SNR{11}$.

Next, we compare $R_1\leq I_{12b}$ with the bound~\eqref{eq:OB-IS_IRC2}:
\begin{subequations}
\begin{align}
\Delta_{R_1} &= \IC{X_1 X_3}{Y_1}{X_2 Q} - [ \IC{X_1 X_3}{Y_1}{V_2 Q} -I_1 ] 
\nonumber\\
 &\leq \frac{1}{2} \log_2 \{ 1 +\SNR{11} +\SNR{13} \} +\frac{1}{2} \nonumber\\
&\quad -\frac{1}{2}\log_2\left\{ \frac{ N(1 +\SNR{11} +\SNR{13})}{(1+N)(1
+\bar{\alpha}_2\SNR{12})} \right\} \displaybreak[2]\label{eq:AP-CG-CF-B1}\\
 &\leq \frac{1}{2} + \frac{1}{2} \log_2\left\{ \frac{2(1+N)}{N} \right\}
\nonumber\\
 &= 1 + \C{ \frac{1}{N} } \label{eq:AP-CG-CF-B2}
\end{align}
\end{subequations}
where in~\eqref{eq:AP-CG-CF-B1} we have already reduced the expression of the
inner bound by eliminating the term $\bar{\alpha}_2\SNR{12}$. If
$\SNR{31}<\SNR{11}$, the gap for $R_1$ is dominated
by~\eqref{eq:AP-CG-CF-B2}, since it is always greater
than~\eqref{eq:AP-CG-CF-A2}, otherwise, the gap is the maximum of both.

Upper bounds on the gap of single rates and sum-rates can be derived using the
expressions from the outer bound \eqref{eq:OB-IS_IRC1}--\eqref{eq:OB-IS_IRC3},
\eqref{eq:OB-IS_IRC6}--\eqref{eq:OB-IS_IRC11},
\eqref{eq:OB-IS_IRC14}--\eqref{eq:OB-IS_IRC17}, and
\eqref{eq:OB-IS_IRC19}--\eqref{eq:OB-IS_IRC20},
and the assumption $\SNR{31}<\SNR{21}$ is needed for the gap to be bounded.
These upper bounds on the gap were analyzed numerically, due to their
complexity, and after cumbersome calculations the largest gap comes from the
sum-rate:
\begin{align*}
\MoveEqLeft[1]
\Delta_{R_1 +R_2} \leq \min\{ \eqref{eq:OB-IS_IRC8}, \eqref{eq:OB-IS_IRC11} \}
-[\, I_{13}+I_{23}\, ] \\
 &\leq \max\{ \eqref{eq:OB-IS_IRC11} -[\, I_{13b}+I_{23a}\, ], \,
\eqref{eq:OB-IS_IRC8} -[\, I_{13b}+I_{23b}\, ] \} \\
 &\leq 1 \!+\C{ \frac{1}{N} } \!+ \max\left\{ \C{N} \!+\C{\frac{1+2N}{2+N}}\!,
\, 1\!+\C{ \frac{1}{N} } \right\}.
\end{align*}
The value of $N$ that minimizes this gap is $N\approx 1.81$, with the gap per
real dimension being approximately $1.32$ bits.

\section{Proof of Proposition~\ref{th:CG-NoRelay} (Limited Relaying Benefit)}
\label{sec:AP-Proof-CG-NoRelay}

Let us define $\mathcal{R}_{o'}(P_1)$ as the outer bound region composed by the
bounds \eqref{eq:OB-IS_IRC1}, \eqref{eq:OB-IS_IRC3},
\eqref{eq:OB-IS_IRC9}--\eqref{eq:OB-IS_IRC11}, \eqref{eq:OB-IS_IRC17}, and
\eqref{eq:OB-IS_IRC20}. This new outer bound is analogous to the outer bound
presented by Telatar and Tse~\cite{telatar_bounds_2007} with the addition of the
\emph{antenna} $Y_3$. If the quality of the source-to-relay link is really low,
this extra antenna does not provide much information and thus, both outer bounds
should be within a constant gap. Since the gap between Han-Kobayashi's inner
bound and Telatar-Tse's outer bound is half a bit, it follows that Han-Kobayashi
scheme is within a constant gap to our outer bound under the aforementioned
conditions.

We only show one of these gaps here, but all of them can be derived similarly.
The expression for~\eqref{eq:OB-IS_IRC10} in the Gaussian case, i.e.,
\eqref{eq:OB-Gaussian10}, is
\begin{align}
\MoveEqLeft[1]
(R_1+R_2)_{IS-IRC} \nonumber\\
&= \IC{X_1}{Y_1 Y_3}{\ve{V_1} X_2 X_3} + \IC{X_1 X_2}{Y_2 Y_3}{X_3} \nonumber\\
&\leq \C{ \frac{\SNR{11}+\SNR{31}}{1\!+\!\SNR{21}\!+\!\SNR{31}} } \!+\! \C{
\SNR{21} \!+\!\SNR{22}\!+\!\SNR{31}(1\!+\!\SNR{22}) }, \label{eq:AP-NoRelay-1}
\end{align}
while the analogous bound in Telatar-Tse's outer bound is
\begin{align}
\MoveEqLeft
(R_1+R_2)_{IC} &\nonumber\\
&= \IC{X_1}{Y_1}{V_1 X_2} + \I{X_1 X_2}{Y_2} \nonumber\\
&= \C{ \frac{\SNR{11}}{1+\SNR{21}} } + \C{ \SNR{21} +\SNR{22} }.
\label{eq:AP-NoRelay-2}
\end{align}
Then, we calculate the gap between~\eqref{eq:AP-NoRelay-1} and
\eqref{eq:AP-NoRelay-2}
\begin{align*}
\Delta_{ob} &= (R_1+R_2)_{IS-IRC} - (R_1+R_2)_{IC} \nonumber\\
&= \C{ \frac{2\SNR{31}}{1+\SNR{11}+\SNR{21}} }
-\C{ \frac{\SNR{31}}{1+\SNR{21}} } +\C{
\frac{\SNR{31}}{1+\frac{\SNR{21}}{1+\SNR{22}}} }
\nonumber\\
&\leq \C{ \frac{2\SNR{31}}{1+\SNR{11}+\SNR{21}} }
+\C{ \frac{\SNR{31}}{1+\frac{\SNR{21}}{1+\SNR{22}}} }.
\end{align*}
The gap in this sum-rate can be upper bounded by $1$ bit given that $\SNR{31}
\leq \SNR{11}$ and $\SNR{31} \leq \SNR{21}/(1+\SNR{22})$. Further analysis of
the other bounds assures that the gap between outer bounds is half a bit per
rate if $\SNR{31} \leq \SNR{11}/(1+\SNR{12})$ and $\SNR{31} \leq
\SNR{21}/(1+\SNR{22})$ hold. Therefore, the use of the relay can improve
the rate by at most $1$ bit per real dimension compared to the Han-Kobayashi
scheme without the relay.

\addtolength{\textheight}{-5.5cm}

%

\bibliographystyle{IEEEtran}
\bibliography{IEEEabrv,biblio}

\end{document}